\documentclass[usenatbib]{mn2e}
\usepackage[pdftex]{graphicx}
\usepackage{amssymb}
\usepackage{amsmath}
\usepackage{lscape}

\begin{document}

\title[SLUGGS: Stacked globular cluster spectra]{The SLUGGS survey: Globular cluster stellar population trends from weak absorption lines in stacked spectra}

\author[Usher et al.]{Christopher~Usher,$^1$\thanks{E-mail: cusher@astro.swin.edu.au} Duncan~A.~Forbes,$^{1}$ Jean~P.~Brodie,$^{2}$ Aaron~J.~Romanowsky,$^{2,3}$ \newauthor Jay~Strader,$^{4}$ Charlie~Conroy,$^{2}$ Caroline~Foster,$^{5}$ Nicola~Pastorello,$^{1}$ \newauthor Vincenzo~Pota$^{1,2}$ and Jacob~A.~Arnold$^{2}$ \\
$^1$Centre for Astrophysics and Supercomputing, Swinburne University of Technology, Hawthorn, VIC 3122, Australia\\
$^2$University of California Observatories, 1156 High Street, Santa Cruz, CA 95064, USA\\
$^3$Department of Physics and Astronomy, San Jos\'e State University, One Washington Square, San Jose, CA 95192, USA \\
$^4$Department of Physics and Astronomy, Michigan State University, East Lansing, Michigan 48824, USA \\
$^5$Australian Astronomical Observatory, PO Box 915, North Ryde, NSW 1670, Australia}
\maketitle

\begin{abstract}
As part of the SLUGGS survey, we stack 1137 Keck \textsc{deimos} spectra of globular clusters from 10 galaxies to study their stellar populations in detail.
The stacked spectra have median signal to noise ratios of $\sim 90$ \AA$^{-1}$.
Besides the calcium triplet, we study weaker sodium, magnesium, titanium and iron lines as well as the H$\alpha$ and higher order Paschen hydrogen lines. 
In general, the stacked spectra are consistent with old ages and a Milky Way-like initial mass function.
However, we see different metal line index strengths at fixed colour and magnitude, and differences in the calcium triplet--colour relation from galaxy to galaxy.
We interpret this as strong evidence for variations in the globular cluster colour--metallicity relation between galaxies.
Two possible explanations for the colour--metallicity relation variations are that the average ages of globular clusters vary from galaxy to galaxy or that the average abundances of light elements (i.e. He, C, N and O) differ between galaxies.
Stacking spectra by magnitude, we see that the colours become redder and metal line indices stronger with brighter magnitudes.
These trends are consistent with the previously reported `blue tilts' being mass--metallicity relations.
\end{abstract}


\section{Introduction}

Most studies \citep[e.g.][]{2006ApJ...639...95P} of globular clusters (GCs) beyond the Local Group use integrated optical colours to determine the GC metallicities.
Such studies typically assume that all GCs in all galaxies follow the same colour--metallicity relationship.
However, the stellar population properties of early-type galaxies scale with galaxy mass \citep[e.g.][]{2002MNRAS.330..547T, 2005ApJ...621..673T, 2006MNRAS.370.1106G, 2010MNRAS.408...97K} with age and metallicity increasing with galaxy mass.
The abundances of the $\alpha$-elements (except Ca), C, N and Na increase with stellar mass \citep{2012MNRAS.421.1908J, 2014ApJ...780...33C} while the initial mass function (IMF) becomes more bottom heavy (dominated by dwarf stars) with increasing galaxy mass \citep[e.g.][]{2012ApJ...760...71C, 2013MNRAS.433.3017L} in line with evidence from dynamical studies \citep[e.g.][]{2012Natur.484..485C}.

Since the spatial distribution \citep[e.g.][]{2011ApJS..197...33S, 2012MNRAS.425...66F} and kinematics \citep[e.g.][]{2013MNRAS.428..389P} of red GCs are observed to agree well with those of their host galaxy's starlight, the GC stellar population parameters are also expected to scale with galaxy mass.
The mean GC colour becomes redder with brighter galaxy luminosity, seeming to indicate that the average GC metallicity increases with galaxy mass \citep[e.g.][]{1991ApJ...379..157B, 2006ApJ...639...95P, 2006AJ....132.2333S}.
Since the ages, chemical abundances and IMF of a stellar population affect its integrated colours \citep[e.g.][]{2009ApJ...699..486C}, variations in the ages, chemical abundances or IMF of GCs between galaxies would lead to variations in the colour--metallicity relation between galaxies. 

There have been only a few studies comparing the GC colour--metallicity relation between galaxies.
\citet{2014MNRAS.437.1734V} compared the colours and metallicities of Milky Way GCs with GCs from M49 and M87, finding agreement between the three galaxies.
However, their comparison lacked Milky Way GCs more iron rich than [Fe/H] $= -0.3$ and had only a couple of M49 and M87 GCs more iron poor than [Fe/H] $= -1.5$.
In contrast, \citet{2011MNRAS.415.3393F} found a large difference in the GC relationship between colour and the metallicity sensitive Ca\,\textsc{ii} triplet index (CaT) between NGC 1407 and NGC 4494.
With a larger dataset of 11 galaxies, \citet{2012MNRAS.426.1475U} saw comparable differences in the GC colour--CaT relationship between galaxies.
Assuming that the CaT accurately traces metallicity this would indicate that the colour--metallicity relation varies from galaxy to galaxy.
Even if the colour--metallicity relation was universal, a varying CaT--metallicity relation would indicate varying GC stellar populations.
Additionally, several studies have found variations in the ultraviolet colour--metallicity relation, with GCs in M87 \citep{2006AJ....131..866S} showing bluer ultraviolet colours than Milky Way GCs, M31 GCs \citep{2007ApJS..173..643R} showing similar colours to the GCs of the Milky Way and GCs associated with the Sagittarius dwarf galaxy showing redder ultraviolet colours \citep{2012AJ....144..126D}.

GCs in most massive galaxies (but not all, i.e. NGC 4472, \citealt{2006AJ....132.2333S}) show a colour--magnitude relation known as the `blue tilt' where the blue GCs become redder at brighter magnitudes \citep{2006AJ....132.1593S, 2006AJ....132.2333S, 2006ApJ...636...90H}.
The evidence for the red GCs having a colour--magnitude relation is much weaker with \citet{2009ApJ...699..254H} finding no evidence for a `red tilt' while \citet{2010ApJ...710.1672M} found evidence for a weak red tilt only in the centres of high mass galaxies.
Since colour traces metallicity and absolute magnitude traces mass, the blue tilt has usually been explained as a mass--metallicity relationship where the more massive GCs were able to retain more of the metals from the first generation of their stars and self-enrich \citep{2008AJ....136.1828S, 2009ApJ...695.1082B}.
The population of stars responsible for this self-enrichment should leave a signature in the form of chemical abundance trends with magnitude.

All studied Milky Way GCs show star to star variations in their light element abundances \citep[see review by][]{2012A&ARv..20...50G}.
In each GC, significant numbers of stars show enhanced He, N, Na and Al and reduced O, C and Mg abundances compared to the remaining stars which show a `normal' $\alpha$-element enhanced abundance pattern.
There is some evidence \citep{2006AJ....131.1766C, 2010A&A...516A..55C} that the [O/Na] and [Mg/Al] ranges become wider at brighter GC luminosities, while \citet{2013ApJ...776L...7S} found that metal rich M31 GCs have higher [N/Fe] at higher masses.
These variations in light element are also commonly seen as split or broadened main sequences, sub-giant branches or red giant branches (RGBs) in high quality colour-magnitude diagrams \citep[e.g.][]{2013MNRAS.431.2126M}.
A few massive MW GCs including $\omega$ Cen \citep{1975ApJ...201L..71F, 1995ApJ...447..680N},  M22 \citep{2011A&A...532A...8M}, NGC 1851 \citep{2011A&A...533A..69C} and NGC 5824 \citep{2014MNRAS.438.3507D} show significant [Fe/H] dispersions.
Several models \citep[e.g.][]{2007A&A...464.1029D, 2010MNRAS.407..854D, 2011ApJ...726...36C} have been proposed to explain the observed abundance spreads, but a common feature is that GCs were able to self-enrich, with the more massive GCs being able to retain more of their metals.
What if any connection exists between the self-enrichment process that creates the blue tilt and the origin of the light element anti-correlations observed within GCs is unclear.
However, in $\omega$ Cen the Na-O anti-correlation is observed over a wide range of iron abundances \citep{2011ApJ...731...64M}, suggesting that the origin of the blue tilt and the light element anti-correlations do not share a common origin.

Here we stack large numbers of GC spectra observed with the Deep Imaging Multi-Object Spectrograph (\textsc{deimos}, \citealt{2003SPIE.4841.1657F}) on the Keck II telescope to reach sufficient signal to noise ratios (S/N) to study their stellar populations in greater detail.
A highly multiplexed spectrograph, \textsc{deimos} is most sensitive in the red where the old stellar populations of GCs are brightest.
The strongest spectral feature in the \textsc{deimos} wavelength range is the  CaT at 8498 \AA{}, 8542 \AA{} and 8662 \AA{} which has been used extensively to measure the metallicities of both resolved stars and integrated light (see \citealt{2012MNRAS.426.1475U} and references therein).
Recently interest in the red wavelength range has focused on using IMF sensitive features such as the Na\,\textsc{i} doublet at 8183 \AA{} and 8194 \AA{} and the CaT to study the low mass IMF in early-type galaxies \citep[e.g.][]{2012ApJ...760...71C, 2013MNRAS.433.3017L}.
Other important spectral features include the Mg\,\textsc{i} line at 8807 \AA{} and several TiO molecular bands \citep{2009MNRAS.396.1895C}.
The \textsc{deimos} spectra frequently cover H$\alpha$ as well as the higher order Paschen hydrogen lines.
In addition to studying the CaT strengths of GCs, \citet{2010AJ....139.1566F} studied the strengths of the Paschen lines and a handful of weak Fe and Ti in the vicinity of the CaT.
\citet{2008ApJ...682.1217K, 2011ApJ...727...78K, 2011ApJ...727...79K} have used \textsc{deimos} in a similar setup to ours to measure Fe and $\alpha$-element abundances of RGB stars in Local Group dwarf galaxies using weak metal lines.

In Section \ref{dataset} we describe our sample of galaxies and GCs.
In Section \ref{analysis} we describe our spectral stacking procedure and our measurement of spectral indices. 
In Section \ref{discussion} we compare our colour and spectral index measurements with single stellar population models and consider how these spectral indices and colours vary from galaxy to galaxy as well as how they vary with GC luminosity.
Finally, in Section \ref{conclusion} we give our conclusions.
Through out this paper we will use metallicity ([Z/H]) to refer to the total abundances of all metals.
This is distinct from the iron abundance ([Fe/H]), a distinction not always made by all authors.
Complicating any comparison between studies is the fact that different models  and calibrations are tied to different metallicity or iron abundance scales \citep[for example see][]{2012MNRAS.426.1475U}.

\section{Dataset}
\label{dataset} 

\begin{table}
  \caption{\label{tab:galaxies}Galaxy Properties}
  \begin{tabular}{c c c c c c}
    \hline 
    Galaxy   & $V_{sys}$     & $D$    & $M_{K}$ & Spectra & GCs \\ 
             & [km s$^{-1}$] & [Mpc]  & [mag]   &         &     \\ 
    (1)      & (2)           & (3)    & (4)     & (5)     & (6) \\ \hline
    NGC 1407 & 1779          & 26.8   & $-25.4$ & 223     & 203 \\
    NGC 2768 & 1353          & 21.8   & $-24.7$ & 52      & 36  \\
    NGC 3115 & 663           & 9.4    & $-24.0$ & 135     & 122 \\
    NGC 3377 & 690           & 10.9   & $-22.8$ & 92      & 84  \\
    NGC 4278 & 620           & 15.6   & $-23.8$ & 172     & 145 \\
    NGC 4365 & 1243          & 23.3   & $-25.2$ & 134     & 129 \\
    NGC 4473 & 2260          & 15.3   & $-23.8$ & 63      & 60  \\
    NGC 4494 & 1342          & 16.6   & $-24.1$ & 56      & 51  \\
    NGC 4649 & 1110          & 17.3   & $-25.5$ & 153     & 145 \\
    NGC 5846 & 1712          & 24.2   & $-25.0$ & 57      & 53  \\ \hline

  \end{tabular}
	
\medskip
\emph{Notes} Column (1): Galaxy name.
Column (2): Systemic velocity from \citet{2011MNRAS.413..813C}.
Column (3): Distance from \citet{2011MNRAS.413..813C} based on surface brightness fluctuations (SBF) of \citet{2001ApJ...546..681T}. For NGC 4365, NGC 4473 and NGC 4649, \citet{2011MNRAS.413..813C} used the SBF distance of \citet{2007ApJ...655..144M}.
Column (4): $K$ band absolute magnitude from \citet{2011MNRAS.413..813C} calculated from the 2MASS \citep{2006AJ....131.1163S} total $K$ band apparent magnitude and the distance in Column (4), corrected for foreground extinction.
Column (5): Total number of GC spectra.
Column (6): Number of unique GCs with spectra.
NGC 1407 and NGC 3115 are not part of the \citet{2011MNRAS.413..813C} sample so their radial velocities were taken from NED while their distances are from \citet{2001ApJ...546..681T} with the same -0.06 mag distance moduli offset as \citet{2011MNRAS.413..813C}.
We assume that NGC 1400 and NGC 1407 are part of the same group and use the mean of their distance moduli for NGC 1407.
Their $K$ band absolute magnitude is also calculated from the 2MASS total $K$ apparent magnitude and the distance in Column (4) in the same manner as \citet{2011MNRAS.413..813C}.
\end{table}

Here we study GC spectra from 10 galaxies that are a part of the SAGES Legacy Unifying Globulars and GalaxieS (SLUGGS) survey\footnote{http://sluggs.swin.edu.au} \citep{SLUGGS} sample.
The SLUGGS survey is an ongoing study of 25 massive, nearby, early-type galaxies and their GC systems using Keck \textsc{DEIMOS} spectroscopy and wide-field imaging from Suprime-Cam \citep{2002PASJ...54..833M} on the Subaru telescope.
Details of the galaxies in this study are presented in Table \ref{tab:galaxies}.
We used spectra and CaT strengths from \citet{2012MNRAS.426.1475U} for the 8 galaxies in \citet{2012MNRAS.426.1475U} with more than 40 GCs with CaT index measurements.
For these galaxies we use the radial velocities from \citet{2013MNRAS.428..389P}.
For NGC 4278 we supplemented this with spectra, CaT strengths and radial velocities from \citet{2013MNRAS.436.1172U}.
We add two other galaxies, NGC 4473 \citep[Pota et al. in prep.]{2013MNRAS.433..235P} and NGC 4649 (Pota et al. in prep.), for which we have large numbers of \textsc{deimos} spectra.

The spectra were all observed with \textsc{deimos} in multi-slit mode between 2006 and 2013 with the primary aim of measuring GC radial velocities.
Exposure times averaged two hours per slit mask.
All observations used a central wavelength of 7800 \AA{}, the 1200 line mm$^{-1}$ grating and 1 arcsec slits.
This setup yields a resolution of $\Delta \lambda \sim 1.5$ \AA{} and covers the Na\,\textsc{i} doublet to CaT wavelength region.
In roughly half of the slits H$\alpha$ is also covered.
The \textsc{deimos} data were reduced using the \textsc{idl} \textsc{spec2d} pipeline \citep{2013ApJS..208....5N, 2012ascl.soft03003C}.
We used the radial velocities measured by the sources of the spectra.
These were measured using the \textsc{IRAF}\footnote{\textsc{IRAF} is distributed by the National Optical Astronomy Observatory, which is operated by the Association of Universities for Research in Astronomy (AURA) under cooperative agreement with the National Science Foundation.} procedure \textsc{fxcor}.
Further details of the observations, data reduction and radial velocity measurements are given in \citet{2013MNRAS.428..389P}.

For spectra observed in January of 2013, we found wavelength calibration issues at the bluest wavelengths ($\lambda \sim 6500$ \AA) with offsets of up to $\sim$ 10 \AA{} between the observed wavelengths of sky emission lines and their true wavelengths.
We used the sky emission lines present in the background spectra to correct the wavelength issues of these spectra.
This resulted in 0.1 \AA{} RMS differences between the skylines and the corrected spectra.

For 6 of the 10 galaxies we used the $gi$ photometry from \citet{2012MNRAS.426.1475U} which is based on the Subaru Suprime-Cam, Hubble Space Telescope Wide-Field Planetary Camera 2 and Hubble Space Telescope Advanced Camera for Surveys (ACS) photometry in \citet{2013MNRAS.428..389P}.
For NGC 4278 we used the ACS and Suprime-Cam photometry presented in \citet{2013MNRAS.436.1172U} using equations A1 and A2 of \citet{2012MNRAS.426.1475U} to convert their $gz$ photometry into $gi$. 
For objects in NGC 2768 with ACS photometry of \citet{2014MNRAS.437..273K} we used equations A6 and A7 of \citet{2012MNRAS.426.1475U} to convert their $BI$ photometry into $gi$.
For the remaining GCs in NGC 2768 we used the Suprime-Cam $gi$ photometry of \citet{2014MNRAS.437..273K}.
For both NGC 4473 and NGC 4649 we used the Suprime-Cam $gi$ photometry from Pota et al. (in prep.).
For 7 of the NGC 4649 GCs without Suprime-Cam imaging we used ACS $gz$ photometry from \citet{2012ApJ...760...87S} using equations A1 and A2 of \citet{2012MNRAS.426.1475U} to convert their $gz$ photometry into $gi$.
For 2 NGC 4649 GCs without either Suprime-Cam or ACS photometry we used CFHT MegaCam photometry from Pota et al. (in prep.) and used the photometric transformations provided on the MegaPipe website\footnote{http://www3.cadc-ccda.hia-iha.nrc-cnrc.gc.ca/en/megapipe/docs/filt.html} to transform from the MegaCam filter system to the SDSS filter system used for the Suprime-Cam photometry.
Thus all the GCs in this study have original or transformed $gi$ photometry.

\section{Analysis}
\label{analysis}
For those spectra without CaT measurements from \citet{2012MNRAS.426.1475U} or \citet{2013MNRAS.436.1172U}, we used the method of \citet{2012MNRAS.426.1475U} to measure the strengths of the CaT index.
To reduce the effects of the strong sky line residuals found in the CaT spectral region, this method uses the technique of \citet{2010AJ....139.1566F} to mask the sky line regions before fitting a linear combination of stellar templates to the observed spectra using the \textsc{pPXF} pixel fitting code \citep{2004PASP..116..138C}.
The fitted spectra are then normalised and the strength of the CaT lines measured on the normalised spectra.
A Monte Carlo resampling technique is used to estimate 68 percent confidence intervals for each CaT measurement.
In \citet{2012MNRAS.426.1475U} when a single mask was observed on multiple nights, the spectra were combined from different nights into one spectrum before measuring the CaT.
In this work we considered the spectra from the same mask but different nights separately to allow telluric absorption to be corrected.

We excluded from our analysis those spectra with average S/N less than 8 \AA$^{-1}$ in the wavelength range of 8400 \AA{} to 8500 \AA{}.
Each spectrum was inspected and spectra with possible contamination from the host galaxy starlight or poor sky subtraction were excluded from our analysis.
Out of the 1028 GCs in this study we only have size information for only 391 (NGC 3115 from \citealt{2014AJ....148...32J}, NGC 4278 sizes from \citealt{2013MNRAS.436.1172U}, NGC 4365 from \citealt{2012MNRAS.420...37B}, NGC 4473 from \citealt{2009ApJS..180...54J}, NGC 4649 from \citealt{2012ApJ...760...87S}).
10 of these have half light radii larger than 10 pc, making them ultra compact dwarfs (UCD) according to the definition of \citet{2011AJ....142..199B}.
Given that size information is not available for most of the GCs in this study, we did not distinguish between GCs and UCDs in our analysis.
However, we did exclude the UCD M60-UCD1 \citep{2013ApJ...775L...6S} as it has a noticeably larger velocity dispersion (68 km s$^{-1}$) compared to the other GCs and UCDs in this study ($< 30$ km s$^{-1}$).  

\begin{figure*}
\begin{center}
\includegraphics[width=504pt]{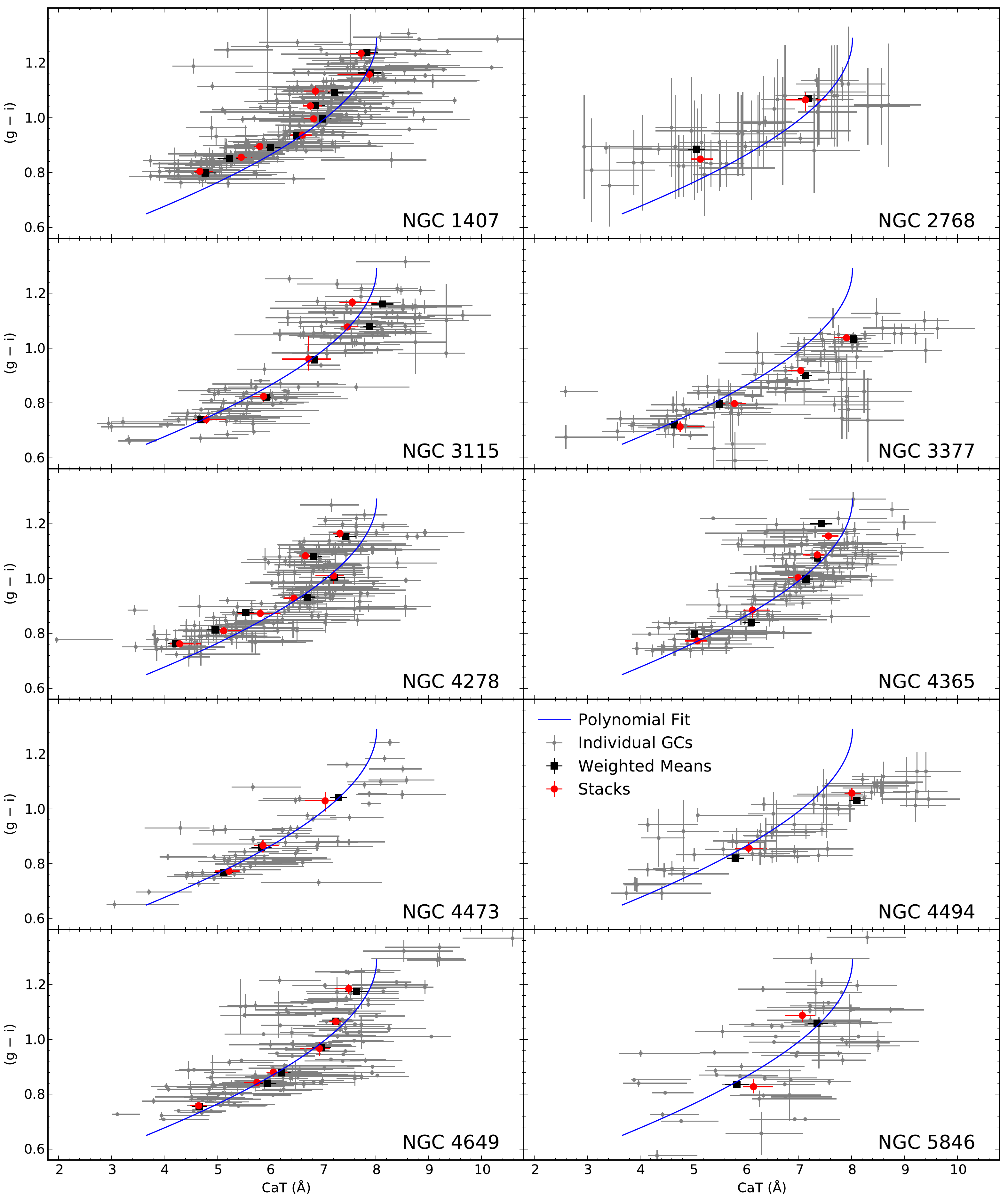}
\caption[CaT--colour diagrams for each galaxy]{CaT--colour diagrams for each of the ten galaxies in this study.
The grey points are the CaT measurements from individual spectra.
The black squares are the weighted means of the individual spectra using the same binning as used for stacking.
The red circles show the measurements from the stacked spectra.
The blue lines show a cubic polynomial fit to the individual spectra in all galaxies combined except NGC 2768 and NGC 5846. 
Note that the blue line is the same in all panels.
In general, the stacked spectra and the mean values agree.
Notice that the shape of the CaT--colour relation is near linear in NGC 3377 and NGC 4494 while in NGC 1407 and NGC 4278 the blue GCs have lower CaT strengths than predicted by the fitted polynomial.
The CaT-colour relation varies galaxy to galaxy.}
\label{fig:CaT_colour}
\end{center}
\end{figure*}

In Figure~\ref{fig:CaT_colour} we plot the observed colours and measured CaT values of the individual GC spectra.
To compare the relationship between colour and CaT between galaxies we fit a cubic polynomial between colour and CaT to all GCs except those in NGC 2768 and NGC 5846, which both suffer from lower quality photometry:
\begin{equation}\label{eq:CaT_colour}
\text{CaT} = -2.30 (g - i)^{3} - 3.32 (g - i)^{2} + 19.96 (g - i) - 7.28 .
\end{equation}
This fit shows good agreement with the observations in NGC 3115, NGC 4365, NGC 4473 and NGC 4649.
In NGC 1407 and NGC 4278 the fit predicts CaT values higher than observed at blue colours while in NGC 3377 and NGC 4494 the fit predicts CaT values lower than observed at red colours.
These results are the same as those found by \citet{2012MNRAS.426.1475U}.

\begin{figure}
\begin{center}
\includegraphics[width=240pt]{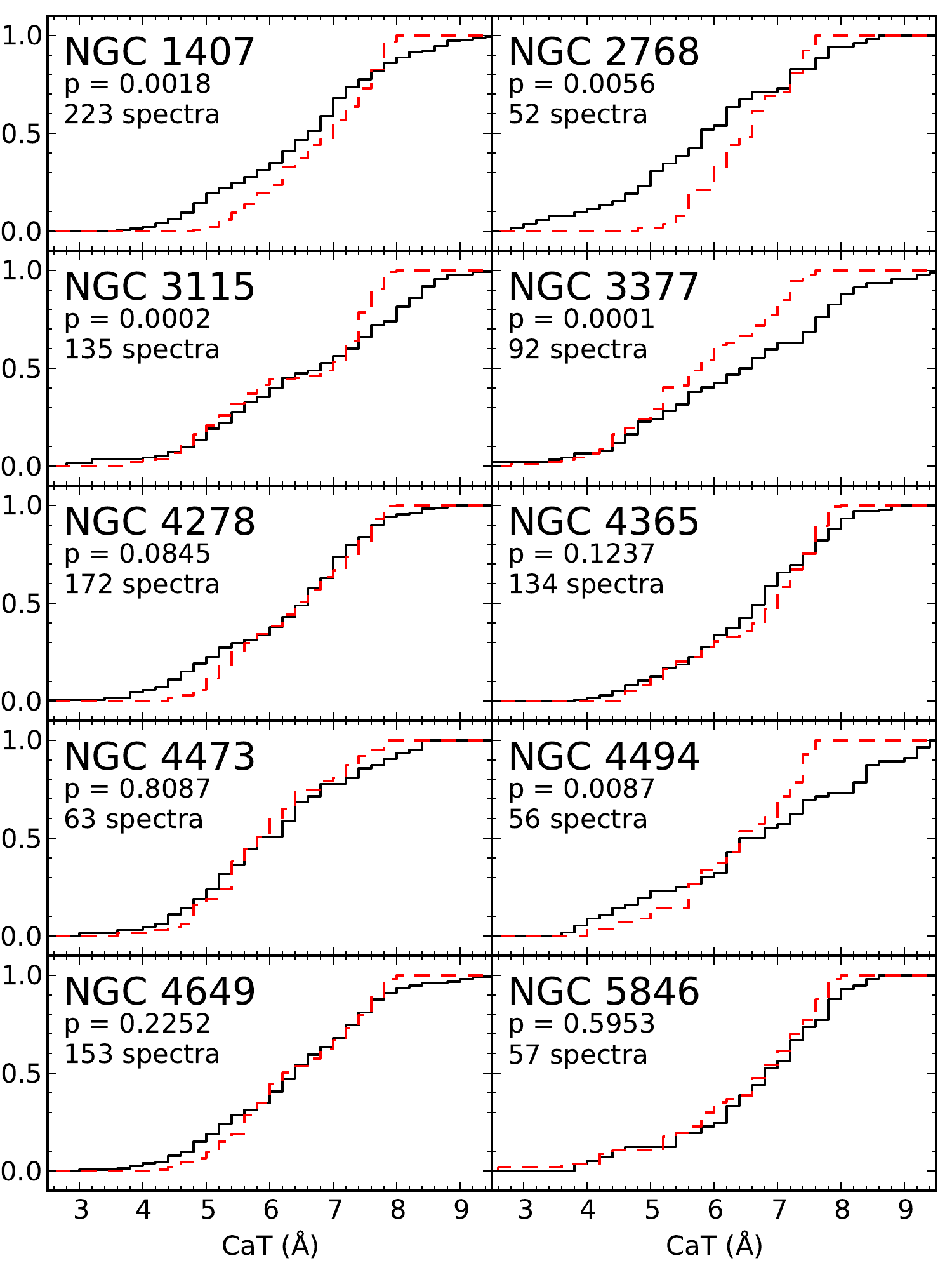}
\caption[Cumulative CaT histograms]{Cumulative CaT histograms for each galaxy.
The solid black lines are the measured CaT distributions while the dashed red lines show the CaT distributions predicted by the colour--CaT relation given in Equation \ref{eq:CaT_colour}.
The Kolmogorov-Smirnov test probability that these are both drawn from the same distribution is printed for each galaxy along with the number of spectra.
Although the CaT distributions of galaxies like NGC 4365 and NGC 4649 are well predicted by their colour distributions, in NGC 1407 the colours over-predict the CaT values for blue colours, while in NGC 3377 and NGC 4494 the colours under-predicts the CaT values for red colours.
This is further evidence that the CaT--colour relation varies galaxy to galaxy.}
\label{fig:CaT_colour_hist}
\end{center}
\end{figure}

In Figure~\ref{fig:CaT_colour_hist} we compare cumulative CaT histograms of the observed CaT distribution with the CaT distribution predicted from the colour distribution using Equation~\ref{eq:CaT_colour}.
We ran Kolmogorov-Smirnov tests to quantify the difference between the two CaT distributions.
Some galaxies such as NGC 4365 and NGC 4649 show agreement between the measured CaT values and those predicted from colours, while other galaxies show disagreement with the colours, predicting too few CaT weak GCs in NGC 1407 and too few CaT strong GCs in NGC 3377.
As in \citet{2012MNRAS.426.1475U}, we see evidence that the colour--CaT relation varies galaxy to galaxy.

\subsection{Stacking the spectra}
\label{stacking}
We studied two spectral regions.
The first, which covers from 8100 \AA{} to 8900 \AA{}, contains the CaT at 8498 \AA{}, 8542 \AA{} and 8662 \AA{}, the sodium doublet at 8183 \AA{} and 8195 \AA{} (Na82), the magnesium line at 8807 \AA{} and a whole host of weaker metal lines.
The second, which covers 6502 \AA{} to 6745 \AA{}, contains the H$\alpha$ line at 6563 \AA{}.
In the blue region we only used spectra with coverage blueward of 6540 \AA{} in the rest frame so that only spectra with observed H$\alpha$ lines were included.
Since the grating setup we used only covers H$\alpha$ in only approximately 60 percent of the slits (674 of 1137), the signal to noise ratios of the blue regions of the stacked spectra are lower than those of the red region of the stacks. 

The spectral region from 8128 \AA{} to 8348 \AA{}, which contains the sodium doublet, is affected by telluric absorption lines.
To remove the effect of these lines, we observed the spectroscopic standard star BD +28 4211 on 2012 October 20 using \textsc{deimos} in long-slit mode with the same spectroscopic setup as the GC spectra for a total of 250 s.
Likewise, the long slit data was also reduced using \textsc{idl} \textsc{spec2d} pipeline.
For each observed mask we stacked all GCs from that mask using the observed wavelengths rather than shifting them to the rest wavelengths.
We then scaled the normalised telluric absorption spectrum to match the telluric lines in the stacked spectra.
Lastly, we divided each observed spectrum by the scaled telluric spectrum to remove the effects of the telluric lines.
Unfortunately NGC 1407's radial velocity places the Na82 feature at the same observed wavelength as the strongest telluric absorption in this wavelength range.
For four masks in NGC 1407 with the strongest telluric absorption we were unable to satisfactorily remove the absorption.
The variances of these spectra in the affected wavelength range were multiplied by a hundred in order to down-weight them in the stacking procedure.
NGC 5846's radial velocity also places the Na82 feature in the same wavelength range as strong telluric lines but we did not down-weight its spectra as the strength of the telluric absorption was not as strong as in the NGC 1407 observations.

Before stacking, the spectra were shifted to the rest frame and re-dispersed to a common linear wavelength scale of 0.33 \AA{} per pixel.
We do not smooth the spectra to a common dispersion before stacking.
We continuum normalised each red spectrum by fitting and dividing by a linear combination of Chebyshev polynomials of order zero to eight while each blue spectrum was normalised using Chebyshev polynomials of order zero to five.
When fitting the polynomials we masked wavelength ranges identified from the \citet{2012MNRAS.424..157V} stellar population models as containing strong absorption lines.
Next we grouped the spectra and stacked the spectra in each group.
At each pixel the flux from each spectrum in a group was weighted by its variance and summed.
For each group we calculated a mean colour and a mean magnitude by using the pixel weights in the 8400 \AA{} to 8500 \AA{} range.

\begin{figure}
\begin{center}
\includegraphics[width=240pt]{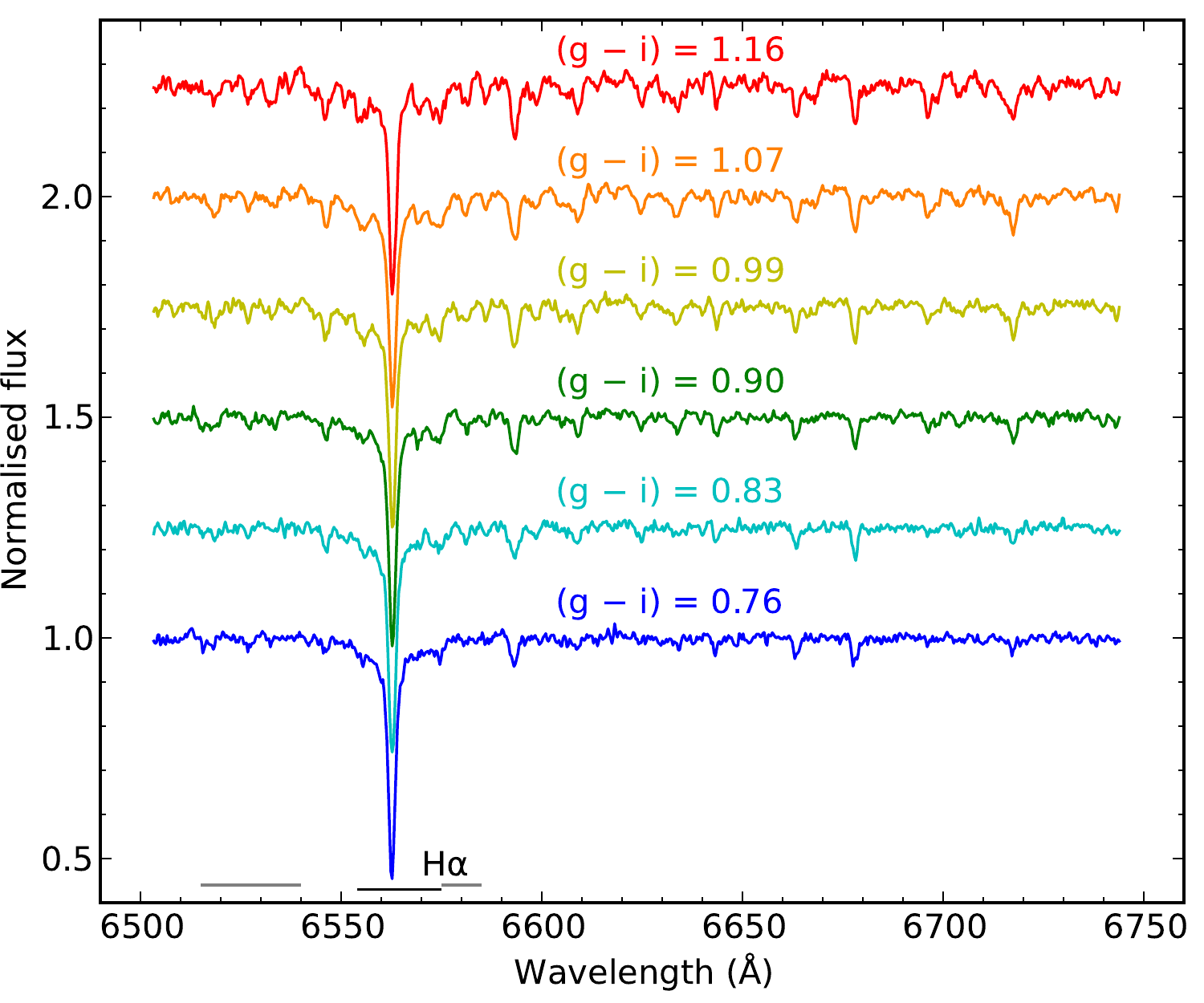}
\caption[Stacked spectra in the H$\alpha$ region]{Normalised and offset stacks of all 674 GC spectra in the H$\alpha$ region.
The spectra have been grouped and colour coded by their $(g - i)$ colour.
The horizontal lines show the definition of the H$\alpha$ index.
The black line at the bottom shows the feature passband while the grey lines show the pseudo-continuum passbands.
Most noticeable is the strengthening of the weak metal lines (e.g. at 6593 \AA{} and 6717 \AA{}) with redder colours.
The H$\alpha$ line itself becomes slightly weaker with redder colours.}
\label{fig:every_blue}
\end{center}
\end{figure}

\begin{figure*}
\begin{center}
\includegraphics[width=504pt]{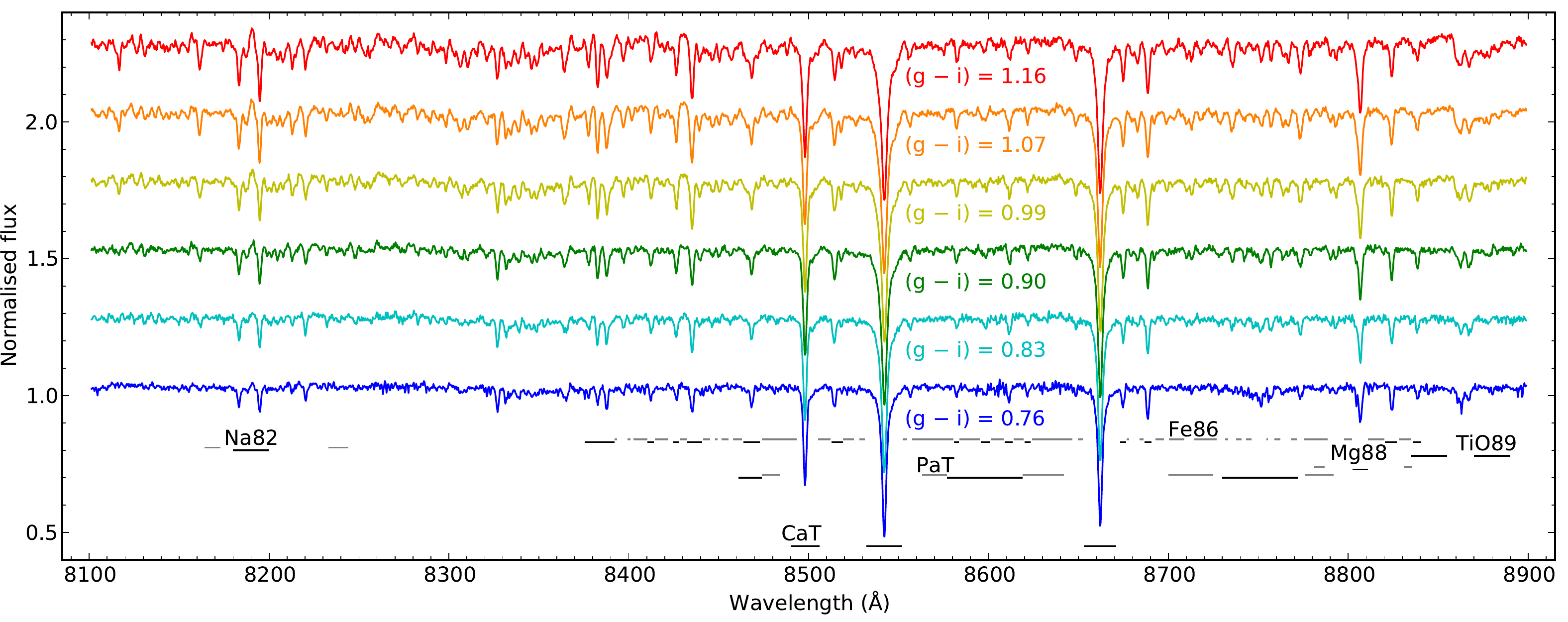}
\caption[Stacked spectra in the Na82 and CaT region]{Normalised and offset stacks of all 1137 GC spectra in the region around Na82 and CaT features.
The spectra have been grouped and colour coded by their $(g - i)$ colour.
The horizontal lines show the index definitions.
The black line at the bottom shows the feature passband while the grey lines show the pseudo-continuum passbands.
Going from the bluest spectra to the reddest the strengths of the metal lines increase.
At the reddest colours TiO molecular features (such as at $\sim 8190$, $\sim 8430$ and $\sim 8860$ \AA{}) appear.}
\label{fig:every_red}
\end{center}
\end{figure*}

First we stacked the spectra from all galaxies in six bins of equal size by colour as seen in Figures \ref{fig:every_blue} and \ref{fig:every_red}.
For the red spectral region, these stacks have S/N of 248 to 320 \AA$^{-1}$ while in the blue they have S/N of 159 to 239 \AA$^{-1}$.
Second, for each galaxy, the spectra were grouped by $(g - i)$ colour in bins of $\sim 25$ spectra.
For these stacks, the S/N in the red region is between 68 \AA$^{-1}$ and 155 \AA$^{-1}$ with a median of 94 \AA$^{-1}$ while in the blue it is between 32 \AA$^{-1}$ and 140 \AA$^{-1}$ with a median of 70 \AA$^{-1}$.

\subsection{Spectral feature measurements}
\label{indices}
To compare the stacked spectra in a quantitative manner we measured several spectral indices from the stacked spectra.
As noted by \citet{2012ApJ...747...69C}, spectral indices suffer from several shortcomings.
The strength of an absorption feature is measured relative to a pseudo-continuum.
The level of the pseudo-continuum is also affected by absorption lines so the variations in the measured index can be due to the feature of interest, the pseudo-continuum or both.
Spectral indices rarely contain absorption lines from just one species, complicating their interpretation.
Since a spectral index collapses all information about a feature into a single value, useful information from the shape of a spectral feature is lost. 
However, spectral indices do allow for spectra to be compared in a model independent manner.

We measured the strength of the CaT feature in the stacked spectra using the technique in \citet{2012MNRAS.426.1475U} (see above for more details).
As part of the CaT measurement process, the velocity dispersion of each spectrum is fitted.
We measured the strength of the Mg\,\textsc{i} line at 8807 \AA{} using the index definition of \citet{2009MNRAS.396.1895C} (here denoted Mg88).
We defined a weak metal line index, Fe86, consisting of Fe and Ti lines in the spectral region of the CaT.
The index was defined by identifying Fe and Ti lines that changed with metallicity in the \citet{2012MNRAS.424..157V} models in the CaT spectral region and by examining the stacks of all GCs plotted in Figure \ref{fig:every_red}.
A subset of these lines were studied by \citet{2010AJ....139.1566F}.
We used the H$\alpha_{A}$ index of \citet{2005ApJ...632..137N} to measure the strength of H$\alpha$ (here denoted H$\alpha$). 
To measure the strength of the Na\,\textsc{i} doublet at 8183 \AA{} and 8195 \AA{}, we used the Na\,\textsc{i}8200 index of \citet{2012MNRAS.424..157V} (here denoted Na82).
We used the PaT index of \citet{2001MNRAS.326..959C} to measure the strength of the hydrogen Paschen lines in the region of the CaT.
To measure the strength of the TiO band head at 8859 \AA{} we used the flux ratio defined by \citet{2012ApJ...747...69C} (here denoted TiO89).
Except for the CaT strengths and the TiO89 flux ratios, we used the method of \citet{2001MNRAS.326..959C} to calculate the index values.
The index definitions are given in Table \ref{tab:indices} and plotted in Figures \ref{fig:every_blue} and \ref{fig:every_red}.
These measurements are given in Table \ref{tab:by_colour} and will be discussed in Sections \ref{model_comparison}, \ref{sec:SFe} and \ref{with_galaxies}.

To estimate the uncertainty of the measured spectral features, as well as the colours and magnitudes of the stacks, we used bootstrapping.
For each galaxy, 256 random samples of the GC spectra from that galaxy were created with replacement with sizes equal to the number of GC spectra in that galaxy.
The stacking process and line index measurements were repeated for each of these samples.
For the colour, the magnitude and each of the measured spectral features, the distribution of that property in the bootstrap samples was smoothed with a Gaussian kernel.
For each of these distributions, a 68.3 \% confidence interval was calculated by integrating the likelihood function over all values where the likelihood function is larger than some constant value and varying the constant value until the integral of the likelihood equals 0.683.
The limits of the integration are then the confidence interval \citep{2010arXiv1009.2755A}.

\begin{table}
\caption{\label{tab:indices} Spectral index definitions}
\begin{tabular}{c c c c} \hline
Name        & Feature          & Pseudo-continuum  & Reference                        \\
            & passbands        & passbands         &                                  \\
            & \AA{}            & \AA{}             &                                  \\
(1)         & (2)              & (3)               & (4)                              \\ \hline
CaT         & 8490.0 -- 8506.0 & Polynomial        & U12 \\
            & 8532.0 -- 8552.0 & normalisation     &                                  \\
            & 8653.0 -- 8671.0 &                   &                                  \\
Mg88        & 8802.5 -- 8811.0 & 8781.0 -- 8787.9  & C09 \\
            &                  & 8831.0 -- 8835.5  &                                  \\
Fe86  & 8375.5 -- 8382.0 & 8392.0 -- 8393.5  & This                             \\
            & 8410.4 -- 8414.0 & 8399.4 -- 8400.9  & work                             \\
            & 8424.5 -- 8428.0 & 8402.7 -- 8410.3  &                                  \\
            & 8432.5 -- 8440.0 & 8414.5 -- 8422.1  &                                  \\
            & 8463.7 -- 8473.0 & 8428.6 -- 8432.3  &                                  \\
            & 8512.8 -- 8519.0 & 8441.4 -- 8445.2  &                                  \\
            & 8580.8 -- 8583.5 & 8447.9 -- 8449.4  &                                  \\
            & 8595.7 -- 8601.0 & 8451.5 -- 8455.4  &                                  \\
            & 8609.0 -- 8613.5 & 8458.0 -- 8463.0  &                                  \\
            & 8620.2 -- 8623.3 & 8474.0 -- 8493.3  &                                  \\
            & 8673.2 -- 8676.5 & 8505.3 -- 8512.1  &                                  \\
            & 8686.8 -- 8690.7 & 8519.2 -- 8525.2  &                                  \\
            & 8820.5 -- 8827.0 & 8528.3 -- 8531.3  &                                  \\
            & 8836.0 -- 8840.6 & 8552.3 -- 8554.9  &                                  \\
            &                  & 8557.5 -- 8580.4  &                                  \\
            &                  & 8583.9 -- 8595.3  &                                  \\
            &                  & 8601.2 -- 8608.4  &                                  \\
            &                  & 8613.9 -- 8619.4  &                                  \\
            &                  & 8624.3 -- 8646.6  &                                  \\
            &                  & 8649.8 -- 8652.5  &                                  \\
            &                  & 8676.9 -- 8678.1  &                                  \\
            &                  & 8684.0 -- 8686.1  &                                  \\
            &                  & 8692.7 -- 8697.6  &                                  \\
            &                  & 8700.3 -- 8708.9  &                                  \\
            &                  & 8714.5 -- 8726.8  &                                  \\
            &                  & 8731.5 -- 8733.2  &                                  \\
            &                  & 8737.6 -- 8740.8  &                                  \\
            &                  & 8743.3 -- 8746.1  &                                  \\
            &                  & 8754.5 -- 8755.4  &                                  \\
            &                  & 8759.0 -- 8762.2  &                                  \\
            &                  & 8768.0 -- 8771.5  &                                  \\
            &                  & 8775.5 -- 8788.7  &                                  \\
            &                  & 8797.6 -- 8802.2  &                                  \\
            &                  & 8811.0 -- 8820.0  &                                  \\
            &                  & 8828.0 -- 8835.0  &                                  \\
TiO89       &                  & 8835.0 -- 8855.0  & CvD12 \\
            &                  & 8870.0 -- 8890.0  &                                  \\
Na82        & 8180.0 -- 8200.0 & 8164.0 -- 8173.0  & V12 \\
            &                  & 8233.0 -- 8244.0  &                                  \\
H$\alpha$   & 6554.0 -- 6575.0 & 6515.0 -- 6540.0  & N05 \\
            &                  & 6575.0 -- 6585.0  &                                  \\
            &                  & 8776.0 -- 8792.0  &                                  \\
PaT         & 8461.0 -- 8474.0 & 8474.0 -- 8484.0  & C01 \\
            & 8577.0 -- 8619.0 & 8563.0 -- 8577.0  &                                  \\
            & 8730.0 -- 8772.0 & 8619.0 -- 8642.0  &                                  \\
            &                  & 8700.0 -- 8725.0  &                                  \\ \hline
\end{tabular}

\defcitealias{2012MNRAS.426.1475U}{U12}
\defcitealias{2009MNRAS.396.1895C}{C09}
\defcitealias{2005ApJ...632..137N}{N05}
\defcitealias{2012MNRAS.424..157V}{V12}
\defcitealias{2001MNRAS.326..959C}{C01}
\defcitealias{2012ApJ...747...69C}{CvD12}

\emph{Notes}
Column (1): Index name.
Column (2): Feature passbands in \AA.
Column (3): Pseudo-continuum passbands in \AA.
Column (4): Reference for index definition. U12 : \citet{2012MNRAS.426.1475U}, C09 : \citet{2009MNRAS.396.1895C}, CvD12 : \citet{2012ApJ...747...69C}, V12 : \citet{2012MNRAS.424..157V}, N05: \citet{2005ApJ...632..137N}, C01 : \citet{2001MNRAS.326..959C}.
\end{table}

In Figure \ref{fig:CaT_colour} we compare the colours and CaT values of the stacks with those of individual GC spectra.
We also calculated the variance weighted means of colour and CaT in the same bins as the stacks.
We see good agreement between the stack colours and CaT values and the means of the individual measurements in the same bin.

\begin{figure}
\begin{center}
\includegraphics[width=240pt]{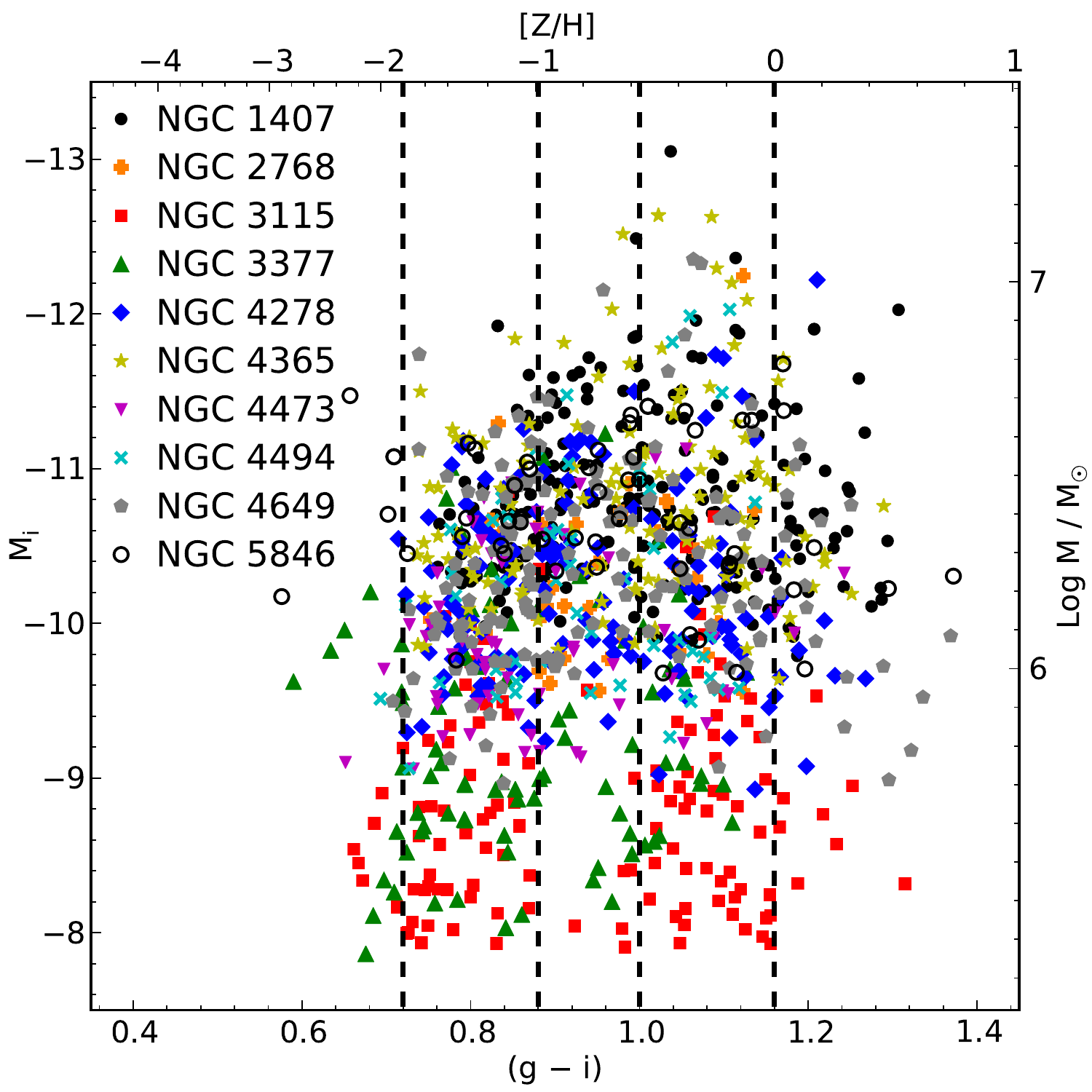}
\caption[Colour--magnitude diagram]{Colour--magnitude diagram for all the GC spectra used in this study.
$i$-band absolute magnitude is plotted versus $(g - i)$ colour.
The dashed vertical lines show the colour cuts used when stacking GCs by magnitude.
On the top axis the corresponding metallicities based on the colour--metallicity relation of \citet{2012MNRAS.426.1475U} (their equation 10) are shown.
For illustrative purposes, the mass corresponding to the absolute magnitude and a constant mass-to-light ratio of 2 is shown on the right.
More distant and luminous galaxies such as NGC 1407 and NGC 4365  contribute GCs predominantly brighter than $M_{i} \sim -10$, while closer galaxies such as NGC 3115 and NGC 3377 contribute fainter GCs.
Note that the GC selection varies galaxy to galaxy so that the colour distribution of GCs with CaT measurements may not be representative of the overall GC colour distribution within that galaxy.
A colour--magnitude relation can be seen for the blue GCs ($g - i < 0.94$, a `blue tilt') which is much stronger for GCs brighter than $M_{i} = -11$.
Almost all GCs brighter than $M_{i} = -12$ are red.}
\label{fig:cmd}
\end{center}
\end{figure}

\subsection{Trends with magnitude}
\label{magnitude}
As can be seen in the colour--magnitude diagram in Figure~\ref{fig:cmd}, the absolute magnitudes of the GCs in this study range from $M_{i} = -13.1$ to $M_{i} = -7.9$.
Assuming a mass-to-light ratio of 2 (see below) this corresponds to a mass range of $2.2 \times 10^{7}$ M$_{\odot}$ to $1.8 \times 10^{5}$ M$_{\odot}$.
The bright GCs tend to come from different galaxies than the faint GCs.
This makes it difficult to tell if any observed trends with magnitude are really due to magnitude or are due to differences between galaxies (i.e. the variations in colour--CaT relations seen between galaxies, Figure \ref{fig:CaT_colour}).
To attempt to disentangle galaxy to galaxy variation from trends with luminosity, we split the galaxies into two groups by the shape of their colour-CaT relation and stacked the spectra by magnitude within each group.
We included NGC 3115, NGC 3377, NGC 4473 and NGC 4494 in one group and NGC 1407, NGC 4278, NGC 4365 and NGC 4649 in another.
The galaxies in the first group show bluer colours at low and at high CaT strengths compared with the second group (see Figure \ref{fig:CaT_colour}).
Since both the GC and galaxy luminosities are higher on average in the second group than in the first, we will refer to the groups as the faint and bright groups respectively.

We combined GC spectra from each group in two colour ranges: blue with colours in the range $0.72 < (g - i) < 0.88$ and red with colours in the range $1.00 < (g - i) < 1.16$.
The limits of the colour ranges can be seen in the Figure \ref{fig:cmd}.
We stacked the faint group spectra in four bins by magnitude and the bright group in six bins by magnitude.
The S/N of faint group stacks ranged from 67 \AA$^{-1}$ to 198 \AA$^{-1}$ with a median of 97 \AA{}$^{-1}$ in red and from 44 \AA$^{-1}$ to 168 \AA$^{-1}$ with a median of 73 \AA$^{-1}$ in the blue while the bright group stacks have a red S/N between 69 \AA$^{-1}$ and 244 \AA$^{-1}$ with a median of 99 \AA{}$^{-1}$ and a blue S/N between 39 \AA$^{-1}$ and 161 \AA$^{-1}$ with a median of 76 \AA$^{-1}$.
We measured the colours, magnitudes and line indices strengths as before.
1024 bootstrap iterations were used to estimate the uncertainties on these measurements.
The measured colours, magnitudes and spectral indices in Table \ref{tab:by_mag}.
Using Equation \ref{eq:CaT_trans} the mean CaT value of the blue stacks corresponds to [Z/H] $= -1.21$ while the CaT value of the red stacks corresponds to [Z/H] $= -0.45$.

We used the \citet{2009ApJ...699..486C} simple stellar population models with an age of 12.6 Gyr and a \citet{2001MNRAS.322..231K} IMF to estimate the $i$-band mass-to-light ratio.
The model mass-to-light ratio increases from 1.8 at [Fe/H] $= -2.0$ to 2.3 at [Fe/H] $= -0.5$ before rising sharply to 3.7 at [Fe/H] $= 0.2$.
Beyond age and metallicity, the IMF and the horizontal branch morphology also affects the mass-to-light ratios \citep{2009ApJ...699..486C}.
In addition to stellar population effects, dynamical evolution preferentially removes low mass stars from GCs \citep[e.g.][]{1969A&A.....2..151H, 2003MNRAS.340..227B}.
The relative fraction of low mass stars lost is dependent on the mass of the GC and on the IMF \citep{2014ApJ...780...43G}.
\citet{2011AJ....142....8S} found that although the mass-to-light ratios of metal poor GCs in M31 are in agreement with the predictions of simple stellar population models with a \citet{2001MNRAS.322..231K} IMF from \citet{2009ApJ...699..486C}, the metal rich GCs in M31 have lower than predicted mass-to-light ratios.
Due to the uncertainty regarding the correct IMF to use, for illustrative proposes we adopt a mass-to-light ratio of 2 for all metallicities and study trends with luminosity not mass.
The stacks range in luminosity from $4.7 \times 10^{6}$ L$_{\odot}$ to $1.5 \times 10^{5}$ L$_{\odot}$.

\begin{figure}
\begin{center}
\includegraphics[width=240pt]{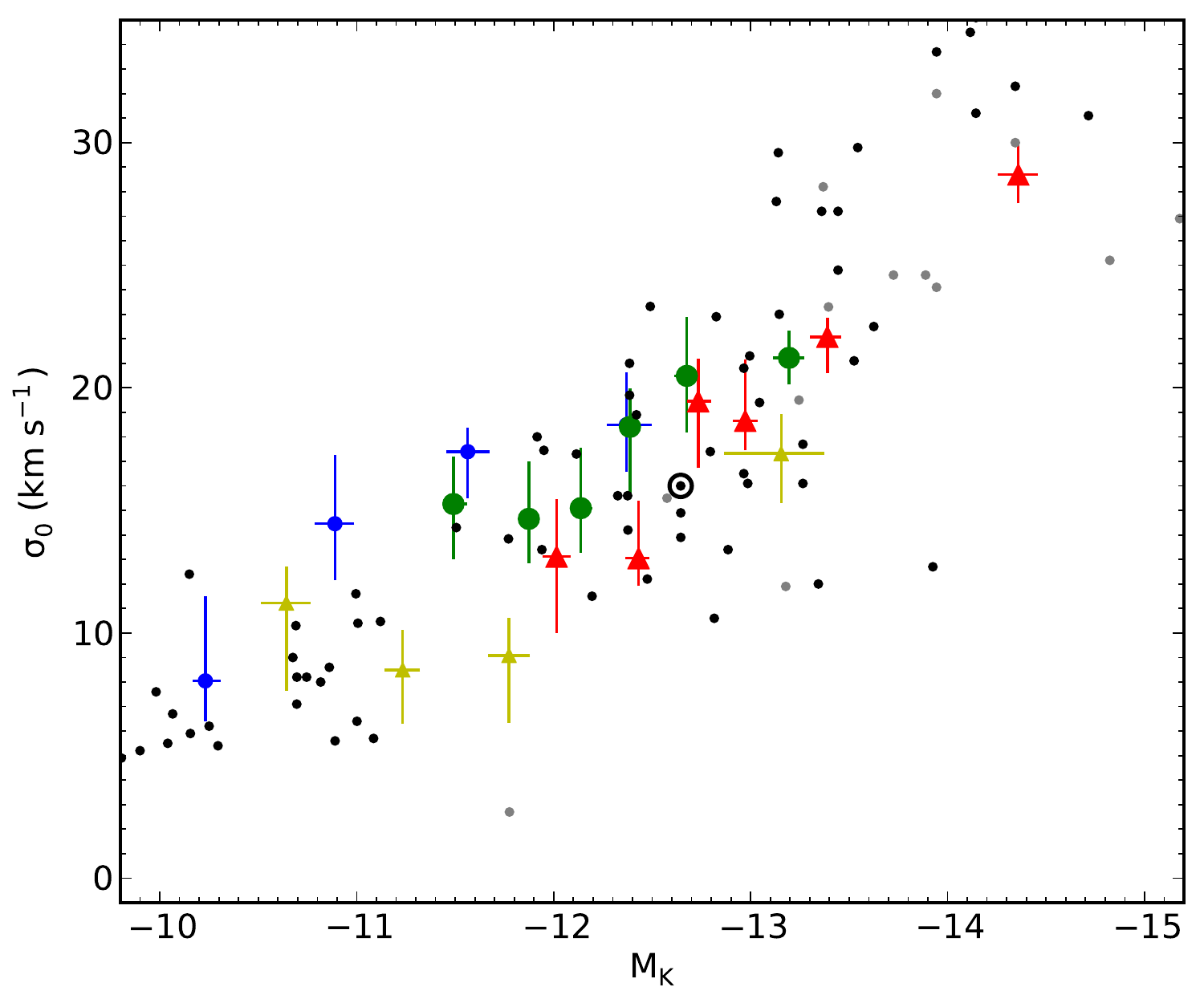}
\caption[Velocity dispersion versus absolute magnitude]{Central velocity dispersion versus absolute $K$-band magnitude.
The stacks of blue GC spectra in the faint galaxy group are plotted as small blue circles, blue GC spectra in the bright galaxy group as large green circles, red GC spectra in the faint group as yellow small triangles and red GC spectra in the bright group as red large triangles.
Black points are GCs from \citet{2008MNRAS.389.1924F} (half light radii $< 10$ pc) while grey points are UCDs (half light radii $> 10$ pc) from the same source.
The circled point is $\omega$ Centauri.
Equation \ref{eq:i-K} was used to transform the stacks' mean $i$-band magnitudes into $K$-band magnitudes.
In line with \citet{2008MNRAS.389.1924F}, the velocity dispersions measured from the stacked spectra was increased by 10 percent to estimate the central velocity dispersion.
The stacks' velocity dispersion trends agree with that from \citet{2008MNRAS.389.1924F}, suggesting that our stacking procedure is reliable.
The data suggest that metal poor (blue) stacks have higher velocity dispersions at fixed magnitude in line with the results of \citet{2011AJ....142....8S} who found that metal poor GCs have significantly higher $K$-band mass-to-light ratios compared to metal rich GCs.}
\label{fig:stacking_sigma}
\end{center}
\end{figure}

As a check on the stacking procedure, in Figure \ref{fig:stacking_sigma} we compare the velocity dispersions measured from the stacked spectra by \textsc{pPXF} with the \citet{2008MNRAS.389.1924F} compilation of GC and UCD kinematics.
To convert our $i$-band absolute magnitudes into $K$-band magnitudes, we use the \citet{2009ApJ...699..486C} simple stellar population models with an age of 12.6 Gyr and a \citet{2001MNRAS.322..231K} IMF to find the following relation between $(i - K)$ and $(g - i)$:
\begin{equation}\label{eq:i-K}
(i - K) = 1.364 \times (g - i) + 0.756 .
\end{equation}
This relation has an RMS of 0.034 mag. 
Since the GCs in this study are unresolved, we follow \citet{2008MNRAS.389.1924F} and multiply the measured velocity dispersions by 10 percent to estimate the central velocity dispersions.
As can be seen in Figure \ref{fig:stacking_sigma}, the velocity dispersion trends of the stacks agree with the trend for GCs and UCDs from \citet{2008MNRAS.389.1924F}, suggesting that our stacking procedure is reliable.
The metal poor stacked spectra appear to have higher velocity dispersions at fixed absolute magnitudes compared to the metal rich stacks.
This is in agreement with the results of \citet{2011AJ....142....8S} who found that metal poor GCs have significantly higher $K$-band mass-to-light ratios compared to metal rich GCs.
We plan to further investigate GC velocity dispersions and mass-to-light ratios in a future paper.

To investigate trends with magnitude we fit linear relations between the absolute magnitude and the various spectral features as well as the colours.
We used 1024 bootstrap iterations to estimate 68 \% confidence intervals for the slope and intercept as well as the probability that the slope is non-zero.
We list the fitted slopes and intercepts as well as the probability that the fitted slope is less than zero in Table \ref{tab:mag_relations}.
We use the bootstrap probability that the fitted slope is less than zero being greater than 0.95 for negative slopes or less than 0.05 for positive slopes as criteria for a significant relation.
Significant relations are bold in Table \ref{tab:mag_relations}.

To allow the two galaxy groups to be compared at fixed luminosity, we used the fitted relations with magnitude to calculate the value of each spectral feature and colour at $M_{i} = 10$. 
We also used the bootstrap samples to calculate the probability that difference between each feature at this absolute magnitude is greater than zero.
As before we consider probabilities above 0.95 for positive differences and below 0.05 for negative differences to be significant.
The values of each feature at $M_{i} = 10$ and the probabilities that the differences between the two groups are greater than zero is given in \ref{tab:mag_relations}.
We will discuss the colours and index strengths of the GC spectra stacked by magnitude in Sections \ref{with_galaxies} and \ref{with_lumin}.

\section{Discussion}
\label{discussion}
\subsection{Comparison with stellar population models}
\label{model_comparison}
\begin{figure*}
\begin{center}
\includegraphics[width=504pt]{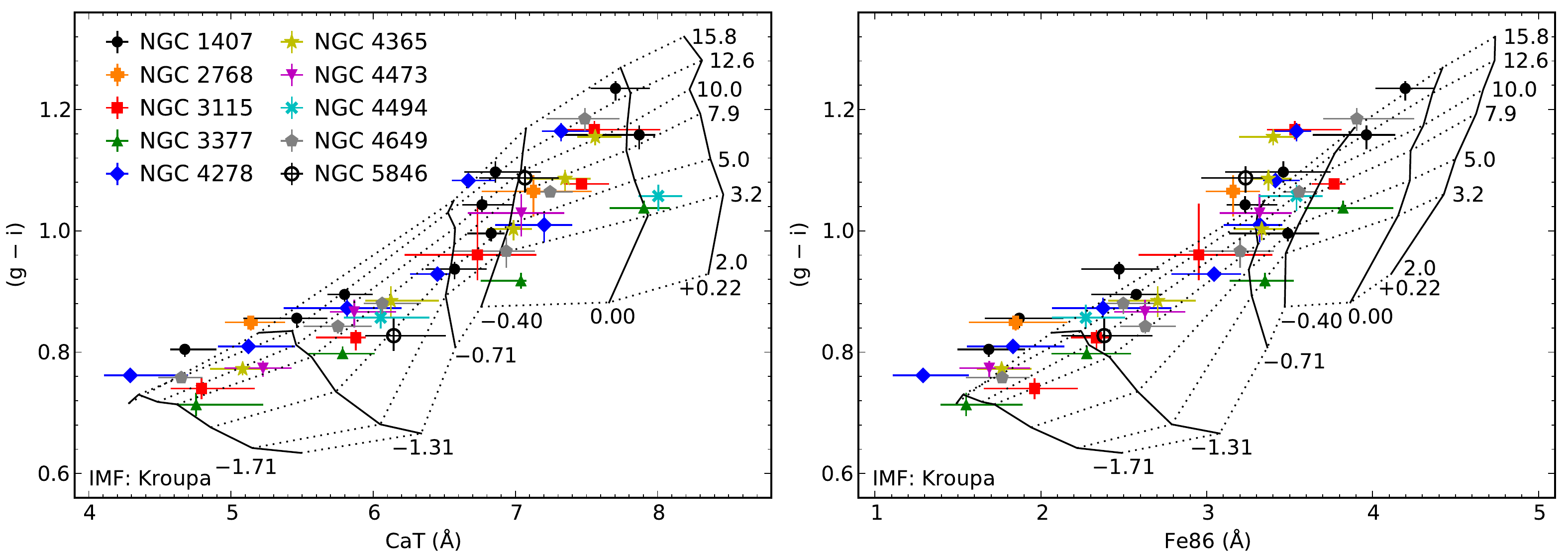}
\caption{Colour versus metal line indices for GC spectra stacked by colour.
On the left, $(g - i)$ colour is plotted as a function of the CaT index, while on the right, $(g - i )$ colour is plotted as a function of the Fe86 index.
In both panels the \citet{2012MNRAS.424..157V} model predictions for a \citet{2001MNRAS.322..231K} IMF and varying age and metallicity are over-plotted.
For each model grid, solid lines are lines of constant metallicity; dotted lines are constant age.
The points are colour coded by their host galaxy using the same colours and shapes as in Figure \ref{fig:cmd}.
Assuming universally old ages, the colour--CaT plot shows poor agreement between the observations and the models with the models predicting redder colours than observed at intermediate to strong CaT values.
However, the colour--Fe86 plot shows good agreement between the models and the stacked spectra.
At high CaT values, NGC 4494 is an outlier in the colour--CaT plot but is consistent with the other galaxies in the colour--Fe86 plot.}
\label{fig:stacking_CaT_Fe}
\end{center}
\end{figure*}

\begin{figure*}
\begin{center}
\includegraphics[width=504pt]{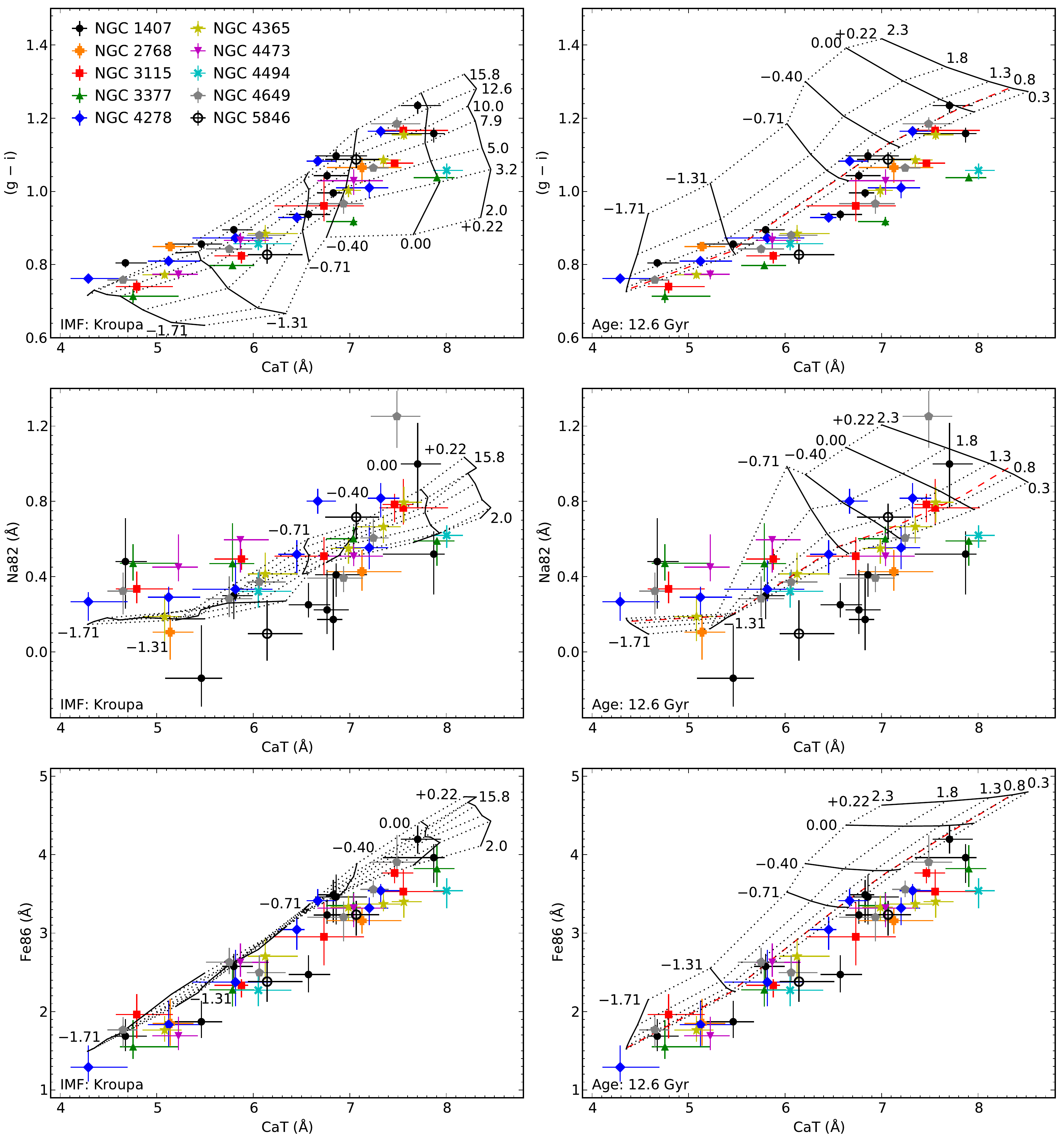}
\caption[Spectra stacked by colour]{Colours and line indices for GC spectra stacked by colour.
In the left column the \citet{2012MNRAS.424..157V} model predictions for a \citet{2001MNRAS.322..231K} IMF and varying age and metallicity are over-plotted.
In the right column the \citet{2012MNRAS.424..157V} model predictions for varying IMF slope and metallicity at a fixed age of 12.6 Gyr are over-plotted in black.
A slope of 1.3 corresponds to a \citet{1955ApJ...121..161S} IMF.
The 12.6 Gyr \citet{2001MNRAS.322..231K} IMF model is plotted with a red dashed line.
For each model grid, solid lines are lines of constant metallicity; dotted lines are constant age or IMF slope.
\emph{Top} $(g - i)$ colour as a function of the CaT index.
At low CaT values NGC 3115, NGC 3377 and NGC 4473 show bluer colours than the other galaxies while NGC 3377 and NGC 4494 show bluer colours at high CaT values.
Assuming universally old ages, the \citet{2012MNRAS.424..157V} models do not predict the observed relation between CaT and colour.
The colours and CaT values do suggest a bottom light IMF.
\emph{Middle} The Na82 sodium index as a function of the CaT index.
The Na82 values are likely affected by systematics due in the removal of telluric absorption lines.
At low CaT values NGC 3115, NGC 3377 and NGC 4473 show higher Na82 strengths than other galaxies  and while at high CaT values both NGC 3377 and NGC 4494 appear to have lower Na82 valves.
At high metallicities the Na82 and the CaT respond in opposite directions to the slope of the IMF.
However, the Na82 is affected by the Na abundance which is known to be non-solar in GCs.
At lower metallicities the \citet{2012MNRAS.424..157V} models are less reliable.
\emph{Bottom} The Fe86 weak metal line index as a function of the CaT index.
The similar relations between the indices for different galaxies suggest that both indices are measuring metallicity.
Unlike the CaT, Fe86 is insensitive to the IMF at high metallicities.}
\label{fig:stacking_CaT1}
\end{center}
\end{figure*}

\begin{figure*}
\begin{center}
\includegraphics[width=504pt]{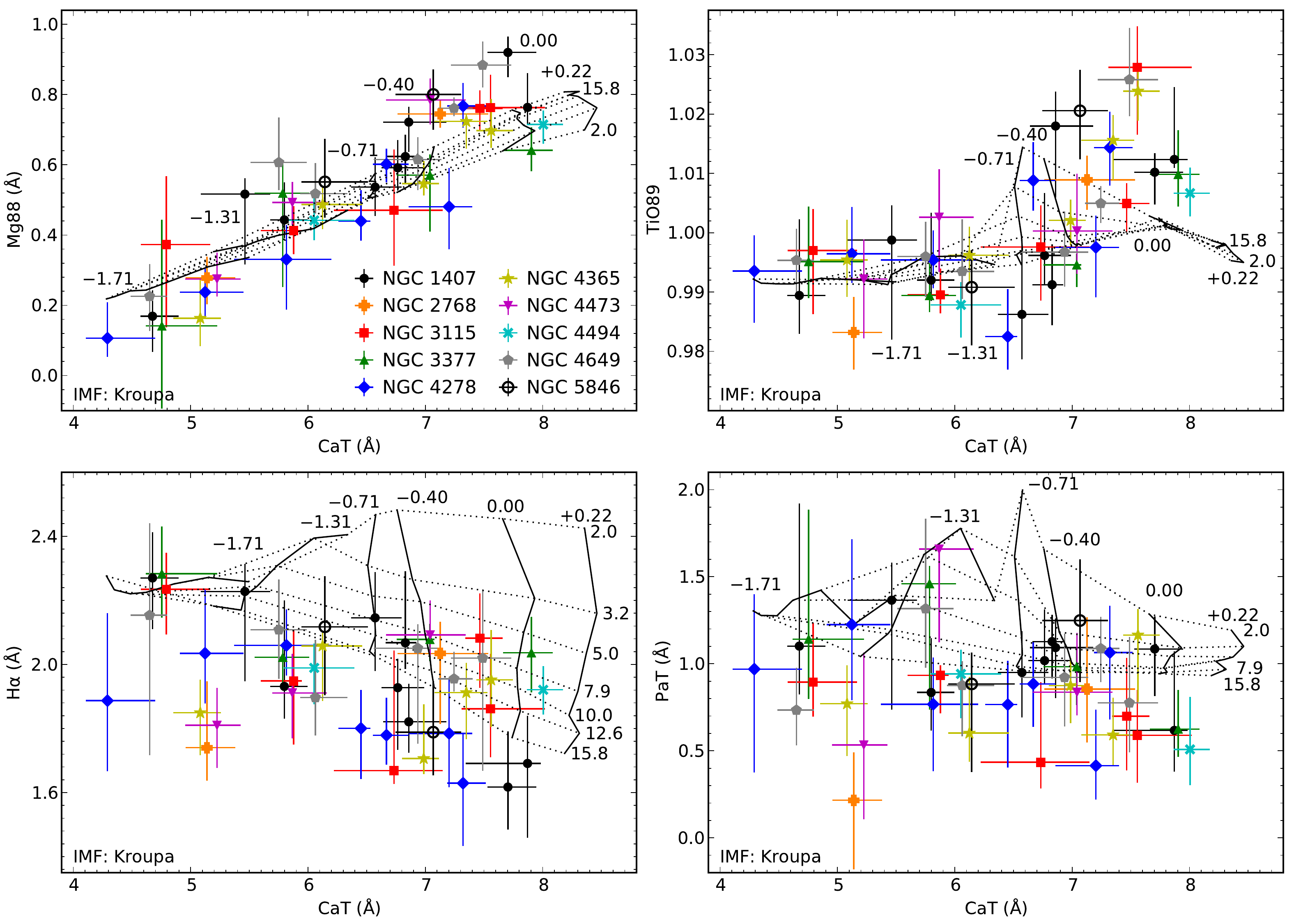}
\caption[Spectra stacked by colour]{Line indices for GC spectra stacked by colour.
The points are colour coded by by their host galaxy using the same colours and shapes as in Figure \ref{fig:stacking_CaT1}.
In each plot the \citet{2012MNRAS.424..157V} model predictions for a \citet{2001MNRAS.322..231K} IMF and varying age and metallicity are over-plotted.
For each model grid, solid lines are lines of constant metallicity; dotted lines are of constant age.
\emph{Top left} The Mg88 magnesium index as a function of CaT.
All galaxies show similar Mg88--CaT trends, suggesting that the [Mg/Fe]--[Z/H] relation is similar for all galaxies.
However, the observed relation is tilted compared to the \citet{2012MNRAS.424..157V} models.
\emph{Top right} The TiO89 TiO index versus the CaT index.
At low CaT values, the measured strength of this molecular feature is independent of the CaT value.
At high CaT values the strength increases with the CaT values.
The \citet{2012MNRAS.424..157V} models poorly predict the strength of TiO89 at high metallicities.
\emph{Bottom left} The H$\alpha$ index versus the CaT index.
Overall H$\alpha$ strength decreases with CaT strength in line with the model predictions but there is a wide scatter in H$\alpha$ strengths.
Many of the stacks show weaker H$\alpha$ strengths than the predictions of the oldest models.
NGC 3377 and NGC 4494, which show the bluest colours at high CaT, show the strongest H$\alpha$ at high CaT, suggesting younger ages.
\emph{Bottom right} The PaT Paschen line index versus the CaT index.
Although the strengths of the Paschen lines show large scatter, the mean values are weaker than the oldest models.}
\label{fig:stacking_CaT2}
\end{center}
\end{figure*}

To gain some understanding on how these indices and the $(g - i)$ colour are predicted to respond to changes in metallicity, age and the IMF, we analysed the simple stellar population models of \citet{2012MNRAS.424..157V}.
The spectral resolution of these models (1.5 \AA{} in the CaT spectral region, 2.5 \AA{} in the regions of H$\alpha$ and Na82) closely matches that of the \textsc{deimos} spectra and the models span a wide range of metallicities, unlike the more advanced models of \citet{2012ApJ...747...69C}, as well as a range of IMFs.
Additionally, the \citet{2012MNRAS.424..157V} model spectral energy distributions are publicly available.
\citet{2012MNRAS.426.1475U} found that the \citet{2012MNRAS.424..157V} models predicted a CaT--[Z/H] relation in good agreement with observations.

The \citet{2012MNRAS.424..157V} models use the scaled solar isochrones from \citet{2000A&AS..141..371G} with use the stellar libraries of \citet{2001MNRAS.326..959C} in the CaT region (8350 to 9020 \AA), MILES \citep{2006MNRAS.371..703S} from 3525 to 7500 \AA{} and Indo-US \citep{2004ApJS..152..251V} from 7500 to 8350 \AA.
These spectral libraries are dominated by field stars but contain a number of open cluster and metal poor GC stars.
Since the abundance pattern of stars in the solar neighbourhood is not that of GCs in the Milky Way, the models likely do not accurately predict the strengths of some spectral features.
For the CaT, Mg88, Fe86, PaT and TiO89 indices we used the \citet{2012MNRAS.424..157V} models based on the stellar library of \citet{2001MNRAS.326..959C}. 
For the H$\alpha$ and Na82 indices as well as the $(g - i)$ colour we used the \citet{2012MNRAS.424..157V} models based on the MIUSCAT stellar library \citep{2012MNRAS.424..157V}.
To minimise systematic effects due to flat fielding, we applied the same normalisation that we applied to the observed spectra to the model spectral energy distributions before measuring the spectral features in an identical fashion.
Due to the similarity of the model spectral resolution to that of the obsereved spectra we did not consider what effect spectral resolution has on the index measurement.

We compare the colours and spectral indices of the stacked spectra with the \citet{2012MNRAS.424..157V} models in Figures \ref{fig:stacking_CaT_Fe}, \ref{fig:stacking_CaT1} and \ref{fig:stacking_CaT2}.  
Although the models predict the general colour and index--CaT trends observed, the models disagree with the observations in several details.
In Figure \ref{fig:stacking_CaT_Fe} we plot colour as functions of CaT and Fe86 index strengths.
Assuming that all GCs have similar old ages, the \citet{2012MNRAS.424..157V} models predict redder colours for GCs at intermediate CaT strengths than are observed.
Some of the disagreement between the models and the observed values may be due to differences in $\alpha$-element abundances between the models and the observed GCs.
The observed colour--Fe86 relation generally agrees with the model predictions.
As seen the top right panel of Figure \ref{fig:stacking_CaT1}, the observed colour--CaT relation suggests a bottom light IMF.

As seen in the middle row of Figure \ref{fig:stacking_CaT1}, the predicted relationship between the Na82 and CaT indices is harder to assess due to the likely non-solar [Na/Fe] abundance in the GCs and the possible systematics due to correcting the telluric absorption present in this wavelength region.
Additionally, \citet{2012MNRAS.424..157V} state that their models for the Na82 wavelength region are unreliable for metallicities below [Z/H] = $-0.71$.
At high CaT strengths, the majority of the stacked spectra suggest a Milky Way like IMF \citep[i.e. ][ like]{2001MNRAS.322..231K}.
If the stacked spectra show enhanced [Na/Fe] like Milky Way GCs \citep[e.g.][]{2014ApJS..210...10R}, the Na82 index should be stronger \citep{2012ApJ...747...69C}, strengthening the evidence for bottom light IMF.

The Fe86 index strengths show a tight relation with the CaT index strengths that is parallel but offset from the predictions of \citet{2012MNRAS.424..157V}.
Interestingly, the models of \citet{2012ApJ...747...69C} under-predict the observed strength of the CaT in \emph{galaxy} spectra \citep{2012ApJ...760...71C} by a similar amount ($\sim 0.5$ \AA).
Fitting the Fe86 index strength as a function of CaT index strength we find:
\begin{equation}\label{eq:SFe_fit}
\text{Fe86} = (0.724 \pm 0.032) \times \text{CaT} + (-1.802 \pm 0.209) .
\end{equation}
This fit has a RMS of 0.212 \AA{} and a $\chi^{2}$ value of 34.7 for 43 degrees of freedom.
\citet{2010AJ....139.1566F} observed a similar relation between a subset of the lines in the Fe86 index and the CaT.
However, they were only able to measure the strength of the weak metal lines on the spectra fitted as part of their CaT measurement process.
The relation between the Mg88 and CaT indices is also linear but shows more scatter than the Fe86--CaT index relation.
Fitting the Mg88 index strengths we find:
\begin{equation}\label{eq:Mg88_fit}
\text{Mg88} = (0.184 \pm 0.014) \times \text{CaT} + (-0.643 \pm 0.091) .
\end{equation}
This fit has a RMS of 0.092 \AA{} and a $\chi^{2}$ value of 45.3 for 43 degrees of freedom.
Although most Mg88 strengths agree with the \citet{2012MNRAS.424..157V} models, at low CaT values the Mg88 strengths are lower than the models while at high CaT values, the Mg88 values are higher than the models.
These differences between the observations and the \citet{2012MNRAS.424..157V} models are likely due to differences in the abundance patterns of the stellar libraries used in the models compared to the GC abundance patterns.

The strength of the TiO89 molecular feature appears flat at low CaT values but increases with CaT strength at higher CaT values.
Although the \citet{2012MNRAS.424..157V} models predict the behaviour of the TiO89 index at weak CaT strengths, the models under-predict this TiO index at stronger CaT values.
This is likely due to the solar abundance pattern of the stellar library used by \citet{2012MNRAS.424..157V} at high metallicities as the TiO molecular bands are strongly sensitive to both the O and Ti abundances \citep{2014ApJ...780...33C}.
Since the TiO89 index is defined as a flux ratio it is sensitive to differences in flux calibration.
However, differences in the TiO89 index due to flux calibration should be minimised since we normalised the continuum in an identical fashion for both the observed spectra and the model spectra.
The increase in strength of the TiO89 feature with metallicity is apparent in Figure \ref{fig:every_red}.
Although there is a larger amount of scatter, the observed H$\alpha$ values decrease with increasing CaT strength in line with the model predictions.
Likewise the PaT values show a large amount of scatter but no trend with CaT, again in agreement with the models.
The lack of significant PaT absorption in the stacked spectra allows us to rule out truly young ages for any of the stacked spectra.
For both H$\alpha$ and PaT the mean index measurements are systematically weaker than the oldest \citet{2012MNRAS.424..157V} model predictions.
\citet{2010AJ....139.1566F} observed significant Paschen line strengths in their fitted spectra but not in the raw spectra.
\citet{2012MNRAS.426.1475U} refit these spectra and observed no Paschen absorption in the fitted spectra.
Since we do not observe Paschen line absorption in our stacks, the absorption observed by \citet{2010AJ....139.1566F} was likely an artefact of their fitting procedure.
Combining the $(g - i)$ colours as well as the CaT, Na82, Fe86 and H$\alpha$ strengths with the \citet{2012MNRAS.424..157V} models, the stacked spectra suggest old ages and a bottom light IMF.

\citet{2012MNRAS.424..157V} state that their models are unreliable at young ages for model metallicities below [Z/H] = -0.40 and that the models in the Na82 spectral region are unreliable for metallicities below [Z/H] = -0.71 for any age.
The inability to vary chemical abundances from the Solar neighbourhood pattern or vary the horizontal branch morphology are limitations to using the \citet{2012MNRAS.424..157V} models to further interpret our data.
Medium resolution stellar population models with the ability to fully vary chemical abundances and to adopt different horizontal branch morphologies would be necessary to recover more information from the \textsc{deimos} spectra.
 
\subsection{Fe86 based metallicities}

\begin{figure}
\begin{center}
\includegraphics[width=240pt]{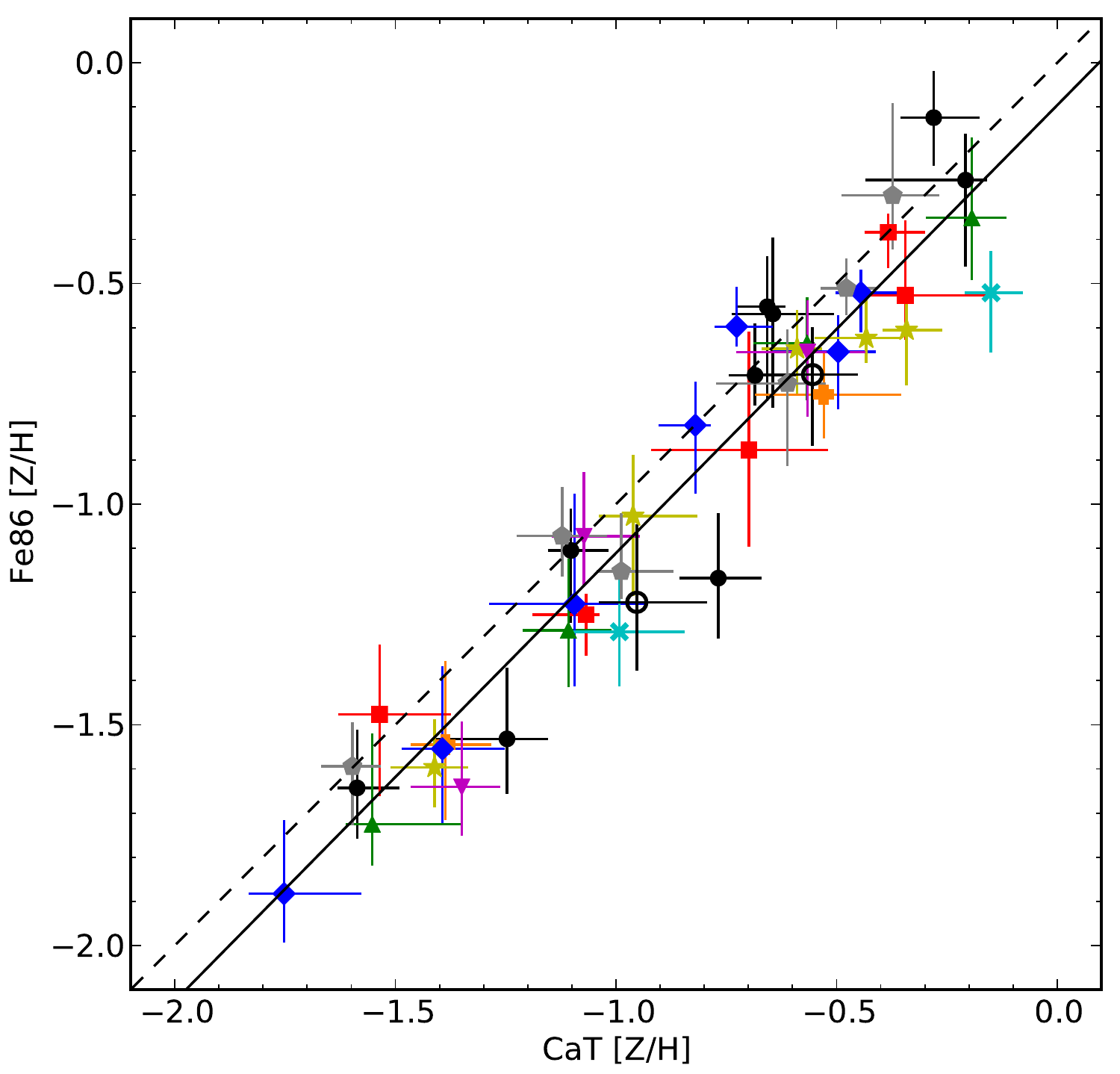}
\caption[Fe86 based metallicities versus CaT based metallicities]{Fe86 based metallicities versus CaT based metallicities.
For both indices the 12.6 Gyr old, \citet{2001MNRAS.322..231K} IMF \citet{2012MNRAS.424..157V} model was used to convert the index strengths into metallicities.
For the CaT based metallicity, equation 9 of \citet{2012MNRAS.426.1475U} was used to correct the model metallicities for the effects of the [$\alpha$/Fe] pattern of the solar neighbourhood.
The points are colour coded by their host galaxy using the same colours and shapes as in Figure \ref{fig:stacking_CaT1}.
The dashed line is 1 to 1.
The solid line is a fit to the two metallicity determinations.
The slope of the fit is consistent with unity ($1.00 \pm 0.04$) although the intercept ($-0.11 \pm 0.04$) is significantly offset from zero.
The RMS of the fit is 0.13 dex; the $\chi^{2}$ value is 34.1 for 43 degrees of freedom.}
\label{fig:weak_CaT_metal}
\end{center}
\end{figure}

\label{sec:SFe}

Although \citet{2012MNRAS.426.1475U} showed that the CaT can be used to derive metallicities as well as the traditionally employed Lick indices \citep{1994ApJS...94..687W}, it is desirable to further check the reliability of the CaT as a metallicity indicator.
As noted before, \citet{2008ApJ...682.1217K} used the weak metal lines in \textsc{deimos} spectra of RGB stars to derive iron and $\alpha$-element abundances.
Our Fe86 index measures some of the stronger lines fit by \citet{2008ApJ...682.1217K} and as seen in the lower left corner of Figure \ref{fig:stacking_CaT1} appears to correlate well with the strength of the CaT.

While the CaT is sensitive to [Ca/H] \citep{2012ApJ...747...69C}, its strength seems to be more closely linked to the overall metallicity [Z/H] \citep{2012ApJ...759L..33B}.
As noted before, the chemical abundance pattern of the \citet{2012MNRAS.424..157V} stellar library at high metallicities is that of the stellar neighbourhood, while at lower metallicities many of the stars in the stellar library are GC members.
In \citet{2012MNRAS.426.1475U}, the \citet{2010MNRAS.404.1639V} version of the \citet{2003MNRAS.340.1317V} models were used to derive a relationship between CaT strength and metallicity.
There, better agreement was found between the CaT based metallicities and Lick index based metallicities \citep[derived using the models of][]{2003MNRAS.339..897T} if the model metallicities were corrected for the differences between $\alpha$ element enhancement in the solar neighbourhood and the expected $\alpha$ element enhancement of GCs ([$\alpha$/Fe] = 0.3).
After applying the same corrections to the model metallicities as \citet{2012MNRAS.426.1475U} (their equation 9) we found the following relation between CaT and metallicity using the \citet{2012MNRAS.424..157V} models with an age of 12.6 Gyr and a \citet{2001MNRAS.322..231K} IMF:
\begin{equation}\label{eq:CaT_trans}
\text{[Z/H]} = (0.431 \pm 0.011) \times \text{CaT} + (-3.604 \pm 0.072) .
\end{equation}
The RMS of this relation is 0.027 dex.

Since the Fe86 index is dominated by a mixture of Fe and Ti lines it is unclear whether and how metallicities derived from this index should be corrected for the abundance pattern of the models.
Assuming that the index is more strongly sensitive to [Fe/H] than [Z/H] we do not adjust the model metallicity.
We derived the following relation between the strength of the Fe86 index and metallicity using the \citet{2012MNRAS.424..157V} models with an age of 12.6 Gyr and a \citet{2001MNRAS.322..231K} IMF:
\begin{equation}\label{eq:weak_trans}
\text{[Z/H]} = (0.604 \pm 0.012) \times \text{Fe86} + (-2.663 \pm 0.041) .
\end{equation}
The RMS of this relation is 0.029 dex.

In Figure \ref{fig:weak_CaT_metal} the CaT and Fe86 metallicities are compared.
Although the slope of the relation between the two metallicities is consistent with unity ($1.00 \pm 0.04$), the Fe86 metallicities are offset to lower metallicities by $0.11 \pm 0.04$ dex.
As mentioned above, the models of \citet{2012ApJ...747...69C} also under-predict the observed strength of the CaT by a similar amount.
The RMS of this fit is 0.13 dex and the $\chi^{2}$ value is 34.1 for 43 degrees of freedom.
Although the CaT index measures the strength of the CaT feature only, the wavelength range that the stellar templates are fit to as part of the CaT measurement process includes most of the lines in the Fe86 index as well as the Mg88 index so the CaT index is not completely independent of either the Fe86 index or the Mg88 index.
It is reassuring that spectral features dominated by different species (Ca\,\textsc{ii} for the CaT, Fe\,\textsc{i} and Ti\,\textsc{i} for Fe86) give consistent results modulo a 0.1 dex offset.
This suggests that both are measuring metallicity accurately. 
It would be desirable to use weak Fe and other metal lines from across the \textsc{deimos} wavelength range to measure metallicities of individual GCs as \citet{2008ApJ...682.1217K} has done for individual RGB stars. 

\subsection{Stellar population variations between galaxies}

\begin{figure*}
\begin{center}
\includegraphics[width=504pt]{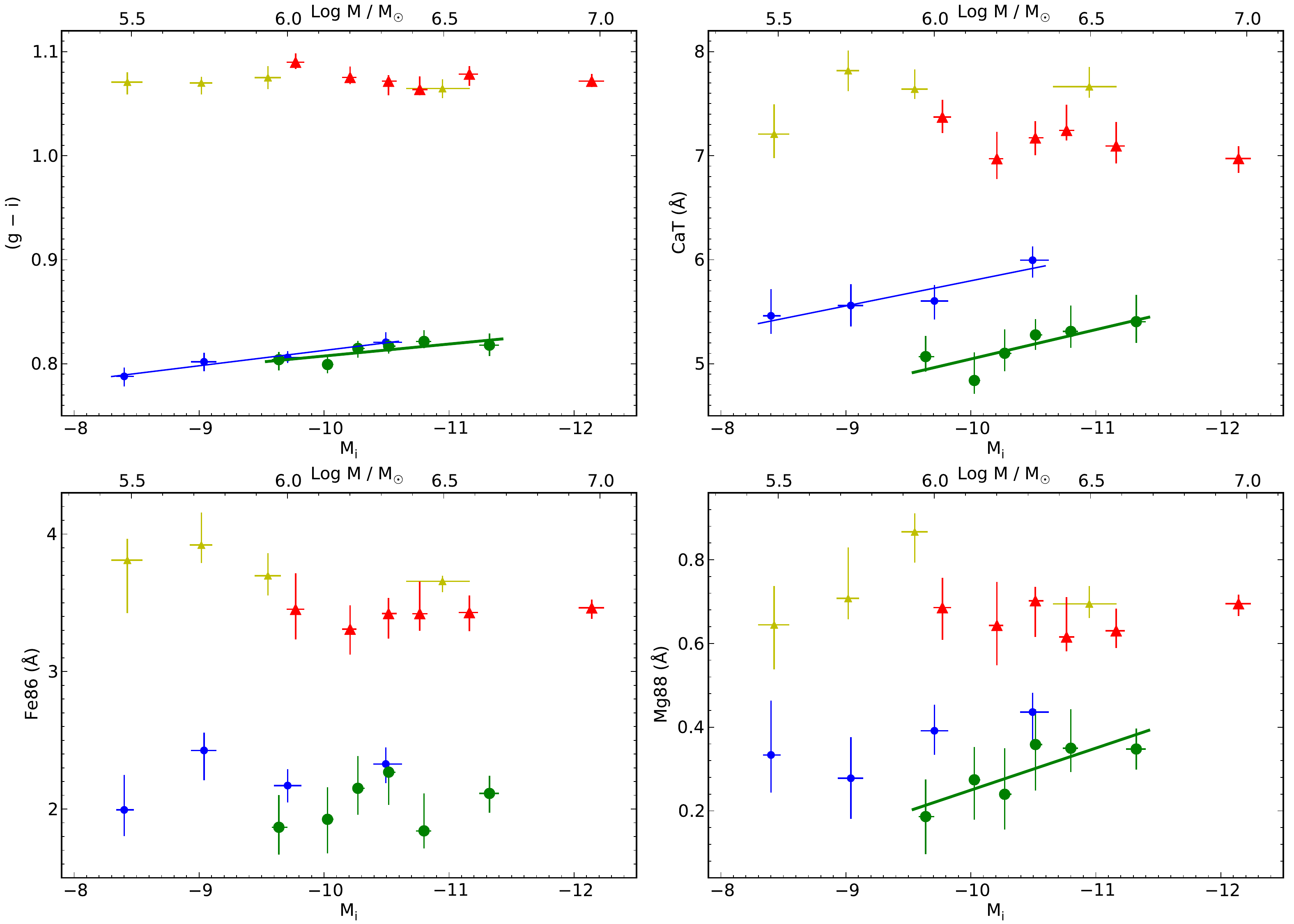}
\caption[Spectra stacked by magnitude]{Stellar population properties of GCs stacked by magnitude in fixed colour ranges.
The stacks of blue GC spectra in the faint galaxy group are plotted as small blue circles, blue GC spectra in the bright galaxy group as large green circles, red GC spectra in the faint group as yellow small triangles and red GC spectra in the bright group as red large triangles.
The colours and symbols are the same as in Figure \ref{fig:stacking_sigma}.
For illustrative purposes the mass corresponding to the absolute magnitude and a consistent mass-to-light ratio of 2 is shown on the top of each plot.
Where a significant relation is observed with magnitude, the relation is plotted with a line (thicker for the bright group).
\emph{Top left} $(g - i)$ colour as a function of $i$-band absolute magnitude.
For both red and blue, the two groups have consistent colours at fixed magnitude.
The blue stacks from both groups become significantly redder at brighter luminosities (the blue tilt).
\emph{Top right} The CaT index as a function of $i$-band absolute magnitude.
Both red and blue stacks show significant CaT offsets between the faint and bright groups.
Both the blue faint group and bright group stacks have significantly stronger CaT values at brighter magnitudes than at fainter magnitudes.
\emph{Bottom left} The Fe86 weak iron line index as a function of $i$-band absolute magnitude.
Like the CaT index, both red and blue stacks show significant Fe86 offsets between the faint group and bright group.
\emph{Bottom right} The Mg88 magnesium index as a function of $i$-band absolute magnitude.
Like the CaT and Fe86 indices, both red and blue stacks show significant Mg88 offsets between the faint group and bright group.
The Mg88 values of the bright blue stacks become significantly stronger with brighter magnitudes.}
\label{fig:mag_0}
\end{center}
\end{figure*}

\begin{figure*}
\begin{center}
\includegraphics[width=504pt]{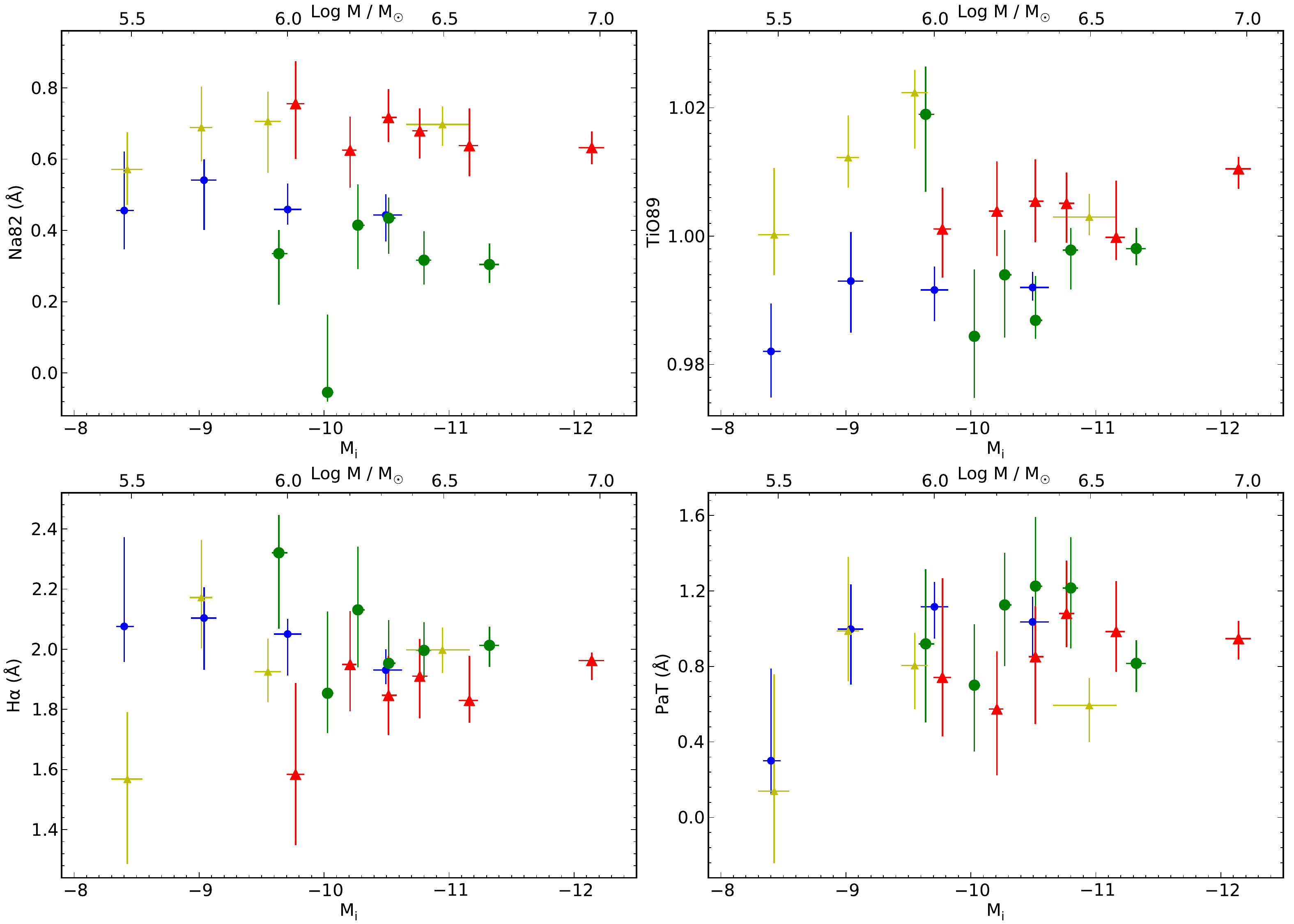}
\caption[Spectra stacked by magnitude]{GCs stacked by magnitude in fixed colour ranges.
The stacks of blue GC spectra in the faint group are plotted as small blue circles, blue GC spectra in the bright group as large green circles, red GC spectra in the faint group as yellow small triangles and red GC spectra in the bright group as red large triangles.
For illustrative purposes the mass corresponding to the absolute magnitude and a constant mass-to-light ratio of 2 is shown on the top of each plot.
\emph{Top left} The Na82 sodium index as a function of $i$-band absolute magnitude.
There is weak evidence that the faint group blue stacks have stronger Na82 index strengths than the bright group blue stacks.
\emph{Top right} The TiO89 TiO index as a function of $i$-band absolute magnitude.
\emph{Bottom left} The H$\alpha$ index as a function of $i$-band absolute magnitude.
\emph{Bottom right} The PaT Paschen line index as a function of $i$-band absolute magnitude.
None of the Na82, TiO89, H$\alpha$ or PaT indices have any significant index--magnitude relationships.}
\label{fig:mag_1}
\end{center}
\end{figure*}

\label{with_galaxies}
As in \citet{2012MNRAS.426.1475U} we see variations in the $(g - i)$--CaT relation between galaxies (Figures \ref{fig:CaT_colour} and \ref{fig:CaT_colour_hist}).
At blue colours the GCs of NGC 1407 and NGC 4278 have weaker CaT values than predicted by Equation \ref{eq:CaT_colour}.
For red colours, NGC 1407, NGC 4278, NGC 4365 and NGC 4649 GCs have weaker CaT strengths than predicted from their colours while NGC 3115, NGC 3377, NGC 4473 and NGC 4494 have stronger than predicted CaT values.
The shapes of the colour--CaT relations of NGC 1407 and NGC 4494 are radically different as first noticed by \citet{2011MNRAS.415.3393F}.

As seen in Figure \ref{fig:stacking_CaT1}, at low CaT values NGC 3115, NGC 3377 and NGC 4473 show bluer colours compared to the other galaxies while at high CaT values NGC 3377 and NGC 4494 also show bluer colours than the other galaxies.
Additionally, at low CaT values NGC 3115, NGC 3377 and NGC 4473 show higher Na82 strengths than the other galaxies.
At high CaT values, NGC 3377 and NGC 4494 show slightly weaker Na82, Fe86, Mg88 and TiO89 strengths than those of the other galaxies and stronger H$\alpha$ strengths.
In the \citet{2012MNRAS.424..157V} models at high metallicities, $(g - i)$ becomes bluer, Na82, Fe86 and Mg88 index strengths become weaker and H$\alpha$ becomes stronger at younger ages while the CaT strength is unchanged.
Since an extremely blue horizontal branch produces bluer colours, stronger H$\beta$ indices and weaker metal index strengths in the Lick index spectral regions \citep{2013ApJ...769L...3C}, it is also possible that blue horizontal branches are responsible.
We note that while the high CaT index stack of NGC 4494 is an outlier in the in colour--CaT relation, it is consistent with other galaxies in the colour--Fe86 relation (Figure \ref{fig:stacking_CaT_Fe}).
Without knowing whether the CaT or Fe86 measurement better reflects the metallicity of NGC 4494's metal rich spectral stack, we cannot tell whether it is truly outlier.

Although their colours at fixed magnitude are consistent, in the spectra stacked by magnitude (Figure \ref{fig:mag_0}), the CaT and Fe86 indices of the faint group are all significantly stronger ($p > 0.95$, see Table \ref{tab:mag_relations}) at $M_{i} = -10$ than those of the bright group for both the blue and red colours.
If anything the red GCs in the faint group are bluer than those in the bright group although this difference is not significant ($p = 0.936$).
Using Equation \ref{eq:CaT_colour}, this small colour difference of $(g-i) = -0.0101^{+0.0072}_{-0.0071}$ would correspond to a small CaT difference of $-0.049^{+0.035}_{-0.035}$ \AA{} which is in the opposite direction to and much smaller in magnitude than the observed CaT difference of $0.416_{-0.136}^{+0.138}$ \AA. 
The Mg88 indices of the faint group are also stronger than those of the bright group although the difference for the red GCs barely falls below our significance limit ($p = 0.946$). 
There is also evidence that the faint group has stronger Na82 indices than the bright group stacks only at blue colours (Figure \ref{fig:mag_1}).
NGC 1407's radial velocity places its Na82 index in the same observed wavelength as the strongest telluric absorption lines potentially biasing Na82 measurements.
We restacked the bright group without NGC 1407 and measured consistent Na82 values to those of stacks including this galaxy.
There is no evidence that there is a Na82 index difference between the groups for red colours.
Nor is there evidence that the TiO89, H$\alpha$ or PaT index strengths are significantly different between the faint and bright groups at either blue or red colours.
Since the metal index offsets between the two galaxy groups are seen at all luminosities, the offsets between the groups are not due to the index--metallicity relations changing with magnitude.

Since the strengths of Fe86 and Mg88 correlate strongly with the strength of the CaT, we checked to see if the Fe86 and Mg88 offsets at $M_{i} = -10$ are consistent with being caused by the CaT offsets at the same magnitude ($0.75_{-0.18}^{+0.10}$ and $0.42_{-0.14}^{+0.14}$ \AA{} for the blue and red stacks respectively).
Using Equation \ref{eq:SFe_fit} the CaT offsets predict Fe86 offsets of $0.54_{-0.13}^{+0.07}$ and $0.30_{-0.10}^{+0.10}$ \AA, consistent with the observed values of $0.31_{-0.15}^{0.14}$ and $0.33_{-0.15}^{+0.11}$ \AA.
Likewise, using Equation \ref{eq:Mg88_fit}, the CaT offsets predict Mg88 offsets of $0.138_{-0.032}^{+0.018}$ \AA{} and $0.077_{-0.025}^{+0.025}$ \AA, again consistent with observed values of $0.146_{-0.058}^{+0.065}$ \AA{} and $0.077_{-0.050}^{+0.056}$ \AA{} respectively.
The consistency of these three metal indices and the strong correlation of the CaT and Fe86 indices with metallicity points to a metallicity difference between the two groups.
It also suggests that there is no large variation in the [Mg/Fe]--[Fe/H] or [Ca/Fe]--[Fe/H] relations between galaxies.
At low CaT strengths, the top right panel of Figure \ref{fig:stacking_CaT2} shows that the TiO89 strength is largely independent of CaT strength in line with the lack of a significant difference in TiO89 between the two groups.
However, at CaT strengths above $\sim 6.3$ \AA{}, the TiO89 strength increases with CaT strength.
Fitting the slope of the CaT--TiO89 relation above CaT $= 6.3$ \AA{} to be $0.0166 \pm 0.0048$ \AA$^{-1}$, the CaT difference predicts a TiO89 difference of $0.0069_{-0.0023}^{+0.0023}$ which is consistent with the observed, non significant ($p = 0.939$), TiO89 difference of $0.0073_{-0.0045}^{+0.0043}$.
The behaviour of the TiO89 index, which is sensitive to the O and Ti abundances, provides further evidence that the observed metal index differences are caused by variations in the colour--metallicity relation.
Since the H$\alpha$ and PaT indices show weak or no dependence on CaT at low CaT values, the lack of offsets between the faint and bright groups for these indices is consistent with the CaT, Fe86 and Mg88 offsets being due to a metallicity offset.

The behaviour of the Na82 index is difficult to interpret with existing models and data.
Besides the Na\,\textsc{i} doublet at 8183 \AA{} and 8195 \AA{}, the index definition contains a TiO molecular bandhead and is adjacent to the Paschen limit at 8205 \AA{}.
At solar metallicity the index is sensitive to age, the IMF and [Na/Fe] \citep{2012ApJ...747...69C}.
As mentioned above, the \citet{2012MNRAS.424..157V} models are unreliable in the Na82 spectral region at low metallicities.
Besides the model uncertainties, the Na82 index likely suffers from measurement systematics due to uncertain telluric corrections.
As seen in the lower left panel of Figure \ref{fig:stacking_CaT1}, at low CaT strengths, there is no significant relation (a slope of $0.02 \pm 0.07$) between the CaT and Na82 indices while above CaT $= 6.3$ \AA{} a relation is seen with a slope of $0.29 \pm 0.12$. 
These CaT--Na82 relations predict Na82 differences of $0.016_{-0.004}^{0.002}$ and $0.121_{-0.039}^{+0.040}$ for blue and red colours respectively yet differences of $0.201_{-0.074}^{+0.063}$ and $-0.019_{-0.075}^{+0.079}$ are observed.
The larger than expected difference in Na82 between galaxy groups for blue GCs could be be explained by higher Na abundances in the faint group while for red GCs younger ages in the faint group could explain the smaller than expected difference.  
The behaviour of the Na82 index does not seem to be following the behaviour of the other metal indices. 

Although the spectral indices have been measured in a homogeneous manner using the same spectroscopic setup, data reduction pipeline and analysis code, the photometry is heterogeneous with data from different telescopes, analysed in different ways.
If the differences between galaxies were caused by incorrect photometric zero points or by incorrectly removing the foreground extinction, then the observed colours would be shifted bluer or redder.
However, the differences in colour--CaT relationships cannot be simply resolved by shifting individual galaxies' GC colours.
If a colour term has been been neglected in the zero points, the slope of the colour--CaT relationship could change.
While this could improve the agreement for NGC 3377 and NGC 4494, changing the slope would not help NGC 1407.
It is unlikely that reddening would be correlated with metallicity nor can extinction make an object bluer.
For NGC 3115 \citep{2014AJ....148...32J}, NGC 4278 \citep{2012MNRAS.426.1475U}, NGC 4365 \citep{2012MNRAS.420...37B}, NGC 4473 \citep{2009ApJS..180...54J} and NGC 4649 \citep{2012ApJ...760...87S} where ACS $gz$ photometry is available, the same colour--CaT relations are seen for both the ACS and Suprime-Cam photometry.
Furthermore, photometric issues cannot explain why CaT weak spectra from NGC 3115, NGC 3377 and NGC 4473 have stronger Na82 strengths than other galaxies and why CaT strong NGC 3377 and NGC 4494 spectra have weaker metal and stronger H$\alpha$ indices than other galaxies.
Although it is highly desirable to improve the uniformity of the SLUGGS GC photometry, we do not believe heterogeneous photometry to be the source of the colour--metal index variations.

An important caveat that the median magnitude of GCs in this study is $M_{i} = -10.3$ which is significantly brighter than the turnover magnitude ($M_{i} = -8.4$, \citealt{2010ApJ...717..603V}) so that stellar populations of the observed sample of GCs may not be representative of all GCs in the sample galaxies.
Since in several galaxies including NGC 1407 \citep{2009AJ....137.4956R}, M87 \citep{2011ApJS..197...33S} and NGC 5128 \citep{2010AJ....139.1871W}, GCs with magnitudes comparable to or brighter than $\omega$ Cen ($M_{i} = -11.3$, \citealt{2014MNRAS.437.1725V}) show different kinematics than fainter GCs, it is quite possible that the very brightest GCs are a distinct population with different stellar population trends than the majority of GCs.
In this study, 9 \% of the GCs are brighter than $\omega$ Cen.

Although the mean magnitude of GCs in the galaxies with bluer, more linear colour--CaT relations is fainter, we do not believe that the differences in colour--CaT relations between galaxies are due to stellar population variations with GC magnitude.
If we restrict the CaT measurements of individual GCs in Figure \ref{fig:CaT_colour} to GCs in the magnitude range $-10.5 < M_{i} < -9.5$, we see the same colour--CaT relationships.
In Figure \ref{fig:mag_0}, stacked spectra with the same colours and absolute magnitudes from the two galaxy groups have significantly different metal index strengths.
This is also seen in the two groups having consistent colour-- and index--magnitude slopes.
Together these suggest that the differences in the luminosities of observed GCs between galaxies are not responsible for the observed colour--CaT variations.

With the small number of galaxies in this study, it is difficult to reliably study how the colour-metal index variations correlate with host galaxy properties. 
The three most luminous galaxies in this study (NGC 1407, NGC 4365 and NGC 4649) belong to the bright group while least luminous galaxy in our sample (NGC 3377) belongs to the faint group.
The remaining four galaxies all have similar luminosities ($M_{K} \sim -24$) with only NGC 4278 belonging to the bright group.
This hints, but does not prove, that the shape of the colour--CaT relation is linked to galaxy mass or a property that depends on galaxy mass.

Possible stellar population differences that could explain the observed variations in metal index strength at fixed colour between galaxies include different ages, chemical abundance patterns or initial mass functions.
Before discussing each of these in turn, it should noted that one way that the age and abundance pattern of a stellar population manifest themselves is through the morphology of its horizontal branch.
In Milky Way GCs, metallicity, age and helium abundance seem to determine the horizontal branch morphology \citep{2010A&A...517A..81G, 2014ApJ...785...21M}. 
It should also be noted that our results only indicate differences in the mean index--colour relations from galaxy to galaxy and that there likely exists spreads in some of these properties within each galaxy (e.g. a range of GC ages at fixed metallicity as is seen in Milky Way GCs \citealt{2011ApJ...738...74D, 2013ApJ...775..134V}).

\subsubsection{Age differences?}
One possible explanation for the colour--metal index variations is differences in the age--metallicity relation between galaxies.
As can be seen in the top left panel of Figure \ref{fig:stacking_CaT1}, younger ages lead to bluer colours.
The \citet{2012MNRAS.424..157V} models predict that age has little effect on the CaT at high metallicities but at low metallicities predict stronger CaT values at younger ages.
Both the \citet{2009ApJ...699..486C} and \citet{2013ApJS..204....3C} models predict that for solar iron abundances 8 Gyr ($z \sim 1$) populations are 0.1 mag bluer in $(g - i)$ than 13 Gyr ($z \sim 8$) populations.
At low metallicities, the morphology of horizontal branch strongly depends on  
age \citep[e.g.][]{2010ApJ...708..698D}.
The horizontal branch morphology has a large effect on the integrated colours while having a relatively minor effect on CaT strengths \citep{2005MNRAS.362..799M, 2009ApJ...699..486C, 2012ApJ...759L..33B}.
For example, the \citet{2013ApJS..204....3C} models, which include a detailed treatment of the horizontal branch, predict that 12 Gyr ($z \sim 4$) populations are 0.05 mag bluer in $(g - i)$ than 13 Gyr ($z \sim 8$) populations at [Fe/H] $= -2$.

The bluer colours of the GCs in the faint group of galaxies suggest that less massive galaxies might have younger GCs on average.
The concept of GC starting to form slightly later and continuing to form over a period of time is consistent with observations of single stellar population equivalent ages being younger in less massive galaxies \citep[e.g.][]{2005ApJ...621..673T, 2010MNRAS.408...97K, 2012MNRAS.419.3167S}.
\citet{2007ApJ...659..188F} compared galaxy ages with the mean colours of their GC subpopulations. 
After removing the known relationships between GC colour and host galaxy luminosity, they found no relationships between the galaxy central age and GC colour. 
They interpret this as either GCs in younger galaxies are more metal rich than those in old galaxies or that GCs are uniformly old.
However, by removing the GC colour--galaxy luminosity relations to subtract the GC metallicity--galaxy luminosity relations, they also take away any GC age--galaxy luminosity relations that might be present.

Spectroscopic studies \citep[e.g.][]{1998ApJ...496..808C, 2001ApJ...563L.143F, 2005A&A...439..997P, 2005AJ....130.1315S, 2008MNRAS.385...40N} of GC ages have shown that the majority of GCs have old ($\sim 10$ Gyr) ages and that there is no significant difference in age between the metal poor and metal rich populations.
Evidence for a population of young or intermediate age metal rich GCs have been observed in galaxies such as NGC 5128 \citep[e.g.][]{2010ApJ...708.1335W} and NGC 4649 \citep{2006MNRAS.368..325P}.
In one of the only studies of the stellar populations of GCs in a sub $L^{*}$ galaxy (NGC 7457, $M_{K} = -22.4$),  \citet{2008AJ....136..234C} found that the two most metal rich GCs they observed display emission lines which likely indicates young ages. 
In their study of GCs in Local Group dwarf ellipticals, \citet{2006MNRAS.372.1259S} derived significantly younger ages for metal rich GCs compared to metal poor GCs in the same galaxies.
A caveat to all these studies is the difficulty in determining accurate ages for old stellar populations from integrated spectra, especially due to the effects of the horizontal branch.   
Additionally, the Milky Way GCs thought to have formed in dwarf galaxies show young ages at fixed metallicity than GCs thought to have formed in situ \citep{2010MNRAS.404.1203F, 2013MNRAS.436..122L}.
Taken together, this suggests that it is plausible that less massive galaxies have slightly younger GCs systems which would be bluer.

We can use the CaT index strengths and $(g-i)$ colours of the two galaxy groups to estimate the mean age differences required to explain the differences in CaT index observed between the groups.
Using Equation \ref{eq:CaT_trans} and the \citet{2013ApJS..204....3C} models, we estimate ages of 7.3 and 13.1 Gyr and metallicities of [Z/H] $= -1.1$ and $-1.4$ for the blue GCs in the faint and bright galaxy groups respectively.
For the red GCs, we estimate 7.8 and 10.6 Gyr and metallicities of [Z/H]  $-0.3$ and $-0.5$ for the faint and bright groups respectively. 
Using the \citet{2009ApJ...699..486C} models we get ages of 6.6 and 8.3 Gyr for the blue colour range, and of 6.5 and 12.0 Gyr for the red colour range.
These age and metallicities assume that the relationship between CaT strength and metallicity does not vary with age and that the observed differences in the colour--CaT relationship are solely due to age. 
The uncertainties inherent in stellar population modelling are highlighted by the differences in ages derived using the \citet{2013ApJS..204....3C} and \citet{2009ApJ...699..486C} models.
It should be noted that since the S/N cut we apply to the spectra effectively imposes a magnitude limit and since younger GCs are brighter than old GCs due to the effects of stellar and dynamical evolution, our sample is biased towards younger GCs and the true difference in mean ages may be smaller.

As less luminous galaxies have on average younger stellar populations, it is quite possible that host on average younger stellar populations.
Under the assumption that the differences in CaT strength are solely due to an age difference, an age difference of only a few Gyr is required between the two galaxy groups.
Differences in the GC age--metallicity relation between galaxies are a possible explanation for at least some of the variation in the colour--metal index relation.  

\subsubsection{Abundance differences?}
The abundance pattern of a stellar population also affects its colours.
Helium enhanced populations have hotter isochrones and horizontal branches which produce bluer optical colours.
In models of \citet{2013ApJ...769L...3C}, a He enhanced population (Y $= 0.33$) is $\sim 0.25$ mag bluer in $(g - z)$ than a He normal population (Y $= 0.23$) at solar iron abundance but no significant difference in colour is seen below [Fe/H] $= -1$.
The effect of $\alpha$-element enhancement depends greatly on the colour considered.
In the models of \citet{2007MNRAS.382..498C}, $\alpha$-element enhanced populations are bluer than scaled solar populations for the bluest optical colours (e.g. $\Delta(u-g) = -0.04$ for [Fe/H] = 0 and an age of 12 Gyr) but are redder than scaled solar populations at fixed [Fe/H] for the redder colours (e.g. $\Delta(r-z) = 0.07$).
Different elements also affect affect optical colours in different ways.
\citet{2009ApJ...694..902L} found that increased abundances of heavier elements such as Mg, Ti and Fe leads to redder colours while enhanced C, N or O leads to bluer colours.

While we do not believe that [Mg/Fe] or [Ca/Fe] varies significantly between galaxies due the lack of differences in the Mg88--CaT or Fe86--CaT relations, there still could be variations in the abundances of He, C, N or O between the two groups.
In the Milky Way, GCs contain stars with both `normal' $\alpha$-element enhanced abundance patterns and stars that are relatively He, N and Na rich and C and O poor.
It is possible that the process that produces these abundance anti-correlations could vary galaxy to galaxy, producing variations in the mean abundances of these light elements. 
\citet{2013ApJ...762..107G} found that a solar metallicity population with a normal $\alpha$-element enhanced abundance pattern was 0.05 mag redder in $(g-i)$ than a He, N and Na enhanced and C and O depleted population. 
However, they did not consider the effects of the abundance variations on the horizontal branch.
Although CaT strengths are relatively insensitive to the horizontal branch morphology, the effects of helium on the isochrones or the effects of CNO abundances on the pseudo-continuum by CN and TiO molecular bands could affect the CaT strengths.
In the lower left panel of Figure \ref{fig:stacking_CaT1}, the Na82 index is stronger in galaxies in the faint group compared to the bright group at low CaT values.
This suggests that the Na abundance is higher in the faint group than in the bright group.
Variations in the abundances of elements such as He, C, N and O are capable of producing colour variations of $\Delta(g - i) \sim 0.1$ and could be responsible for colour--metal index variations observed.

\subsubsection{Initial mass function differences?}
As can be seen in the upper left panel of Figure \ref{fig:stacking_CaT1}, populations rich in low mass stars have redder colours and weaker CaT strengths.
If, like the IMF of their host galaxies, the IMF of GCs was more dwarf rich in more massive galaxies then more massive galaxies would have GCs with redder colours and weaker CaT values compared to less massive galaxies in agreement with the colour--CaT variations seen.
However from the combination of CaT, Na82 and Fe86 indices, we do not see evidence that any of the galaxies host low mass star rich IMFs. 
Due to the effects of internal dynamical evolution, GCs preferentially lose low mass stars \citep[e.g.][]{1969A&A.....2..151H, 2003MNRAS.340..227B, 2009A&A...507.1409K}.
Thus their present day stellar mass distributions are different from those of idealised single stellar populations with the same IMF.
Removing low mass stars would make a GC bluer and appear to have a bottom light IMF which would translate into a stronger CaT value.
However, as can be seen in the upper left panel of Figure \ref{fig:stacking_CaT1}, the removal of low mass stars does not have a large enough effect on colour or CaT strength to explain the observed galaxy to galaxy variations.
We consider it unlikely that variations in the IMF or the present day mass function between galaxies can explain the observed metal line index--colour variations.

\subsubsection{Summary}
From the colour--CaT relations of individual GCs and from the differences in metal index strengths of stacked spectra with the same colours and luminosities, there is strong evidence that the GC colour--metallicity relation varies from galaxy to galaxy.
The data hints that lower mass galaxies have bluer, more linear colour--metallicity relations than higher mass galaxies. 
We do not believe that these variations are due to heterogeneous photometry or differences in GC luminosities between galaxies.
Homogeneous colour and metallicity data from more galaxies is required to confirm whether the variations in the colour--metallicity relation are real and to establish whether they are linked to galaxy mass or a related galaxy property such as stellar age.
We consider a variation in GC ages between galaxies to be a possible explanation for the observed colour--metallicity relations.
Alternatively, variations in the abundances of He, C, N or O between galaxies could cause the colour--metal line index relations to vary.
These two scenarios are not exclusive.
We do not consider differences in the IMF to be a likely cause.

It would be highly desirable to repeat the spectral stacking procedure with spectra at blue to try understand if the colour--metal index relation variations are caused by GC age variations.
Blue spectra would also allow us to look for variations in C or N abundance by studying the strengths of the associated molecular bands.
Acquiring photometry over a wider wavelength range would also be useful in understanding the origin of the colour-metal index variations.
For example, Optical-near-infrared colours such as $(i - K)$ are less sensitive to the effects of age and the horizontal branch morphology compared to $(g - i)$ while near-ultraviolet colours such as $(u - g)$ are more sensitive \citep[e.g.][]{2009ApJ...699..486C}.
Since abundance variations in MW GC stars are most apparent in the bluest filters \citep[e.g.][]{2013MNRAS.431.2126M}, $u$-band photometry would also be useful to distinguish abundance variations from age variations. 
For example, the calculations of \citet{2013ApJ...762..107G} predict a He, N and Na rich, C and O poor population to be 0.05 mag redder in $(u - g)$ and 0.05 mag bluer in $(g - i)$ than an population with a normal $\alpha$-enhanced abundance pattern.
In the\citet{2009ApJ...699..486C} models, a 9 Gyr solar metallicity population is 0.10 mag and 0.05 mag bluer in $(u - g)$ and $(g - i)$ respectively than a 12 Gyr population.

\subsection{Stellar population variation with luminosity}
\label{with_lumin}
As seen in Figure \ref{fig:mag_0}, both the bright and faint group blue stacks are significantly redder at brighter magnitudes.
This is the blue tilt that is seen in Figure \ref{fig:cmd}.
The slopes seen in the two groups are consistent with one another.
The observed blue tilt is smaller but consistent with the mean blue tilt observed by \citet{2010ApJ...710.1672M} despite the narrow colour range used for stacking.
The red stacks show no evidence for a colour--magnitude relation which is consistent with its non-detection in photometric studies such as \citet{2009ApJ...699..254H}, although \citet{2010ApJ...710.1672M} do observe a weak red tilt in the centres of their most massive galaxies. 

Both the blue faint group and the blue bright group CaT values become significantly stronger with magnitude.
Although the relations are offset from one another, the slopes are consistent.
No trend with magnitude is seen for the red spectra stacks.
Although no significant Fe86--magnitude trends are observed, both the bright and faint group blue stacks show slightly higher Fe86 at brighter magnitudes.
Both the blue faint groups and bright groups show higher Mg88 values at brighter magnitudes although only the bright groups shows a significant relation.
The red stacks show no Mg88--magnitude relation.
As seen in Figure \ref{fig:mag_1}, none of the Na82, Ti89, H$\alpha$ or PaT indices show any significant relations with magnitude.

Using Equation \ref{eq:CaT_colour} and the observed colour--magnitude relations, we predict CaT--magnitude relation slopes of $-0.148^{+0.051}_{-0.062}$ and $-0.115^{+0.064}_{-0.086}$ for the faint and bright groups respectively.
These predicted slopes are consistent with the observed CaT--magnitude slopes.
Using the Equation \ref{eq:SFe_fit} and the observed CaT--magnitude relation, we predict that the Fe86--magnitude slopes to be $-0.175^{+0.099}_{-0.061}$ and $-0.204^{+0.125}_{-0.101}$ for the faint and bright groups respectively, consistent with the observed Fe86--magnitude slopes.
Likewise, using Equation \ref{eq:Mg88_fit}, the observed CaT--magnitude relations predict Mg88--magnitude slopes of $-0.045^{+0.025}_{-0.015}$ and $-0.052^{+0.032}_{-0.026}$, again consistent with the observed relations.
Since at low CaT values we do not see any significant relations between CaT index strength and any of the Na82, Ti89, H$\alpha$ or PaT indices in the spectra stacked by colour, the lack of significant trends with magnitude for these indices is consistent with the CaT--magnitude trends we observe.
As the CaT index strongly correlates with metallicity, the colour-- and index--magnitude relations are all consistent with a metallicity--magnitude relation.

Since the slopes of the metal line index--magnitude relations are consistent with being explained by the observed colour--metallicity relations, we do not see any evidence for a changing colour--metallicity relation with magnitude.
The colour-- and index--magnitude slopes are consistent between the faint and bright groups, indicating that the blue tilt does not greatly vary from galaxy to galaxy.
However, photometric studies \citep[e.g.][]{2010ApJ...710.1672M} have found that the blue tilt is stronger in centres of giant galaxies than in the outskirts of giant galaxies or in dwarf galaxies.
The GCs in this study are on average further away from centres of their host galaxies than those in \citet{2010ApJ...710.1672M} so it is not surprising that our colour--magnitude slopes are closer to the those measured by \citet{2010ApJ...710.1672M} in the outer regions of giant galaxies.
For the same reason, the lack of a red tilt in our data is also consistent with \citet{2010ApJ...710.1672M} as they only find evidence for a red tilt in the centres of massive galaxies.

The observed trends are all consistent with the blue tilt being purely a metallicity--mass trend with no variations in age, element enhancement or IMF with magnitude. 
However, the uncertainties in our trends with magnitude are large so weak trends could still exist.
The lack of a bottom heavy IMF in the most luminous GCs (Section \ref{model_comparison}) and the consistency of the colour--magnitude trend with being a metallicity--magnitude trend suggest that internal dynamical evolution does not play a major role in producing the observed colour--magnitude trends contrary to the suggestions by \citet{2014ApJ...780...43G}.

This is not the first time the blue tilt has been observed spectroscopically; \citet{2010AJ....139.1566F} found that brighter blue GCs in NGC 1407 have stronger CaT values than fainter blue GCs.
We see the same CaT trend extending to 2 magnitudes fainter along with consistent trends in other metal indices.
\citet{2013ApJ...776L...7S} studied the variation of line indices and abundances with magnitude in metal rich GCs in M31.
They saw no variation in [Fe/H], [Mg/Fe] or [C/Fe] with mass but saw higher [N/Fe] abundances and stronger NaD indices at higher masses.  
We note that our sample of GCs are brighter on average than those in the \citet{2013ApJ...776L...7S} study.

Making the assumption that Mg88 follows [Mg/H], that the Mg88--magnitude relation is consistent with the CaT--magnitude relation suggests that [Mg/Fe] does not vary significantly with magnitude, consistent with the results of \citet{2013ApJ...776L...7S}.
As mention above, the Na82 index is difficult to interpret.
Under the assumption of no IMF or age variation with magnitude, the lack of a Na82--magnitude gradient could be seen as evidence of a lack of any [Na/Fe]--magnitude relationship.
However, the NaD--mass relation observed by \citet{2013ApJ...776L...7S} likely corresponds to a Na82--magnitude relation weak enough to be consistent with our observations.
If the blue tilt is caused by an extension of the light element variations seen in Milky Way GCs to higher masses, then we would expect light element enhancements to vary with luminosity.
The [N/Fe] and NaD trends observed by \citet{2013ApJ...776L...7S} at high metallicities are consistent with this picture.
Repeating the our stacking procedure with blue optical spectra would allow us to measure how [C/Fe] and [N/Fe] vary with magnitude by studying their molecular bands.

\section{Conclusions}
\label{conclusion}
Here we have stacked 1137 globular cluster (GC) spectra from 10 galaxies to understand their stellar populations.
Within each galaxy we stacked spectra by GC colour.
To disentangle the effects of the varying GC colour--calcium triplet (CaT) relation from trends with magnitude, we combined spectra from the four galaxies (NGC 3115, NGC 3377, NGC 4473 and NGC 4494) with bluer colour--CaT relations together in one group (which we name the faint group) and spectra from the four galaxies with redder colour--CaT relations (NGC 1407, NGC 4278, NGC 4365 and NGC 4649) in another group (the bright group).
This was done before stacking red and blue spectra from each group by magnitude.
For each stacked spectrum we calculated the mean colour and magnitude and measured the velocity dispersion and strengths of 7 spectral indices.
Since the colour and CaT values of the spectra stacked by colour agree with the mean values of individual spectra (Figure \ref{fig:CaT_colour}) and the velocity dispersions of the spectra stacked by magnitude agree with literature values (Figure \ref{fig:stacking_sigma}), we are confident of the stacking procedure.

Comparing our measurements with the simple stellar population models of \citet{2012MNRAS.424..157V} we find that the models generally agree with the observed relations between the spectral indices and $(g - i)$ colour.
However, the models do not predict the shapes of the observed colour--CaT or TiO89--CaT relations and predict stronger H$\alpha$ and PaT indices than observed.
Using the colours plus the CaT, Na82, Fe86, H$\alpha$ and PaT strengths of the stacked spectra, we find the observed GCs to be consistent with old ages and a bottom light IMF. 
Using the strengths of weak metal lines in the CaT region (i.e. the Fe86 index) to measure metallicity shows promise (Figure \ref{fig:weak_CaT_metal}, Section \ref{sec:SFe}).
However, better stellar population models at low metallicity which include the ability to vary chemical abundances and the horizontal branch morphology are needed to fully utilise our data.

In line with the previous studies by \citet{2011MNRAS.415.3393F} and \citet{2012MNRAS.426.1475U}, in Figure \ref{fig:CaT_colour} large variations in the colour--CaT relation can be seen.
Despite having consistent colours, the CaT, Fe86 and Mg88 indices are stronger in stacked spectra from the faint group compared to the bright group stacked spectra at both red and blue colours at fixed luminosity.
The sizes of these offsets are consistent with a metallicity difference between the two groups.
The combination of the variation in colour--CaT index relations and the metal index offsets seen between the groups of galaxies at fixed colours and luminosities makes it clear that the colour--metallicity relation varies galaxy to galaxy.
Variations in GC ages might explain the observed colour--metallicity variations.
Interestingly, the galaxies with bluer colour--metallicity relations have on average fainter luminosities.
This could indicate that GCs in less massive galaxies started forming slightly later and continued to form a few Gyr longer than in more massive galaxies.
Alternatively, differences in the average abundances of light elements such as He, C, N and O could produce the variations in colour--metallicity relations observed.
To understand if age or abundance variations are the explanation for the observed variations between optical colour and metal index strengths, a similar stacking study with spectra at blue wavelengths would be useful as would ultraviolet and near-infrared photometry.

As seen in Figure \ref{fig:mag_0}, at low metallicities we see that GC colours become redder and the CaT, Fe86 and Mg88 metal indices become stronger with brighter magnitudes.
The colour-- and metal index--magnitude relations are all consistent with the blue tilt observed in photometric studies such as \citet{2006AJ....132.2333S}, \citet{2009ApJ...699..254H} and \citet{2010ApJ...710.1672M}.
The blue tilt appears to be just a mass--metallicity relationship with no major change in the IMF or chemical abundances with mass.

In summary, by stacking Keck \textsc{deimos} spectra we find that most GCs are old and have Milky Way-like IMFs.
We see significant differences between galaxies in the strengths of several metal indices at fixed colour and luminosity which strongly suggests that the GC colour--metallicity relation is not universal.
The origin of this variation is presently unknown although age or abundance variations are possible explanations.
For metal poor GCs we also see that GC colours become redder and metal line indices become stronger with brighter magnitudes in line with the frequently observed blue tilt.  

\section*{Acknowledgements}
We thank Michael Murphy for valuable discussions as well as Gary Da Costa and John Blakeslee for their useful comments.
We thank Sreeja Kartha for her assistance with the NGC 2768 photometry.
We thank the referee, Guy Worthey, for his valuable comments which greatly improved the manuscript.
This research was based on data from the MILES project.
This research has made use of the NASA/IPAC Extragalactic Database (NED) which is operated by the Jet Propulsion Laboratory, California Institute of Technology, under contract with the National Aeronautics and Space Administration. 
The analysis pipeline used to reduce the DEIMOS data was developed at UC Berkeley with support from NSF grant AST-0071048.
This publication made use of \textsc{PyRAF} and \textsc{PyFITS} which are products of the Space Telescope Science Institute, which is operated by AURA for NASA.
This research made use of \textsc{TOPCAT} \citep{2005ASPC..347...29T}.
Some of the data presented herein were obtained at the W. M. Keck Observatory, operated as a scientific partnership among the California Institute of Technology, the University of California and the National Aeronautics and Space Administration, and made possible by the generous financial support of the W. M. Keck Foundation.
The authors wish to recognize and acknowledge the very significant cultural role and reverence that the summit of Mauna Kea has always had within the indigenous Hawaiian community.  We are most fortunate to have the opportunity to conduct observations from this mountain. 
This research is based in part on data collected at Subaru Telescope, which is operated by the National Astronomical Observatory of Japan. 
This research is based on observations made with the NASA/ESA Hubble Space Telescope, obtained from the data archive at the Space Telescope Science Institute. STScI is operated by the Association of Universities for Research in Astronomy, Inc. under NASA contract NAS 5-26555.
This material is based upon work supported by the Australian Research Council under grant DP130100388 and by the National Science Foundation under grants AST-1109878 and AST-1211995.
Part of this work was performed on the gSTAR national facility and the swinSTAR supercomputer at Swinburne University of Technology.
gSTAR is funded by Swinburne University of Technology and the Australian Government’s Education Investment Fund.

\bibliographystyle{mn2e}
\bibliography{stacking}

\begin{thebibliography}{117}
\expandafter\ifx\csname natexlab\endcsname\relax\def\natexlab#1{#1}\fi

\bibitem[{{Andrae}(2010)}]{2010arXiv1009.2755A}
{Andrae} R., 2010, ArXiv 1009.2755

\bibitem[{{Bailin} \& {Harris}(2009)}]{2009ApJ...695.1082B}
{Bailin} J., {Harris} W.~E., 2009, \apj, 695, 1082

\bibitem[{{Baumgardt} \& {Makino}(2003)}]{2003MNRAS.340..227B}
{Baumgardt} H., {Makino} J., 2003, \mnras, 340, 227

\bibitem[{{Blom} {et~al.}(2012){Blom}, {Spitler}, \&
  {Forbes}}]{2012MNRAS.420...37B}
{Blom} C., {Spitler} L.~R., {Forbes} D.~A., 2012, \mnras, 420, 37

\bibitem[{{Brodie} \& {Huchra}(1991)}]{1991ApJ...379..157B}
{Brodie} J.~P., {Huchra} J.~P., 1991, \apj, 379, 157

\bibitem[{{Brodie} {et~al.}(2011){Brodie}, {Romanowsky}, {Strader}, \&
  {Forbes}}]{2011AJ....142..199B}
{Brodie} J.~P., {Romanowsky} A.~J., {Strader} J., {Forbes} D.~A., 2011, \aj,
  142, 199

\bibitem[{{Brodie} {et~al.}(2014){Brodie}, {Romanowsky}, {Strader}, {Forbes},
  {Foster}, {Jennings}, {Pastorello}, {Pota}, {Usher}, {Blom}, {Kader},
  {Roediger}, {Spitler}, {Villaume}, {Arnold}, {Kartha}, \& {Woodley}}]{SLUGGS}
{Brodie} J.~P., {Romanowsky} A.~J., {Strader} J., {Forbes} D.~A., {Foster} C.,
  {Jennings} Z.~G., {Pastorello} N., {Pota} V., {Usher} C., {Blom} C., {Kader}
  J., {Roediger} J.~C., {Spitler} L.~R., {Villaume} A., {Arnold} J.~A.,
  {Kartha} S.~S., {Woodley} K.~A., 2014, {ApJ in press, arXiv:1405.2079}

\bibitem[{{Brodie} {et~al.}(2012){Brodie}, {Usher}, {Conroy}, {Strader},
  {Arnold}, {Forbes}, \& {Romanowsky}}]{2012ApJ...759L..33B}
{Brodie} J.~P., {Usher} C., {Conroy} C., {Strader} J., {Arnold} J.~A., {Forbes}
  D.~A., {Romanowsky} A.~J., 2012, \apjl, 759, L33

\bibitem[{{Cappellari} \& {Emsellem}(2004)}]{2004PASP..116..138C}
{Cappellari} M., {Emsellem} E., 2004, \pasp, 116, 138

\bibitem[{{Cappellari} {et~al.}(2011){Cappellari}, {Emsellem}, {Krajnovi{\'c}},
  {McDermid}, {Scott}, {Verdoes Kleijn}, {Young}, {Alatalo}, {Bacon}, {Blitz},
  {Bois}, {Bournaud}, {Bureau}, {Davies}, {Davis}, {de Zeeuw}, {Duc},
  {Khochfar}, {Kuntschner}, {Lablanche}, {Morganti}, {Naab}, {Oosterloo},
  {Sarzi}, {Serra}, \& {Weijmans}}]{2011MNRAS.413..813C}
{Cappellari} M., {Emsellem} E., {Krajnovi{\'c}} D., {McDermid} R.~M., {Scott}
  N., {Verdoes Kleijn} G.~A., {Young} L.~M., {Alatalo} K., {Bacon} R., {Blitz}
  L., {Bois} M., {Bournaud} F., {Bureau} M., {Davies} R.~L., {Davis} T.~A., {de
  Zeeuw} P.~T., {Duc} P.-A., {Khochfar} S., {Kuntschner} H., {Lablanche} P.-Y.,
  {Morganti} R., {Naab} T., {Oosterloo} T., {Sarzi} M., {Serra} P., {Weijmans}
  A.-M., 2011, \mnras, 413, 813

\bibitem[{{Cappellari} {et~al.}(2012){Cappellari}, {McDermid}, {Alatalo},
  {Blitz}, {Bois}, {Bournaud}, {Bureau}, {Crocker}, {Davies}, {Davis}, {de
  Zeeuw}, {Duc}, {Emsellem}, {Khochfar}, {Krajnovi{\'c}}, {Kuntschner},
  {Lablanche}, {Morganti}, {Naab}, {Oosterloo}, {Sarzi}, {Scott}, {Serra},
  {Weijmans}, \& {Young}}]{2012Natur.484..485C}
{Cappellari} M., {McDermid} R.~M., {Alatalo} K., {Blitz} L., {Bois} M.,
  {Bournaud} F., {Bureau} M., {Crocker} A.~F., {Davies} R.~L., {Davis} T.~A.,
  {de Zeeuw} P.~T., {Duc} P.-A., {Emsellem} E., {Khochfar} S., {Krajnovi{\'c}}
  D., {Kuntschner} H., {Lablanche} P.-Y., {Morganti} R., {Naab} T., {Oosterloo}
  T., {Sarzi} M., {Scott} N., {Serra} P., {Weijmans} A.-M., {Young} L.~M.,
  2012, \nat, 484, 485

\bibitem[{{Carretta}(2006)}]{2006AJ....131.1766C}
{Carretta} E., 2006, \aj, 131, 1766

\bibitem[{{Carretta} {et~al.}(2010){Carretta}, {Bragaglia}, {Gratton},
  {Recio-Blanco}, {Lucatello}, {D'Orazi}, \& {Cassisi}}]{2010A&A...516A..55C}
{Carretta} E., {Bragaglia} A., {Gratton} R.~G., {Recio-Blanco} A., {Lucatello}
  S., {D'Orazi} V., {Cassisi} S., 2010, \aap, 516, A55

\bibitem[{{Carretta} {et~al.}(2011){Carretta}, {Lucatello}, {Gratton},
  {Bragaglia}, \& {D'Orazi}}]{2011A&A...533A..69C}
{Carretta} E., {Lucatello} S., {Gratton} R.~G., {Bragaglia} A., {D'Orazi} V.,
  2011, \aap, 533, A69

\bibitem[{{Cenarro} {et~al.}(2001){Cenarro}, {Cardiel}, {Gorgas}, {Peletier},
  {Vazdekis}, \& {Prada}}]{2001MNRAS.326..959C}
{Cenarro} A.~J., {Cardiel} N., {Gorgas} J., {Peletier} R.~F., {Vazdekis} A.,
  {Prada} F., 2001, \mnras, 326, 959

\bibitem[{{Cenarro} {et~al.}(2009){Cenarro}, {Cardiel}, {Vazdekis}, \&
  {Gorgas}}]{2009MNRAS.396.1895C}
{Cenarro} A.~J., {Cardiel} N., {Vazdekis} A., {Gorgas} J., 2009, \mnras, 396,
  1895

\bibitem[{{Chomiuk} {et~al.}(2008){Chomiuk}, {Strader}, \&
  {Brodie}}]{2008AJ....136..234C}
{Chomiuk} L., {Strader} J., {Brodie} J.~P., 2008, \aj, 136, 234

\bibitem[{{Chung} {et~al.}(2013{\natexlab{a}}){Chung}, {Lee}, {Yoon}, \&
  {Lee}}]{2013ApJ...769L...3C}
{Chung} C., {Lee} S.-Y., {Yoon} S.-J., {Lee} Y.-W., 2013{\natexlab{a}}, \apjl,
  769, L3

\bibitem[{{Chung} {et~al.}(2013{\natexlab{b}}){Chung}, {Yoon}, {Lee}, \&
  {Lee}}]{2013ApJS..204....3C}
{Chung} C., {Yoon} S.-J., {Lee} S.-Y., {Lee} Y.-W., 2013{\natexlab{b}}, \apjs,
  204, 3

\bibitem[{{Coelho} {et~al.}(2007){Coelho}, {Bruzual}, {Charlot}, {Weiss},
  {Barbuy}, \& {Ferguson}}]{2007MNRAS.382..498C}
{Coelho} P., {Bruzual} G., {Charlot} S., {Weiss} A., {Barbuy} B., {Ferguson}
  J.~W., 2007, \mnras, 382, 498

\bibitem[{{Cohen} {et~al.}(1998){Cohen}, {Blakeslee}, \&
  {Ryzhov}}]{1998ApJ...496..808C}
{Cohen} J.~G., {Blakeslee} J.~P., {Ryzhov} A., 1998, \apj, 496, 808

\bibitem[{{Conroy} {et~al.}(2014){Conroy}, {Graves}, \& {van
  Dokkum}}]{2014ApJ...780...33C}
{Conroy} C., {Graves} G.~J., {van Dokkum} P.~G., 2014, \apj, 780, 33

\bibitem[{{Conroy} {et~al.}(2009){Conroy}, {Gunn}, \&
  {White}}]{2009ApJ...699..486C}
{Conroy} C., {Gunn} J.~E., {White} M., 2009, \apj, 699, 486

\bibitem[{{Conroy} \& {Spergel}(2011)}]{2011ApJ...726...36C}
{Conroy} C., {Spergel} D.~N., 2011, \apj, 726, 36

\bibitem[{{Conroy} \& {van Dokkum}(2012{\natexlab{a}})}]{2012ApJ...747...69C}
{Conroy} C., {van Dokkum} P., 2012{\natexlab{a}}, \apj, 747, 69

\bibitem[{{Conroy} \& {van Dokkum}(2012{\natexlab{b}})}]{2012ApJ...760...71C}
{Conroy} C., {van Dokkum} P.~G., 2012{\natexlab{b}}, \apj, 760, 71

\bibitem[{{Cooper} {et~al.}(2012){Cooper}, {Newman}, {Davis}, {Finkbeiner}, \&
  {Gerke}}]{2012ascl.soft03003C}
{Cooper} M.~C., {Newman} J.~A., {Davis} M., {Finkbeiner} D.~P., {Gerke} B.~F.,
  2012, in Astrophysics Source Code Library, record ascl:1203.003, p. 3003

\bibitem[{{Da Costa} {et~al.}(2014){Da Costa}, {Held}, \&
  {Saviane}}]{2014MNRAS.438.3507D}
{Da Costa} G.~S., {Held} E.~V., {Saviane} I., 2014, \mnras, 438, 3507

\bibitem[{{Dalessandro} {et~al.}(2012){Dalessandro}, {Schiavon}, {Rood},
  {Ferraro}, {Sohn}, {Lanzoni}, \& {O'Connell}}]{2012AJ....144..126D}
{Dalessandro} E., {Schiavon} R.~P., {Rood} R.~T., {Ferraro} F.~R., {Sohn}
  S.~T., {Lanzoni} B., {O'Connell} R.~W., 2012, \aj, 144, 126

\bibitem[{{Decressin} {et~al.}(2007){Decressin}, {Meynet}, {Charbonnel},
  {Prantzos}, \& {Ekstr{\"o}m}}]{2007A&A...464.1029D}
{Decressin} T., {Meynet} G., {Charbonnel} C., {Prantzos} N., {Ekstr{\"o}m} S.,
  2007, \aap, 464, 1029

\bibitem[{{D'Ercole} {et~al.}(2010){D'Ercole}, {D'Antona}, {Ventura},
  {Vesperini}, \& {McMillan}}]{2010MNRAS.407..854D}
{D'Ercole} A., {D'Antona} F., {Ventura} P., {Vesperini} E., {McMillan}
  S.~L.~W., 2010, \mnras, 407, 854

\bibitem[{{Dotter} {et~al.}(2011){Dotter}, {Sarajedini}, \&
  {Anderson}}]{2011ApJ...738...74D}
{Dotter} A., {Sarajedini} A., {Anderson} J., 2011, \apj, 738, 74

\bibitem[{{Dotter} {et~al.}(2010){Dotter}, {Sarajedini}, {Anderson},
  {Aparicio}, {Bedin}, {Chaboyer}, {Majewski}, {Mar{\'{\i}}n-Franch}, {Milone},
  {Paust}, {Piotto}, {Reid}, {Rosenberg}, \& {Siegel}}]{2010ApJ...708..698D}
{Dotter} A., {Sarajedini} A., {Anderson} J., {Aparicio} A., {Bedin} L.~R.,
  {Chaboyer} B., {Majewski} S., {Mar{\'{\i}}n-Franch} A., {Milone} A., {Paust}
  N., {Piotto} G., {Reid} I.~N., {Rosenberg} A., {Siegel} M., 2010, \apj, 708,
  698

\bibitem[{{Faber} {et~al.}(2003){Faber}, {Phillips}, {Kibrick}, {Alcott},
  {Allen}, {Burrous}, {Cantrall}, {Clarke}, {Coil}, {Cowley}, {Davis}, {Deich},
  {Dietsch}, {Gilmore}, {Harper}, {Hilyard}, {Lewis}, {McVeigh}, {Newman},
  {Osborne}, {Schiavon}, {Stover}, {Tucker}, {Wallace}, {Wei}, {Wirth}, \&
  {Wright}}]{2003SPIE.4841.1657F}
{Faber} S.~M., {Phillips} A.~C., {Kibrick} R.~I., {Alcott} B., {Allen} S.~L.,
  {Burrous} J., {Cantrall} T., {Clarke} D., {Coil} A.~L., {Cowley} D.~J.,
  {Davis} M., {Deich} W.~T.~S., {Dietsch} K., {Gilmore} D.~K., {Harper} C.~A.,
  {Hilyard} D.~F., {Lewis} J.~P., {McVeigh} M., {Newman} J., {Osborne} J.,
  {Schiavon} R., {Stover} R.~J., {Tucker} D., {Wallace} V., {Wei} M., {Wirth}
  G., {Wright} C.~A., 2003, in Presented at the Society of Photo-Optical
  Instrumentation Engineers (SPIE) Conference, Vol. 4841, Society of
  Photo-Optical Instrumentation Engineers (SPIE) Conference Series, {M.~Iye \&
  A.~F.~M.~Moorwood}, ed., pp. 1657--1669

\bibitem[{{Forbes} {et~al.}(2001){Forbes}, {Beasley}, {Brodie}, \&
  {Kissler-Patig}}]{2001ApJ...563L.143F}
{Forbes} D.~A., {Beasley} M.~A., {Brodie} J.~P., {Kissler-Patig} M., 2001,
  \apjl, 563, L143

\bibitem[{{Forbes} \& {Bridges}(2010)}]{2010MNRAS.404.1203F}
{Forbes} D.~A., {Bridges} T., 2010, \mnras, 404, 1203

\bibitem[{{Forbes} {et~al.}(2008){Forbes}, {Lasky}, {Graham}, \&
  {Spitler}}]{2008MNRAS.389.1924F}
{Forbes} D.~A., {Lasky} P., {Graham} A.~W., {Spitler} L., 2008, \mnras, 389,
  1924

\bibitem[{{Forbes} {et~al.}(2012){Forbes}, {Ponman}, \&
  {O'Sullivan}}]{2012MNRAS.425...66F}
{Forbes} D.~A., {Ponman} T., {O'Sullivan} E., 2012, \mnras, 425, 66

\bibitem[{{Forbes} {et~al.}(2007){Forbes}, {Proctor}, {Strader}, \&
  {Brodie}}]{2007ApJ...659..188F}
{Forbes} D.~A., {Proctor} R., {Strader} J., {Brodie} J.~P., 2007, \apj, 659,
  188

\bibitem[{{Foster} {et~al.}(2010){Foster}, {Forbes}, {Proctor}, {Strader},
  {Brodie}, \& {Spitler}}]{2010AJ....139.1566F}
{Foster} C., {Forbes} D.~A., {Proctor} R.~N., {Strader} J., {Brodie} J.~P.,
  {Spitler} L.~R., 2010, \aj, 139, 1566

\bibitem[{{Foster} {et~al.}(2011){Foster}, {Spitler}, {Romanowsky}, {Forbes},
  {Pota}, {Bekki}, {Strader}, {Proctor}, {Arnold}, \&
  {Brodie}}]{2011MNRAS.415.3393F}
{Foster} C., {Spitler} L.~R., {Romanowsky} A.~J., {Forbes} D.~A., {Pota} V.,
  {Bekki} K., {Strader} J., {Proctor} R.~N., {Arnold} J.~A., {Brodie} J.~P.,
  2011, \mnras, 415, 3393

\bibitem[{{Freeman} \& {Rodgers}(1975)}]{1975ApJ...201L..71F}
{Freeman} K.~C., {Rodgers} A.~W., 1975, \apjl, 201, L71

\bibitem[{{Gallazzi} {et~al.}(2006){Gallazzi}, {Charlot}, {Brinchmann}, \&
  {White}}]{2006MNRAS.370.1106G}
{Gallazzi} A., {Charlot} S., {Brinchmann} J., {White} S.~D.~M., 2006, \mnras,
  370, 1106

\bibitem[{{Girardi} {et~al.}(2000){Girardi}, {Bressan}, {Bertelli}, \&
  {Chiosi}}]{2000A&AS..141..371G}
{Girardi} L., {Bressan} A., {Bertelli} G., {Chiosi} C., 2000, \aaps, 141, 371

\bibitem[{{Goudfrooij} \& {Kruijssen}(2013)}]{2013ApJ...762..107G}
{Goudfrooij} P., {Kruijssen} J.~M.~D., 2013, \apj, 762, 107

\bibitem[{{Goudfrooij} \& {Kruijssen}(2014)}]{2014ApJ...780...43G}
---, 2014, \apj, 780, 43

\bibitem[{{Gratton} {et~al.}(2012){Gratton}, {Carretta}, \&
  {Bragaglia}}]{2012A&ARv..20...50G}
{Gratton} R.~G., {Carretta} E., {Bragaglia} A., 2012, \aapr, 20, 50

\bibitem[{{Gratton} {et~al.}(2010){Gratton}, {Carretta}, {Bragaglia},
  {Lucatello}, \& {D'Orazi}}]{2010A&A...517A..81G}
{Gratton} R.~G., {Carretta} E., {Bragaglia} A., {Lucatello} S., {D'Orazi} V.,
  2010, \aap, 517, A81+

\bibitem[{{Harris}(2009)}]{2009ApJ...699..254H}
{Harris} W.~E., 2009, \apj, 699, 254

\bibitem[{{Harris} {et~al.}(2006){Harris}, {Whitmore}, {Karakla}, {Oko{\'n}},
  {Baum}, {Hanes}, \& {Kavelaars}}]{2006ApJ...636...90H}
{Harris} W.~E., {Whitmore} B.~C., {Karakla} D., {Oko{\'n}} W., {Baum} W.~A.,
  {Hanes} D.~A., {Kavelaars} J.~J., 2006, \apj, 636, 90

\bibitem[{{Henon}(1969)}]{1969A&A.....2..151H}
{Henon} M., 1969, \aap, 2, 151

\bibitem[{{Jennings} {et~al.}(2014){Jennings}, {Strader}, {Romanowsky},
  {Brodie}, {Arnold}, {Lin}, {Irwin}, {Sivakoff}, \&
  {Wong}}]{2014AJ....148...32J}
{Jennings} Z.~G., {Strader} J., {Romanowsky} A.~J., {Brodie} J.~P., {Arnold}
  J.~A., {Lin} D., {Irwin} J.~A., {Sivakoff} G.~R., {Wong} K.-W., 2014, \aj,
  148, 32

\bibitem[{{Johansson} {et~al.}(2012){Johansson}, {Thomas}, \&
  {Maraston}}]{2012MNRAS.421.1908J}
{Johansson} J., {Thomas} D., {Maraston} C., 2012, \mnras, 421, 1908

\bibitem[{{Jord{\'a}n} {et~al.}(2009){Jord{\'a}n}, {Peng}, {Blakeslee},
  {C{\^o}t{\'e}}, {Eyheramendy}, {Ferrarese}, {Mei}, {Tonry}, \&
  {West}}]{2009ApJS..180...54J}
{Jord{\'a}n} A., {Peng} E.~W., {Blakeslee} J.~P., {C{\^o}t{\'e}} P.,
  {Eyheramendy} S., {Ferrarese} L., {Mei} S., {Tonry} J.~L., {West} M.~J.,
  2009, \apjs, 180, 54

\bibitem[{{Kartha} {et~al.}(2014){Kartha}, {Forbes}, {Spitler}, {Romanowsky},
  {Arnold}, \& {Brodie}}]{2014MNRAS.437..273K}
{Kartha} S.~S., {Forbes} D.~A., {Spitler} L.~R., {Romanowsky} A.~J., {Arnold}
  J.~A., {Brodie} J.~P., 2014, \mnras, 437, 273

\bibitem[{{Kirby} {et~al.}(2011{\natexlab{a}}){Kirby}, {Cohen}, {Smith},
  {Majewski}, {Sohn}, \& {Guhathakurta}}]{2011ApJ...727...79K}
{Kirby} E.~N., {Cohen} J.~G., {Smith} G.~H., {Majewski} S.~R., {Sohn} S.~T.,
  {Guhathakurta} P., 2011{\natexlab{a}}, \apj, 727, 79

\bibitem[{{Kirby} {et~al.}(2008){Kirby}, {Guhathakurta}, \&
  {Sneden}}]{2008ApJ...682.1217K}
{Kirby} E.~N., {Guhathakurta} P., {Sneden} C., 2008, \apj, 682, 1217

\bibitem[{{Kirby} {et~al.}(2011{\natexlab{b}}){Kirby}, {Lanfranchi}, {Simon},
  {Cohen}, \& {Guhathakurta}}]{2011ApJ...727...78K}
{Kirby} E.~N., {Lanfranchi} G.~A., {Simon} J.~D., {Cohen} J.~G., {Guhathakurta}
  P., 2011{\natexlab{b}}, \apj, 727, 78

\bibitem[{{Kroupa}(2001)}]{2001MNRAS.322..231K}
{Kroupa} P., 2001, \mnras, 322, 231

\bibitem[{{Kruijssen}(2009)}]{2009A&A...507.1409K}
{Kruijssen} J.~M.~D., 2009, \aap, 507, 1409

\bibitem[{{Kuntschner} {et~al.}(2010){Kuntschner}, {Emsellem}, {Bacon},
  {Cappellari}, {Davies}, {de Zeeuw}, {Falc{\'o}n-Barroso}, {Krajnovi{\'c}},
  {McDermid}, {Peletier}, {Sarzi}, {Shapiro}, {van den Bosch}, \& {van de
  Ven}}]{2010MNRAS.408...97K}
{Kuntschner} H., {Emsellem} E., {Bacon} R., {Cappellari} M., {Davies} R.~L.,
  {de Zeeuw} P.~T., {Falc{\'o}n-Barroso} J., {Krajnovi{\'c}} D., {McDermid}
  R.~M., {Peletier} R.~F., {Sarzi} M., {Shapiro} K.~L., {van den Bosch}
  R.~C.~E., {van de Ven} G., 2010, \mnras, 408, 97

\bibitem[{{La Barbera} {et~al.}(2013){La Barbera}, {Ferreras}, {Vazdekis}, {de
  la Rosa}, {de Carvalho}, {Trevisan}, {Falc{\'o}n-Barroso}, \&
  {Ricciardelli}}]{2013MNRAS.433.3017L}
{La Barbera} F., {Ferreras} I., {Vazdekis} A., {de la Rosa} I.~G., {de
  Carvalho} R.~R., {Trevisan} M., {Falc{\'o}n-Barroso} J., {Ricciardelli} E.,
  2013, \mnras, 433, 3017

\bibitem[{{Leaman} {et~al.}(2013){Leaman}, {VandenBerg}, \&
  {Mendel}}]{2013MNRAS.436..122L}
{Leaman} R., {VandenBerg} D.~A., {Mendel} J.~T., 2013, \mnras, 436, 122

\bibitem[{{Lee} {et~al.}(2009){Lee}, {Worthey}, {Dotter}, {Chaboyer},
  {Jevremovi{\'c}}, {Baron}, {Briley}, {Ferguson}, {Coelho}, \&
  {Trager}}]{2009ApJ...694..902L}
{Lee} H.-c., {Worthey} G., {Dotter} A., {Chaboyer} B., {Jevremovi{\'c}} D.,
  {Baron} E., {Briley} M.~M., {Ferguson} J.~W., {Coelho} P., {Trager} S.~C.,
  2009, \apj, 694, 902

\bibitem[{{Maraston}(2005)}]{2005MNRAS.362..799M}
{Maraston} C., 2005, \mnras, 362, 799

\bibitem[{{Marino} {et~al.}(2011{\natexlab{a}}){Marino}, {Milone}, {Piotto},
  {Villanova}, {Gratton}, {D'Antona}, {Anderson}, {Bedin}, {Bellini},
  {Cassisi}, {Geisler}, {Renzini}, \& {Zoccali}}]{2011ApJ...731...64M}
{Marino} A.~F., {Milone} A.~P., {Piotto} G., {Villanova} S., {Gratton} R.,
  {D'Antona} F., {Anderson} J., {Bedin} L.~R., {Bellini} A., {Cassisi} S.,
  {Geisler} D., {Renzini} A., {Zoccali} M., 2011{\natexlab{a}}, \apj, 731, 64

\bibitem[{{Marino} {et~al.}(2011{\natexlab{b}}){Marino}, {Sneden}, {Kraft},
  {Wallerstein}, {Norris}, {da Costa}, {Milone}, {Ivans}, {Gonzalez},
  {Fulbright}, {Hilker}, {Piotto}, {Zoccali}, \&
  {Stetson}}]{2011A&A...532A...8M}
{Marino} A.~F., {Sneden} C., {Kraft} R.~P., {Wallerstein} G., {Norris} J.~E.,
  {da Costa} G., {Milone} A.~P., {Ivans} I.~I., {Gonzalez} G., {Fulbright}
  J.~P., {Hilker} M., {Piotto} G., {Zoccali} M., {Stetson} P.~B.,
  2011{\natexlab{b}}, \aap, 532, A8

\bibitem[{{Mei} {et~al.}(2007){Mei}, {Blakeslee}, {C{\^o}t{\'e}}, {Tonry},
  {West}, {Ferrarese}, {Jord{\'a}n}, {Peng}, {Anthony}, \&
  {Merritt}}]{2007ApJ...655..144M}
{Mei} S., {Blakeslee} J.~P., {C{\^o}t{\'e}} P., {Tonry} J.~L., {West} M.~J.,
  {Ferrarese} L., {Jord{\'a}n} A., {Peng} E.~W., {Anthony} A., {Merritt} D.,
  2007, \apj, 655, 144

\bibitem[{{Mieske} {et~al.}(2010){Mieske}, {Jord{\'a}n}, {C{\^o}t{\'e}},
  {Peng}, {Ferrarese}, {Blakeslee}, {Mei}, {Baumgardt}, {Tonry}, {Infante}, \&
  {West}}]{2010ApJ...710.1672M}
{Mieske} S., {Jord{\'a}n} A., {C{\^o}t{\'e}} P., {Peng} E.~W., {Ferrarese} L.,
  {Blakeslee} J.~P., {Mei} S., {Baumgardt} H., {Tonry} J.~L., {Infante} L.,
  {West} M.~J., 2010, \apj, 710, 1672

\bibitem[{{Milone} {et~al.}(2014){Milone}, {Marino}, {Dotter}, {Norris},
  {Jerjen}, {Piotto}, {Cassisi}, {Bedin}, {Recio Blanco}, {Sarajedini},
  {Asplund}, {Monelli}, \& {Aparicio}}]{2014ApJ...785...21M}
{Milone} A.~P., {Marino} A.~F., {Dotter} A., {Norris} J.~E., {Jerjen} H.,
  {Piotto} G., {Cassisi} S., {Bedin} L.~R., {Recio Blanco} A., {Sarajedini} A.,
  {Asplund} M., {Monelli} M., {Aparicio} A., 2014, \apj, 785, 21

\bibitem[{{Miyazaki} {et~al.}(2002){Miyazaki}, {Komiyama}, {Sekiguchi},
  {Okamura}, {Doi}, {Furusawa}, {Hamabe}, {Imi}, {Kimura}, {Nakata}, {Okada},
  {Ouchi}, {Shimasaku}, {Yagi}, \& {Yasuda}}]{2002PASJ...54..833M}
{Miyazaki} S., {Komiyama} Y., {Sekiguchi} M., {Okamura} S., {Doi} M.,
  {Furusawa} H., {Hamabe} M., {Imi} K., {Kimura} M., {Nakata} F., {Okada} N.,
  {Ouchi} M., {Shimasaku} K., {Yagi} M., {Yasuda} N., 2002, \pasj, 54, 833

\bibitem[{{Monelli} {et~al.}(2013){Monelli}, {Milone}, {Stetson}, {Marino},
  {Cassisi}, {del Pino Molina}, {Salaris}, {Aparicio}, {Asplund}, {Grundahl},
  {Piotto}, {Weiss}, {Carrera}, {Cebri{\'a}n}, {Murabito}, {Pietrinferni}, \&
  {Sbordone}}]{2013MNRAS.431.2126M}
{Monelli} M., {Milone} A.~P., {Stetson} P.~B., {Marino} A.~F., {Cassisi} S.,
  {del Pino Molina} A., {Salaris} M., {Aparicio} A., {Asplund} M., {Grundahl}
  F., {Piotto} G., {Weiss} A., {Carrera} R., {Cebri{\'a}n} M., {Murabito} S.,
  {Pietrinferni} A., {Sbordone} L., 2013, \mnras, 431, 2126

\bibitem[{{Nelan} {et~al.}(2005){Nelan}, {Smith}, {Hudson}, {Wegner}, {Lucey},
  {Moore}, {Quinney}, \& {Suntzeff}}]{2005ApJ...632..137N}
{Nelan} J.~E., {Smith} R.~J., {Hudson} M.~J., {Wegner} G.~A., {Lucey} J.~R.,
  {Moore} S.~A.~W., {Quinney} S.~J., {Suntzeff} N.~B., 2005, \apj, 632, 137

\bibitem[{{Newman} {et~al.}(2013){Newman}, {Cooper}, {Davis}, {Faber}, {Coil},
  {Guhathakurta}, {Koo}, {Phillips}, {Conroy}, {Dutton}, {Finkbeiner}, {Gerke},
  {Rosario}, {Weiner}, {Willmer}, {Yan}, {Harker}, {Kassin}, {Konidaris},
  {Lai}, {Madgwick}, {Noeske}, {Wirth}, {Connolly}, {Kaiser}, {Kirby},
  {Lemaux}, {Lin}, {Lotz}, {Luppino}, {Marinoni}, {Matthews}, {Metevier}, \&
  {Schiavon}}]{2013ApJS..208....5N}
{Newman} J.~A., {Cooper} M.~C., {Davis} M., {Faber} S.~M., {Coil} A.~L.,
  {Guhathakurta} P., {Koo} D.~C., {Phillips} A.~C., {Conroy} C., {Dutton}
  A.~A., {Finkbeiner} D.~P., {Gerke} B.~F., {Rosario} D.~J., {Weiner} B.~J.,
  {Willmer} C.~N.~A., {Yan} R., {Harker} J.~J., {Kassin} S.~A., {Konidaris}
  N.~P., {Lai} K., {Madgwick} D.~S., {Noeske} K.~G., {Wirth} G.~D., {Connolly}
  A.~J., {Kaiser} N., {Kirby} E.~N., {Lemaux} B.~C., {Lin} L., {Lotz} J.~M.,
  {Luppino} G.~A., {Marinoni} C., {Matthews} D.~J., {Metevier} A., {Schiavon}
  R.~P., 2013, \apjs, 208, 5

\bibitem[{{Norris} \& {Da Costa}(1995)}]{1995ApJ...447..680N}
{Norris} J.~E., {Da Costa} G.~S., 1995, \apj, 447, 680

\bibitem[{{Norris} {et~al.}(2008){Norris}, {Sharples}, {Bridges}, {Gebhardt},
  {Forbes}, {Proctor}, {Faifer}, {Forte}, {Beasley}, {Zepf}, \&
  {Hanes}}]{2008MNRAS.385...40N}
{Norris} M.~A., {Sharples} R.~M., {Bridges} T., {Gebhardt} K., {Forbes} D.~A.,
  {Proctor} R., {Faifer} F.~R., {Forte} J.~C., {Beasley} M.~A., {Zepf} S.~E.,
  {Hanes} D.~A., 2008, \mnras, 385, 40

\bibitem[{{Peng} {et~al.}(2006){Peng}, {Jord{\'a}n}, {C{\^o}t{\'e}},
  {Blakeslee}, {Ferrarese}, {Mei}, {West}, {Merritt}, {Milosavljevi{\'c}}, \&
  {Tonry}}]{2006ApJ...639...95P}
{Peng} E.~W., {Jord{\'a}n} A., {C{\^o}t{\'e}} P., {Blakeslee} J.~P.,
  {Ferrarese} L., {Mei} S., {West} M.~J., {Merritt} D., {Milosavljevi{\'c}} M.,
  {Tonry} J.~L., 2006, \apj, 639, 95

\bibitem[{{Pierce} {et~al.}(2006){Pierce}, {Bridges}, {Forbes}, {Proctor},
  {Beasley}, {Gebhardt}, {Faifer}, {Forte}, {Zepf}, {Sharples}, \&
  {Hanes}}]{2006MNRAS.368..325P}
{Pierce} M., {Bridges} T., {Forbes} D.~A., {Proctor} R., {Beasley} M.~A.,
  {Gebhardt} K., {Faifer} F.~R., {Forte} J.~C., {Zepf} S.~E., {Sharples} R.,
  {Hanes} D.~A., 2006, \mnras, 368, 325

\bibitem[{{Pota} {et~al.}(2013{\natexlab{a}}){Pota}, {Forbes}, {Romanowsky},
  {Brodie}, {Spitler}, {Strader}, {Foster}, {Arnold}, {Benson}, {Blom},
  {Hargis}, {Rhode}, \& {Usher}}]{2013MNRAS.428..389P}
{Pota} V., {Forbes} D.~A., {Romanowsky} A.~J., {Brodie} J.~P., {Spitler} L.~R.,
  {Strader} J., {Foster} C., {Arnold} J.~A., {Benson} A., {Blom} C., {Hargis}
  J.~R., {Rhode} K.~L., {Usher} C., 2013{\natexlab{a}}, \mnras, 428, 389

\bibitem[{{Pota} {et~al.}(2013{\natexlab{b}}){Pota}, {Graham}, {Forbes},
  {Romanowsky}, {Brodie}, \& {Strader}}]{2013MNRAS.433..235P}
{Pota} V., {Graham} A.~W., {Forbes} D.~A., {Romanowsky} A.~J., {Brodie} J.~P.,
  {Strader} J., 2013{\natexlab{b}}, \mnras, 433, 235

\bibitem[{{Puzia} {et~al.}(2005){Puzia}, {Kissler-Patig}, {Thomas}, {Maraston},
  {Saglia}, {Bender}, {Goudfrooij}, \& {Hempel}}]{2005A&A...439..997P}
{Puzia} T.~H., {Kissler-Patig} M., {Thomas} D., {Maraston} C., {Saglia} R.~P.,
  {Bender} R., {Goudfrooij} P., {Hempel} M., 2005, \aap, 439, 997

\bibitem[{{Rey} {et~al.}(2007){Rey}, {Rich}, {Sohn}, {Yoon}, {Chung}, {Yi},
  {Lee}, {Rhee}, {Bianchi}, {Madore}, {Lee}, {Barlow}, {Forster}, {Friedman},
  {Martin}, {Morrissey}, {Neff}, {Schiminovich}, {Seibert}, {Small}, {Wyder},
  {Donas}, {Heckman}, {Milliard}, {Szalay}, \& {Welsh}}]{2007ApJS..173..643R}
{Rey} S.-C., {Rich} R.~M., {Sohn} S.~T., {Yoon} S.-J., {Chung} C., {Yi} S.~K.,
  {Lee} Y.-W., {Rhee} J., {Bianchi} L., {Madore} B.~F., {Lee} K., {Barlow}
  T.~A., {Forster} K., {Friedman} P.~G., {Martin} D.~C., {Morrissey} P., {Neff}
  S.~G., {Schiminovich} D., {Seibert} M., {Small} T., {Wyder} T.~K., {Donas}
  J., {Heckman} T.~M., {Milliard} B., {Szalay} A.~S., {Welsh} B.~Y., 2007,
  \apjs, 173, 643

\bibitem[{{Roediger} {et~al.}(2014){Roediger}, {Courteau}, {Graves}, \&
  {Schiavon}}]{2014ApJS..210...10R}
{Roediger} J.~C., {Courteau} S., {Graves} G., {Schiavon} R.~P., 2014, \apjs,
  210, 10

\bibitem[{{Romanowsky} {et~al.}(2009){Romanowsky}, {Strader}, {Spitler},
  {Johnson}, {Brodie}, {Forbes}, \& {Ponman}}]{2009AJ....137.4956R}
{Romanowsky} A.~J., {Strader} J., {Spitler} L.~R., {Johnson} R., {Brodie}
  J.~P., {Forbes} D.~A., {Ponman} T., 2009, \aj, 137, 4956

\bibitem[{{Salpeter}(1955)}]{1955ApJ...121..161S}
{Salpeter} E.~E., 1955, \apj, 121, 161

\bibitem[{{S{\'a}nchez-Bl{\'a}zquez} {et~al.}(2006){S{\'a}nchez-Bl{\'a}zquez},
  {Peletier}, {Jim{\'e}nez-Vicente}, {Cardiel}, {Cenarro},
  {Falc{\'o}n-Barroso}, {Gorgas}, {Selam}, \& {Vazdekis}}]{2006MNRAS.371..703S}
{S{\'a}nchez-Bl{\'a}zquez} P., {Peletier} R.~F., {Jim{\'e}nez-Vicente} J.,
  {Cardiel} N., {Cenarro} A.~J., {Falc{\'o}n-Barroso} J., {Gorgas} J., {Selam}
  S., {Vazdekis} A., 2006, \mnras, 371, 703

\bibitem[{{Schiavon} {et~al.}(2013){Schiavon}, {Caldwell}, {Conroy}, {Graves},
  {Strader}, {MacArthur}, {Courteau}, \& {Harding}}]{2013ApJ...776L...7S}
{Schiavon} R.~P., {Caldwell} N., {Conroy} C., {Graves} G.~J., {Strader} J.,
  {MacArthur} L.~A., {Courteau} S., {Harding} P., 2013, \apjl, 776, L7

\bibitem[{{Sharina} {et~al.}(2006){Sharina}, {Afanasiev}, \&
  {Puzia}}]{2006MNRAS.372.1259S}
{Sharina} M.~E., {Afanasiev} V.~L., {Puzia} T.~H., 2006, \mnras, 372, 1259

\bibitem[{{Skrutskie} {et~al.}(2006){Skrutskie}, {Cutri}, {Stiening},
  {Weinberg}, {Schneider}, {Carpenter}, {Beichman}, {Capps}, {Chester},
  {Elias}, {Huchra}, {Liebert}, {Lonsdale}, {Monet}, {Price}, {Seitzer},
  {Jarrett}, {Kirkpatrick}, {Gizis}, {Howard}, {Evans}, {Fowler}, {Fullmer},
  {Hurt}, {Light}, {Kopan}, {Marsh}, {McCallon}, {Tam}, {Van Dyk}, \&
  {Wheelock}}]{2006AJ....131.1163S}
{Skrutskie} M.~F., {Cutri} R.~M., {Stiening} R., {Weinberg} M.~D., {Schneider}
  S., {Carpenter} J.~M., {Beichman} C., {Capps} R., {Chester} T., {Elias} J.,
  {Huchra} J., {Liebert} J., {Lonsdale} C., {Monet} D.~G., {Price} S.,
  {Seitzer} P., {Jarrett} T., {Kirkpatrick} J.~D., {Gizis} J.~E., {Howard} E.,
  {Evans} T., {Fowler} J., {Fullmer} L., {Hurt} R., {Light} R., {Kopan} E.~L.,
  {Marsh} K.~A., {McCallon} H.~L., {Tam} R., {Van Dyk} S., {Wheelock} S., 2006,
  \aj, 131, 1163

\bibitem[{{Smith} {et~al.}(2012){Smith}, {Lucey}, {Price}, {Hudson}, \&
  {Phillipps}}]{2012MNRAS.419.3167S}
{Smith} R.~J., {Lucey} J.~R., {Price} J., {Hudson} M.~J., {Phillipps} S., 2012,
  \mnras, 419, 3167

\bibitem[{{Sohn} {et~al.}(2006){Sohn}, {O'Connell}, {Kundu}, {Landsman},
  {Burstein}, {Bohlin}, {Frogel}, \& {Rose}}]{2006AJ....131..866S}
{Sohn} S.~T., {O'Connell} R.~W., {Kundu} A., {Landsman} W.~B., {Burstein} D.,
  {Bohlin} R.~C., {Frogel} J.~A., {Rose} J.~A., 2006, \aj, 131, 866

\bibitem[{{Spitler} {et~al.}(2006){Spitler}, {Larsen}, {Strader}, {Brodie},
  {Forbes}, \& {Beasley}}]{2006AJ....132.1593S}
{Spitler} L.~R., {Larsen} S.~S., {Strader} J., {Brodie} J.~P., {Forbes} D.~A.,
  {Beasley} M.~A., 2006, \aj, 132, 1593

\bibitem[{{Strader} {et~al.}(2005){Strader}, {Brodie}, {Cenarro}, {Beasley}, \&
  {Forbes}}]{2005AJ....130.1315S}
{Strader} J., {Brodie} J.~P., {Cenarro} A.~J., {Beasley} M.~A., {Forbes} D.~A.,
  2005, \aj, 130, 1315

\bibitem[{{Strader} {et~al.}(2006){Strader}, {Brodie}, {Spitler}, \&
  {Beasley}}]{2006AJ....132.2333S}
{Strader} J., {Brodie} J.~P., {Spitler} L., {Beasley} M.~A., 2006, \aj, 132,
  2333

\bibitem[{{Strader} {et~al.}(2011{\natexlab{a}}){Strader}, {Caldwell}, \&
  {Seth}}]{2011AJ....142....8S}
{Strader} J., {Caldwell} N., {Seth} A.~C., 2011{\natexlab{a}}, \aj, 142, 8

\bibitem[{{Strader} {et~al.}(2012){Strader}, {Fabbiano}, {Luo}, {Kim},
  {Brodie}, {Fragos}, {Gallagher}, {Kalogera}, {King}, \&
  {Zezas}}]{2012ApJ...760...87S}
{Strader} J., {Fabbiano} G., {Luo} B., {Kim} D.-W., {Brodie} J.~P., {Fragos}
  T., {Gallagher} J.~S., {Kalogera} V., {King} A., {Zezas} A., 2012, \apj, 760,
  87

\bibitem[{{Strader} {et~al.}(2011{\natexlab{b}}){Strader}, {Romanowsky},
  {Brodie}, {Spitler}, {Beasley}, {Arnold}, {Tamura}, {Sharples}, \&
  {Arimoto}}]{2011ApJS..197...33S}
{Strader} J., {Romanowsky} A.~J., {Brodie} J.~P., {Spitler} L.~R., {Beasley}
  M.~A., {Arnold} J.~A., {Tamura} N., {Sharples} R.~M., {Arimoto} N.,
  2011{\natexlab{b}}, \apjs, 197, 33

\bibitem[{{Strader} {et~al.}(2013){Strader}, {Seth}, {Forbes}, {Fabbiano},
  {Romanowsky}, {Brodie}, {Conroy}, {Caldwell}, {Pota}, {Usher}, \&
  {Arnold}}]{2013ApJ...775L...6S}
{Strader} J., {Seth} A.~C., {Forbes} D.~A., {Fabbiano} G., {Romanowsky} A.~J.,
  {Brodie} J.~P., {Conroy} C., {Caldwell} N., {Pota} V., {Usher} C., {Arnold}
  J.~A., 2013, \apjl, 775, L6

\bibitem[{{Strader} \& {Smith}(2008)}]{2008AJ....136.1828S}
{Strader} J., {Smith} G.~H., 2008, \aj, 136, 1828

\bibitem[{{Taylor}(2005)}]{2005ASPC..347...29T}
{Taylor} M.~B., 2005, in Astronomical Society of the Pacific Conference Series,
  Vol. 347, Astronomical Data Analysis Software and Systems XIV, {Shopbell} P.,
  {Britton} M., {Ebert} R., eds., p.~29

\bibitem[{{Terlevich} \& {Forbes}(2002)}]{2002MNRAS.330..547T}
{Terlevich} A.~I., {Forbes} D.~A., 2002, \mnras, 330, 547

\bibitem[{{Thomas} {et~al.}(2003){Thomas}, {Maraston}, \&
  {Bender}}]{2003MNRAS.339..897T}
{Thomas} D., {Maraston} C., {Bender} R., 2003, \mnras, 339, 897

\bibitem[{{Thomas} {et~al.}(2005){Thomas}, {Maraston}, {Bender}, \& {Mendes de
  Oliveira}}]{2005ApJ...621..673T}
{Thomas} D., {Maraston} C., {Bender} R., {Mendes de Oliveira} C., 2005, \apj,
  621, 673

\bibitem[{{Tonry} {et~al.}(2001){Tonry}, {Dressler}, {Blakeslee}, {Ajhar},
  {Fletcher}, {Luppino}, {Metzger}, \& {Moore}}]{2001ApJ...546..681T}
{Tonry} J.~L., {Dressler} A., {Blakeslee} J.~P., {Ajhar} E.~A., {Fletcher}
  A.~B., {Luppino} G.~A., {Metzger} M.~R., {Moore} C.~B., 2001, \apj, 546, 681

\bibitem[{{Usher} {et~al.}(2012){Usher}, {Forbes}, {Brodie}, {Foster},
  {Spitler}, {Arnold}, {Romanowsky}, {Strader}, \&
  {Pota}}]{2012MNRAS.426.1475U}
{Usher} C., {Forbes} D.~A., {Brodie} J.~P., {Foster} C., {Spitler} L.~R.,
  {Arnold} J.~A., {Romanowsky} A.~J., {Strader} J., {Pota} V., 2012, \mnras,
  426, 1475

\bibitem[{{Usher} {et~al.}(2013){Usher}, {Forbes}, {Spitler}, {Brodie},
  {Romanowsky}, {Strader}, \& {Woodley}}]{2013MNRAS.436.1172U}
{Usher} C., {Forbes} D.~A., {Spitler} L.~R., {Brodie} J.~P., {Romanowsky}
  A.~J., {Strader} J., {Woodley} K.~A., 2013, \mnras, 436, 1172

\bibitem[{{Valdes} {et~al.}(2004){Valdes}, {Gupta}, {Rose}, {Singh}, \&
  {Bell}}]{2004ApJS..152..251V}
{Valdes} F., {Gupta} R., {Rose} J.~A., {Singh} H.~P., {Bell} D.~J., 2004,
  \apjs, 152, 251

\bibitem[{{VandenBerg} {et~al.}(2013){VandenBerg}, {Brogaard}, {Leaman}, \&
  {Casagrande}}]{2013ApJ...775..134V}
{VandenBerg} D.~A., {Brogaard} K., {Leaman} R., {Casagrande} L., 2013, \apj,
  775, 134

\bibitem[{{Vanderbeke} {et~al.}(2014{\natexlab{a}}){Vanderbeke}, {West}, {De
  Propris}, {Peng}, {Blakeslee}, {Jord{\'a}n}, {C{\^o}t{\'e}}, {Gregg},
  {Ferrarese}, {Takamiya}, \& {Baes}}]{2014MNRAS.437.1725V}
{Vanderbeke} J., {West} M.~J., {De Propris} R., {Peng} E.~W., {Blakeslee}
  J.~P., {Jord{\'a}n} A., {C{\^o}t{\'e}} P., {Gregg} M., {Ferrarese} L.,
  {Takamiya} M., {Baes} M., 2014{\natexlab{a}}, \mnras, 437, 1725

\bibitem[{{Vanderbeke} {et~al.}(2014{\natexlab{b}}){Vanderbeke}, {West}, {De
  Propris}, {Peng}, {Blakeslee}, {Jord{\'a}n}, {C{\^o}t{\'e}}, {Gregg},
  {Ferrarese}, {Takamiya}, \& {Baes}}]{2014MNRAS.437.1734V}
---, 2014{\natexlab{b}}, \mnras, 437, 1734

\bibitem[{{Vazdekis} {et~al.}(2003){Vazdekis}, {Cenarro}, {Gorgas}, {Cardiel},
  \& {Peletier}}]{2003MNRAS.340.1317V}
{Vazdekis} A., {Cenarro} A.~J., {Gorgas} J., {Cardiel} N., {Peletier} R.~F.,
  2003, \mnras, 340, 1317

\bibitem[{{Vazdekis} {et~al.}(2012){Vazdekis}, {Ricciardelli}, {Cenarro},
  {Rivero-Gonz{\'a}lez}, {D{\'{\i}}az-Garc{\'{\i}}a}, \&
  {Falc{\'o}n-Barroso}}]{2012MNRAS.424..157V}
{Vazdekis} A., {Ricciardelli} E., {Cenarro} A.~J., {Rivero-Gonz{\'a}lez} J.~G.,
  {D{\'{\i}}az-Garc{\'{\i}}a} L.~A., {Falc{\'o}n-Barroso} J., 2012, \mnras,
  424, 157

\bibitem[{{Vazdekis} {et~al.}(2010){Vazdekis}, {S{\'a}nchez-Bl{\'a}zquez},
  {Falc{\'o}n-Barroso}, {Cenarro}, {Beasley}, {Cardiel}, {Gorgas}, \&
  {Peletier}}]{2010MNRAS.404.1639V}
{Vazdekis} A., {S{\'a}nchez-Bl{\'a}zquez} P., {Falc{\'o}n-Barroso} J.,
  {Cenarro} A.~J., {Beasley} M.~A., {Cardiel} N., {Gorgas} J., {Peletier}
  R.~F., 2010, \mnras, 404, 1639

\bibitem[{{Villegas} {et~al.}(2010){Villegas}, {Jord{\'a}n}, {Peng},
  {Blakeslee}, {C{\^o}t{\'e}}, {Ferrarese}, {Kissler-Patig}, {Mei}, {Infante},
  {Tonry}, \& {West}}]{2010ApJ...717..603V}
{Villegas} D., {Jord{\'a}n} A., {Peng} E.~W., {Blakeslee} J.~P., {C{\^o}t{\'e}}
  P., {Ferrarese} L., {Kissler-Patig} M., {Mei} S., {Infante} L., {Tonry}
  J.~L., {West} M.~J., 2010, \apj, 717, 603

\bibitem[{{Woodley} {et~al.}(2010{\natexlab{a}}){Woodley}, {G{\'o}mez},
  {Harris}, {Geisler}, \& {Harris}}]{2010AJ....139.1871W}
{Woodley} K.~A., {G{\'o}mez} M., {Harris} W.~E., {Geisler} D., {Harris}
  G.~L.~H., 2010{\natexlab{a}}, \aj, 139, 1871

\bibitem[{{Woodley} {et~al.}(2010{\natexlab{b}}){Woodley}, {Harris}, {Puzia},
  {G{\'o}mez}, {Harris}, \& {Geisler}}]{2010ApJ...708.1335W}
{Woodley} K.~A., {Harris} W.~E., {Puzia} T.~H., {G{\'o}mez} M., {Harris}
  G.~L.~H., {Geisler} D., 2010{\natexlab{b}}, \apj, 708, 1335

\bibitem[{{Worthey} {et~al.}(1994){Worthey}, {Faber}, {Gonzalez}, \&
  {Burstein}}]{1994ApJS...94..687W}
{Worthey} G., {Faber} S.~M., {Gonzalez} J.~J., {Burstein} D., 1994, \apjs, 94,
  687

\end{thebibliography}

\appendix
\section{Colour and spectral index measurements}

\begin{landscape}

\begin{table*}
\caption{\label{tab:by_colour} Colour and spectral index measurements of spectra stacked by colour}
\begin{tabular}{c c c c c c c c c c} \hline
Galaxy & $(g - i)$ & $M_{i}$ & CaT   & Fe86  & Mg88  & Na82  & TiO89 & H$\alpha$ & PaT   \\
       & [mag]     & [mag]   & [\AA] & [\AA] & [\AA] & [\AA] &       & [\AA]     & [\AA] \\
(1)    & (2)       & (3)     & (4)   & (5)   & (6)   & (7)   & (8)   & (9)       & (10)  \\ \hline
NGC 1407 & $0.805_{-0.012}^{+0.005}$ & $-10.865_{-0.012}^{+0.363}$ & $4.672_{-0.113}^{+0.242}$ & $1.669_{-0.245}^{+0.183}$ & $0.176_{-0.102}^{+0.080}$ & $0.248_{-0.241}^{+0.191}$ & $0.9883_{-0.0064}^{+0.0118}$ & $2.270_{-0.165}^{+0.143}$ & $1.100_{-0.232}^{+0.702}$ \\
NGC 1407 & $0.856_{-0.014}^{+0.009}$ & $-10.899_{-0.216}^{+0.185}$ & $5.455_{-0.307}^{+0.214}$ & $1.841_{-0.190}^{+0.273}$ & $0.520_{-0.179}^{+0.052}$ & $-0.165_{-0.115}^{+0.208}$ & $0.9979_{-0.0164}^{+0.0070}$ & $2.228_{-0.222}^{+0.124}$ & $1.370_{-0.566}^{+0.158}$ \\
NGC 1407 & $0.895_{-0.009}^{+0.009}$ & $-11.153_{-0.134}^{+0.084}$ & $5.804_{-0.100}^{+0.204}$ & $2.543_{-0.262}^{+0.123}$ & $0.457_{-0.050}^{+0.090}$ & $-0.012_{-0.090}^{+0.135}$ & $0.9892_{-0.0045}^{+0.0114}$ & $1.931_{-0.130}^{+0.243}$ & $1.008_{-0.237}^{+0.239}$ \\ 
 ...   & ...       & ...     & ...   & ...   & ...   & ...   & ...   & ...       & ...   \\ \hline
\end{tabular}

\medskip
\emph{Notes}
The full version of this table is provided in a machine readable form in the online Supporting Information.
Column (1): Galaxy.
Column (2): $(g - i)$ colour.
Column (3): $i$-band absolute magnitude.
Column (4): CaT index strength.
Column (5): Fe86 index strength.
Column (6): Mg88 index strength.
Column (7): Na82 index strength.
Column (8): TiO89 index strength.
Column (9): H$\alpha$ index strength.
Column (10): PaT index strength.
\end{table*}

\begin{table*}
\caption{\label{tab:by_mag} Colour and spectral index measurements of spectra stacked by colour}
\begin{tabular}{c c c c c c c c c c c} \hline
Galaxy & $(g - i)$ & $M_{i}$ & CaT   & Fe86  & Mg88  & Na82  & TiO89 & H$\alpha$ & PaT   & $\sigma$      \\
       & [mag]     & [mag]   & [\AA] & [\AA] & [\AA] & [\AA] &       & [\AA]     & [\AA] & [km s$^{-1}$] \\
(1)    & (2)       & (3)     & (4)   & (5)   & (6)   & (7)   & (8)   & (9)       & (10)  & (11)          \\ \hline
Faint & $0.8207_{-0.0089}^{+0.0097}$ & $-10.494_{-0.128}^{+0.099}$ & $5.995_{-0.168}^{+0.133}$ & $2.328_{-0.138}^{+0.122}$ & $0.436_{-0.068}^{+0.046}$ & $0.443_{-0.074}^{+0.058}$ & $0.9920_{-0.0021}^{+0.0024}$ & $1.931_{-0.047}^{+0.069}$ & $1.036_{-0.204}^{+0.133}$ & $16.82_{-1.75}^{+1.95}$ \\
Faint & $0.8060_{-0.0055}^{+0.0060}$ & $-9.708_{-0.111}^{+0.108}$ & $5.604_{-0.177}^{+0.155}$ & $2.171_{-0.123}^{+0.118}$ & $0.391_{-0.058}^{+0.063}$ & $0.459_{-0.043}^{+0.073}$ & $0.9916_{-0.0048}^{+0.0036}$ & $2.050_{-0.138}^{+0.050}$ & $1.116_{-0.169}^{+0.133}$ & $15.81_{-1.72}^{+0.89}$ \\
Faint & $0.8019_{-0.0092}^{+0.0085}$ & $-9.039_{-0.098}^{+0.104}$ & $5.560_{-0.201}^{+0.206}$ & $2.426_{-0.216}^{+0.129}$ & $0.278_{-0.097}^{0.099}$ & $0.541_{-0.140}^{+0.058}$ & $0.9930_{-0.0081}^{+0.0076}$ & $2.104_{-0.172}^{+0.103}$ & $0.998_{-0.294}^{+0.237}$ & $13.14_{-2.10}^{+2.55}$ \\
 ...   & ...       & ...     & ...   & ...   & ...   & ...   & ...   & ...       & ...                   \\ \hline
\end{tabular}

\medskip
\emph{Notes}
The full version of this table is provided in a machine readable form in the online Supporting Information.
Column (1): Galaxy.
Column (2): $(g - i)$ colour.
Column (3): $i$-band absolute magnitude.
Column (4): CaT index strength.
Column (5): Fe86 index strength.
Column (6): Mg88 index strength.
Column (7): Na82 index strength.
Column (8): TiO89 index strength.
Column (9): H$\alpha$ index strength.
Column (10): PaT index strength.
Column (11): Velocity dispersion.
\end{table*}

\end{landscape}

\section{Colour and spectral index relations with magnitude}

\begin{table*}
\caption{\label{tab:mag_relations} Colour and spectral index relations with magnitude}
\begin{tabular}{c c c c c c c} \hline
Sample & Feature & Slope & Intercept & $p_{slope}$ & $M_{i} = -10$ & $p_{diff}$\\
(1)    & (2)     & (3)   & (4)       & (5)         & (6)          & (7)       \\ \hline
Faint Blue & $(g - i)$ & \boldmath{ $ -0.0148^{+0.0051}_{-0.0062}$ } & \boldmath{ $ \hphantom{-}0.665^{+0.050}_{-0.058}$ } & \textbf{ 0.999 } &$ 0.8128^{+0.0056}_{-0.0045}$ & 0.823 \\ 
Bright Blue & $(g - i)$ & \boldmath{ $ -0.0115^{+0.0064}_{-0.0086}$ } & \boldmath{ $ \hphantom{-}0.692^{+0.066}_{-0.090}$ } & \textbf{ 0.963 } &$ 0.8074^{+0.0039}_{-0.0044}$ & 0.823 \\ 
Faint Red & $(g - i)$ & $ \hphantom{-}0.0024^{+0.0051}_{-0.0055} $ & $ \hphantom{-}1.093^{+0.049}_{-0.052} $ & 0.320 &$ 1.0688^{+0.0054}_{-0.0055}$ & 0.064 \\ 
Bright Red & $(g - i)$ & $ \hphantom{-}0.0051^{+0.0040}_{-0.0033} $ & $ \hphantom{-}1.130^{+0.043}_{-0.036} $ & 0.075 &$ 1.0789^{+0.0048}_{-0.0038}$ & 0.064 \\ 
Faint Blue & CaT & \boldmath{ $ -0.242^{+0.137}_{-0.084}$ } & \boldmath{ $ \hphantom{-}3.38^{+1.34}_{-0.80}$ } & \textbf{ 0.959 } &\boldmath{ $ 5.797^{+0.094}_{-0.096}$ } & \textbf{ 1.000 } \\ 
Bright Blue & CaT & \boldmath{ $ -0.282^{+0.174}_{-0.140}$ } & \boldmath{ $ \hphantom{-}2.23^{+1.84}_{-1.41}$ } & \textbf{ 0.955 } &\boldmath{ $ 5.046^{+0.127}_{-0.061}$ } & \textbf{ 1.000 } \\ 
Faint Red & CaT & $ -0.117^{+0.097}_{-0.118} $ & $ \hphantom{-}6.47^{+1.00}_{-1.11} $ & 0.876 &\boldmath{ $ 7.643^{+0.108}_{-0.057}$ } & \textbf{ 0.998 } \\ 
Bright Red & CaT & $ \hphantom{-}0.115^{+0.083}_{-0.083} $ & $ \hphantom{-}8.37^{+0.94}_{-0.88} $ & 0.070 &\boldmath{ $ 7.224^{+0.123}_{-0.089}$ } & \textbf{ 0.998 } \\ 
Faint Blue & Fe86 & $ -0.105^{+0.117}_{-0.098} $ & $ \hphantom{-}1.24^{+1.13}_{-0.95} $ & 0.802 &\boldmath{ $ 2.291^{+0.070}_{-0.089}$ } & \textbf{ 0.988 } \\ 
Bright Blue & Fe86 & $ -0.101^{+0.144}_{-0.136} $ & $ \hphantom{-}0.98^{+1.52}_{-1.44} $ & 0.763 &\boldmath{ $ 1.985^{+0.108}_{-0.118}$ } & \textbf{ 0.988 } \\ 
Faint Red & Fe86 & $ \hphantom{-}0.081^{+0.065}_{-0.118} $ & $ \hphantom{-}4.54^{+0.66}_{-1.20} $ & 0.276 &\boldmath{ $ 3.729^{+0.055}_{-0.072}$ } & \textbf{ 0.995 } \\ 
Bright Red & Fe86 & $ -0.027^{+0.097}_{-0.066} $ & $ \hphantom{-}3.12^{+1.09}_{-0.73} $ & 0.570 &\boldmath{ $ 3.395^{+0.127}_{-0.096}$ } & \textbf{ 0.995 } \\ 
Faint Blue & Mg88 & $ -0.062^{+0.061}_{-0.051} $ & $ -0.22^{+0.60}_{-0.49} $ & 0.851 &\boldmath{ $ 0.396^{+0.041}_{-0.039}$ } & \textbf{ 0.997 } \\ 
Bright Blue & Mg88 & \boldmath{ $ -0.100^{+0.051}_{-0.063}$ } & \boldmath{ $ -0.75^{+0.54}_{-0.67}$ } & \textbf{ 0.963 } &\boldmath{ $ 0.250^{+0.045}_{-0.051}$ } & \textbf{ 0.997 } \\ 
Faint Red & Mg88 & $ -0.017^{+0.043}_{-0.036} $ & $ \hphantom{-}0.57^{+0.43}_{-0.36} $ & 0.661 &$ 0.737^{+0.036}_{-0.026}$ & 0.946 \\ 
Bright Red & Mg88 & $ -0.003^{+0.028}_{-0.029} $ & $ \hphantom{-}0.63^{+0.32}_{-0.33} $ & 0.558 &$ 0.660^{+0.042}_{-0.043}$ & 0.946 \\ 
Faint Blue & Na82 & $ \hphantom{-}0.018^{+0.077}_{-0.051} $ & $ \hphantom{-}0.65^{+0.76}_{-0.50} $ & 0.315 &\boldmath{ $ 0.464^{+0.038}_{-0.043}$ } & \textbf{ 0.996 } \\ 
Bright Blue & Na82 & $ -0.067^{+0.061}_{-0.065} $ & $ -0.41^{+0.66}_{-0.68} $ & 0.870 &\boldmath{ $ 0.263^{+0.055}_{-0.053}$ } & \textbf{ 0.996 } \\ 
Faint Red & Na82 & $ -0.040^{+0.047}_{-0.040} $ & $ \hphantom{-}0.29^{+0.46}_{-0.40} $ & 0.801 &$ 0.686^{+0.037}_{-0.048}$ & 0.432 \\ 
Bright Red & Na82 & $ \hphantom{-}0.040^{+0.043}_{-0.053} $ & $ \hphantom{-}1.10^{+0.48}_{-0.60} $ & 0.228 &$ 0.705^{+0.056}_{-0.069}$ & 0.432 \\ 
Faint Blue & TiO89 & $ -0.0040^{+0.0033}_{-0.0035} $ & $ \hphantom{-}0.952^{+0.032}_{-0.035} $ & 0.902 &$ 0.9920^{+0.0017}_{-0.0025}$ & 0.101 \\ 
Bright Blue & TiO89 & $ \hphantom{-}0.0065^{+0.0048}_{-0.0061} $ & $ \hphantom{-}1.065^{+0.051}_{-0.066} $ & 0.152 &$ 0.9995^{+0.0042}_{-0.0054}$ & 0.101 \\ 
Faint Red & TiO89 & $ \hphantom{-}0.0000^{+0.0034}_{-0.0027} $ & $ \hphantom{-}1.010^{+0.034}_{-0.028} $ & 0.427 &$ 1.0094^{+0.0023}_{-0.0028}$ & 0.939 \\ 
Bright Red & TiO89 & $ -0.0029^{+0.0026}_{-0.0028} $ & $ \hphantom{-}0.973^{+0.030}_{-0.030} $ & 0.874 &$ 1.0021^{+0.0037}_{-0.0037}$ & 0.939 \\ 
Faint Blue & H$\alpha$ & $ \hphantom{-}0.072^{+0.124}_{-0.067} $ & $ \hphantom{-}2.72^{+1.23}_{-0.66} $ & 0.138 &$ 1.997^{+0.027}_{-0.064}$ & 0.162 \\ 
Bright Blue & H$\alpha$ & $ \hphantom{-}0.119^{+0.084}_{-0.131} $ & $ \hphantom{-}3.29^{+0.90}_{-1.41} $ & 0.176 &$ 2.096^{+0.070}_{-0.116}$ & 0.162 \\ 
Faint Red & H$\alpha$ & $ -0.106^{+0.078}_{-0.115} $ & $ \hphantom{-}0.91^{+0.77}_{-1.17} $ & 0.920 &$ 1.970^{+0.075}_{-0.064}$ & 0.916 \\ 
Bright Red & H$\alpha$ & $ -0.105^{+0.098}_{-0.077} $ & $ \hphantom{-}0.72^{+1.11}_{-0.88} $ & 0.849 &$ 1.768^{+0.139}_{-0.105}$ & 0.916 \\ 
Faint Blue & PaT & $ -0.32^{+0.23}_{-0.10} $ & $ -2.2^{+2.2}_{-1.0} $ & 0.942 &$ 1.053^{+0.088}_{-0.136}$ & 0.589 \\ 
Bright Blue & PaT & $ -0.06^{+0.24}_{-0.21} $ & $ \hphantom{-}0.4^{+2.5}_{-2.3} $ & 0.602 &$ 0.976^{+0.188}_{-0.192}$ & 0.589 \\ 
Faint Red & PaT & $ -0.09^{+0.22}_{-0.13} $ & $ -0.2^{+2.2}_{-1.3} $ & 0.606 &$ 0.677^{+0.127}_{-0.116}$ & 0.350 \\ 
Bright Red & PaT & $ -0.13^{+0.17}_{-0.12} $ & $ -0.5^{+1.9}_{-1.4} $ & 0.799 &$ 0.764^{+0.238}_{-0.174}$ & 0.350 \\ \hline
\end{tabular}

\medskip
\emph{Notes}
Column (1): Sample. Blue is spectra in the colour range $0.72 < (g -i) < 0.88$; Red is spectra in the colour range $1.00 < (g - i) < 1.16$; Faint is the spectra from NGC 3115, NGC 3377, NGC 4473 and NGC 4494; Bright is the spectra from NGC 1407, NGC 4278, NGC 4365 and NGC 4649.
Column (2): Spectral feature or colour.
Column (3): Slope of the fitted linear relation between the spectral feature in Column (2) and magnitude for the sample in Column (1). The uncertainties for Columns (3), (4) and (6) were estimated using bootstrapping.
Column (4): Intercept for the fitted linear relation.
Column (5): Probability that the slope is less than zero from bootstrapping.
Column (6): Spectral feature value or colour at $M_{i} = -10$.
Column (7): Probability that the value at $M_{i} = -10$ is greater in the faint group than in bright galaxy group.
Samples and features with significant slopes or differences ($p < 0.05$ or $p > 0.95$) are in bold.
\end{table*}

\end{document}